\documentclass[prd,aps,showpacs,epsf,floats,onecolumn]{revtex4}%
\usepackage{amssymb}
\usepackage{amsfonts}
\usepackage{amsmath}
\usepackage{graphicx}%
\providecommand{\U}[1]{\protect\rule{.1in}{.1in}}

\begin{document}
\title{\textbf{Geometric Algebra and Information Geometry for Quantum Computational
Software}}
\author{\textbf{Carlo Cafaro}}
\affiliation{SUNY Polytechnic Institute, 12203 Albany, New York, USA }

\begin{abstract}
The art of quantum algorithm design is highly nontrivial. Grover's search
algorithm constitutes a masterpiece of quantum computational software. In this
article, we use methods of geometric algebra (GA) and information geometry
(IG) to enhance the algebraic efficiency and the geometrical significance of
the digital and analog representations of Grover's algorithm, respectively.
Specifically, GA is used to describe the Grover iterate and the discretized
iterative procedure that exploits quantum interference to amplify the
probability amplitude of the target-state before measuring the query register.
The transition from digital to analog descriptions occurs via Stone's theorem
which relates the (unitary) Grover iterate to a suitable (Hermitian)
Hamiltonian that controls Schrodinger's quantum mechanical evolution of a
quantum state towards the target state. Once the discrete-to-continuos
transition is completed, IG is used to interpret Grover's iterative procedure
as a geodesic path on the manifold of the parametric density operators of pure
quantum states constructed from the continuous approximation of the parametric
quantum output state in Grover's algorithm. Finally, we discuss the
dissipationless nature of quantum computing, recover the quadratic speedup
relation, and identify the superfluity of the Walsh-Hadamard operation from an
IG\ perspective with emphasis on statistical mechanical considerations.

\end{abstract}

\maketitle

\bigskip\pagebreak

\section{Introduction}

The goal of characterizing the problems that can be solved on a quantum
computer \cite{feynman82}, together with the efficiency with which problems
can be solved, is of enormous theoretical interest and practical importance
\cite{divincenzo00}. This goal, however, seems a daunting one. A quantum
computer is a physical device wherein\textbf{ }quantum algorithms are
executed. Quantum algorithms are described in terms of quantum circuits that
are constructed by assembling wires and discrete sets of quantum gates which
carry information around and perform computational procedures, respectively.
In realistic scenarios, algorithms are executed on imperfect physical
components that implement mathematical operations. Imperfections and noise
lead to errors that cause loss of efficiency and this may reduce the
effectiveness of the information processing device. Error models, also known
as channels, provide a mathematical description of errors occurring on bits
being moved around from one point of the computer to another. Quantum error
correcting codes \cite{gottesman10, carlo10, carlo12, carlo14} can be used to
account for errors introduced by the imperfection of realistic devices at the
cost of a reasonable amount of additional resources. It appears self-evident
that imperfect gates lead to imperfect circuits which, in turn, lead to
imperfect implementations of quantum algorithms. The propagation of
imperfections from gates to circuits in realistic physical implementations of
a quantum computer is likely to have consequences on the scaling of the amount
of basic resources required to synthesize an algorithm.

The fundamental objective of theoretical physics is to describe and, to a
certain extent, understand natural phenomena. To clearly describe a
phenomenon, physicists must master a convenient mathematical language and must
be aware of how it connects to physical reality. Improving the language is not
limited to mathematics, it is indeed one of the main tasks of theoretical
physics. In his Nobel Lecture \cite{feynman72}, Feynman pointed out that,
while working toward the development of the spacetime view of quantum
electrodynamics, he always tried to increase the efficiency of his
demonstrations and to see with more and more clarity why they worked. Enhanced
clarity was achieved by reformulating the same physical concept in many
different mathematical forms. Furthermore, observing that the same physical
reality can be described by many different physical ideas, Feynman concluded
that a good theoretical physicist might find it useful to have a wide range of
physical viewpoints and mathematical expressions for the same theory available
to\textbf{ }her/him. Very challenging and unresolved problems in modern
quantum information science require a very broad vision supported by
quantitative knowledge of both applied mathematical methods and theoretical
physics techniques. It is not unusual to seek answers to questions concerning
a given difficult problem only after having recast it in a novel alternative
form where answers can be found, ideally, in a simpler manner. For instance,
uncovering optimal quantum circuits is of fundamental importance since
computationally intensive problems may require the use of quantum computers.
This consideration motivated Nielsen and collaborators to describe and, to a
certain degree, understand quantum computing as\textbf{ }free fall in a curved
geometry \cite{nielsen06} (for a more explicit analysis, see Ref.
\cite{brandt}): finding the optimal circuit is equivalent to finding the
shortest path between initial and final target points in an appropriate curved geometry.

The art of quantum algorithm design is highly nontrivial: it requires special
techniques and special insights to uncover good algorithms \cite{shor97,
grover97}. Part of the challenge resides in the fact that human beings have a
natural intuition that is better suited for the classical world rather than
the quantum world \cite{nielsenbook}.\textbf{ }In \cite{shor97}, Shor
presented his polynomial time quantum algorithm used for factoring integers
into a product of primes (for instance, $15=3\times5$) and for finding
discrete logarithms \cite{shor97}. From an applied perspective, this algorithm
is essential for testing the security of cryptographic codes \cite{gisin02}.
The first experimental realization of this algorithm was performed by means of
nuclear magnetic resonance (NMR) techniques \cite{chuang01}.

One of the basic problems in computational science is that of searching a very
large database \cite{knuth75}. In \cite{grover97}, Grover presented a quantum
algorithm for solving database search problems. Grover's search algorithm
helps searching for an element in a list of $N$ unsorted elements (for
instance, searching a telephone directory when one knows the phone number but
does not know the person's full name \cite{galindo02}). For the sake of
completeness, we remark that factoring is a problem in the computational
complexity class NP (non-deterministic polynomial time) and it is not known to
be NP-complete. If it was NP-complete, all problems in the complexity class NP
could be efficiently solved using quantum computers \cite{nielsenbook}. The
traveling salesman problem is an example of an NP-complete problem.
Furthermore, we also emphasize that Grover's algorithm does not solve
NP-complete problems in polynomial time, nevertheless its range of
applicability is extremely wide since essentially any difficult problem can be
recast into the form of a search problem (Grover's algorithm solves an
unstructured problem where no assumptions are made on the Boolean function
being inverted). The first experimental implementation of Grover's algorithm
was achieved using NMR methods by\textbf{ }Chuang and collaborators in
\cite{chuang98}.

In this article, we aim to present a novel hybrid geometric characterization
of Grover's search algorithm via the joint application of geometric algebra
(GA, \cite{hestenes66, doran}) and information geometry (IG, \cite{amari}). GA
is a very powerful mathematical language with applications in physics,
computer science, and engineering \cite{doran, lasenby00, hestenes03}. It is
essentially the ordinary Clifford algebra with the powerful addition of a neat
geometric interpretation. Applications of GA in physics span from quantum
theory and gravity \cite{gull98, francis04} to classical electrodynamics
\cite{baylis, ali07, carlo07}. IG is essentially differential geometry applied
to probability calculus \cite{amari}. Applications of IG techniques include
phase transitions in statistical mechanical systems \cite{janke04}, quantum
energy levels statistics \cite{carlomplb}, quantum entanglement \cite{kim11},
quantum aspects of chaoticity \cite{lupo12}, and complex network science
\cite{felice14, franzosi15, franzosi16}. To the best of our knowledge, the
first formal application of (non-geometric) Clifford algebra techniques to
represent quantum algorithms appeared in \cite{gregoric08}. In \cite{alves10},
there appears the first brief GA description of the Grover iterate with no
mention to the GA of quantum states and operators. Using GA techniques, this
latter aspect was addressed in part in \cite{chappell11, chappell12} where
Grover's search process was interpreted as the precession of a spin-$\frac
{1}{2}$ particle. In \cite{pati98, alvarez00, wadati01}, using the
\emph{complex} projective Hilbert space ($%
\mathbb{C}
\mathcal{P}$) of qubits, Grover's search algorithm was recast in differential
geometric language with the notion of distance between pure states quantified
in terms of the Fubini-Study metric. In particular, it was shown that Grover's
dynamics is characterized by a geodesic of $%
\mathbb{C}
\mathcal{P}$. More specifically, the consequences of the removal of the
assumption of temporal independence of the time evolution operator in Grover's
algorithm were investigated in differential geometric terms in \cite{pati98}.
In \cite{alvarez00}, it was shown that, from a statistical standpoint,
Grover's algorithm is described by a unitary and adiabatic process that
preserves\textbf{ }the Fisher information function. In \cite{wadati01}, the
notion of Fubini-Study metric was employed to investigate the role of
entanglement in quantum search.

In \cite{cafaro2011}, we investigated the utility of GA methods in quantum
information science. For instance, using the multiparticle spacetime algebra
(MSTA, that is, the geometric algebra of a relativistic configuration space),
we presented an explicit algebraic description of one and two-qubit quantum
states together with a MSTA characterization of one and two-qubit quantum
computational gates. We concluded that the MSTA approach leads to a useful
conceptual unification where the \emph{complex} qubit space and the
\emph{complex} space of unitary operators acting on them become united, with
both being comprised solely of multivectors in \emph{real} space. We also
reported that the GA approach to rotations based on the rotor group does bring
conceptual and computational advantages compared to standard vectorial and
matrix approaches. In \cite{cafaro2012A, cafaro2012B}, we presented a\textbf{
}preliminary IG characterization of Grover's quantum search algorithm
\cite{grover97}. First, we quantified the notion of quantum distinguishability
between parametric density operators by means of the Wigner-Yanase quantum
information metric. We then showed that the quantum search problem can be
recast in an information geometric framework where Grover's dynamics is
characterized by a geodesic on the manifold of the parametric density
operators of pure quantum states constructed from the continuous approximation
of the parametric quantum output state in Grover's algorithm. Combining the
geometric intuition and the algebraic power of both IG and GA in quantum
search algorithms from a unifying standpoint (digital, digital-to-analog, and
analog viewpoints) remains unexplored.

Here, inspired by the findings reported in \cite{gregoric08, alves10,
chappell11, chappell12, pati98, alvarez00, wadati01} and improving on our
lines of investigations presented in \cite{cafaro2011, cafaro2012A,
cafaro2012B}, we aim to improve upon our understanding of Grover's search
algorithm. We start by observing that the two main physical intuitions that
originally lead Grover to the development of his quantum searching algorithm
were the discretization of Schrodinger's equation and the gravitation towards
lower potential energy regions of a uniform quantum superposition
\cite{grover01}. To sharpen our understanding of his two intuitions, we use in
this article GA and IG methods\textbf{ }to enhance the algebraic efficiency
and the geometrical significance of the digital and analog representations of
Grover's algorithm, respectively. Specifically, GA is used to describe both
the Grover iterate, viewed as a rotation defined by the product of two
reflections (corresponding to the inversion about the mean and the oracle
operators, respectively), as well as the discretized iterative procedure that
exploits quantum interference to amplify the probability amplitude of the
target-state before measuring the query register\textbf{.} The transition from
digital to analog descriptions occurs via Stone's theorem which relates the
(unitary) Grover iterate to a suitable (Hermitian) Hamiltonian that controls
Schrodinger's quantum mechanical evolution of a quantum state towards the
target state. Once the discrete-to-continuos transition is completed, IG is
used to interpret Grover's iterative procedure as a geodesic on the manifold
of the parametric density operators of pure quantum states constructed from
the continuous approximation of the parametric quantum output state in
Grover's algorithm. In particular, we discuss the dissipationless nature of
quantum computing, recover the quadratic speedup relation, and confirm the
superfluity of the Walsh-Hadamard operation from an IG\ perspective with
emphasis on statistical mechanical considerations.

The layout of the article is as follows. In Section II, the concepts of
quantum parallelism, quantum interference, and entanglement responsible for
supra-classical performances in quantum computing are briefly introduced. In
Section III, in preparation for the GA\ language translation and the IG
analysis, we reexamine the standard mathematical steps that characterize
Grover's quantum search algorithm. Special focus is devoted to the quadratic
speedup behavior and to the matrix representations of the Grover iterate
viewed as a rotation defined in terms of a product of two reflections. Some
technical details on the inversion about the mean operator appear in Appendix
A. In Section IV, after introducing the concepts of reflection and rotation in
GA\ terms, we provide a GA description of the Grover iterate and recover the
approximate asymptotic quadric speedup relation. Further details on the GA of
physical space, spacetime, and quantum states and quantum operators appear in
Appendix B. In Section V, we consider the continuous-time Hamiltonian version
of the quantum search problem. We focus on a specific Hamiltonian model which
we study by means of both matrix and geometric algebra techniques. Some
technical details concerning the unitary time-evolution operator appear in
Appendix C. In Section VI, after introducing some basic elements of IG, we
recast the quantum search problem in terms of finding geodesic paths on
manifolds of density matrices. In particular, we discuss the dissipationless
nature of quantum computing, recover the quadratic speedup relation, and
verify the superfluity of the Walsh-Hadamard operation from an
IG\ perspective. In Section VII, we also present several insights on Grover's
fixed point phase-$\frac{\pi}{3}$ quantum search algorithm \cite{grover05A,
grover05B} that arise from either GA or IG standpoints. Some technical details
on the Fisher information function, the mechanical kinetic energy, the
minimization of the action functional in differential geometric terms, and the
steps on information geometric paths appear in Appendix D, Appendix E,
Appendix F, and Appendix G, respectively. Finally, a summary of our findings
together with limitations, improvements, and future lines of investigations
appear in our Final Remarks located in Section VIII.

\section{Quantum mechanical concepts}

In this section, we briefly mention several physical quantum effects
responsible for computational speedups: parallelism, interference, and
entanglement \cite{ekert99, jozsa99, lloyd99, braunstein02, jozsa03,
vedral10}. Recall that the dimensionality of the state space of multiparticle
quantum systems grows exponentially with the number of particles. The reason
why this happens is because in order to model the correlations among
particles, the Hilbert space of a system of distinguishable particles must be
taken as the tensor product of the Hilbert spaces of the individual particles.

\subsection{Parallelism}

Quantum parallelism is the ability of a quantum computer to encode multiple
computational results into a quantum state in a single quantum computational
step \cite{kayebook}. Consider a Boolean function $f$ with an $n$-bit domain
and a one-bit range,%
\begin{equation}
f:\left\{  0,1\right\}  ^{n}\ni x\longmapsto f\left(  x\right)  \in\left\{
0,1\right\}  \text{.}%
\end{equation}
Roughly speaking, quantum computers can evaluate the function $f\left(
x\right)  $ for many different values of $x$ simultaneously thanks to quantum
parallelism that manifests itself in the ability of the computer to be in
superpositions of different states. Note that the same task can be
accomplished on a classical computer using classical parallelism. In the
latter case, unlike the single-circuit quantum scenario, the multiple circuits
built to evaluate $f\left(  x\right)  $ for each value of $x$ are executed
simultaneously. Consider a two qubit quantum computer in the initial state
$\left\vert x,y\right\rangle $ where $x$ and $y$ denote the data and target
registers, respectively. Assume a function $f$ with an $n$-bit input $x$ and
one-bit output. The quantum parallel evaluation of $f$ can be achieved as
follows. First, prepare the $n+1$ qubit state $\left\vert x,y\right\rangle
=\left\vert 0\right\rangle ^{\otimes n}\left\vert 0\right\rangle $. Second,
apply the Walsh-Hadamard transform $H^{\otimes n}=H\otimes...\otimes H$ (that
is, the $n$-fold tensor product of $n$ Hadamard transforms) to the first $n$
qubits. Finally, apply a unitary transformation $U_{f}$ such that%
\begin{equation}
U_{f}\left\vert x,y\right\rangle =\left\vert x,y\oplus f\left(  x\right)
\right\rangle \text{,}%
\end{equation}
where $\oplus$ denotes modulo-$2$ bitwise addition (for instance,
$1101\oplus0111=1010$). The sequence of these operations lead to the state
\cite{kayebook},%
\begin{equation}
\left[  U_{f}\circ\left(  H^{\otimes n}\otimes I\right)  \right]  \left\vert
0\right\rangle ^{\otimes n}\left\vert 0\right\rangle =\frac{1}{\sqrt{2^{n}}%
}\sum_{x=0}^{n-1}\left\vert x\right\rangle \left\vert f\left(  x\right)
\right\rangle \text{,} \label{sup}%
\end{equation}
where $\left\vert x\right\rangle \in\mathcal{H}_{2}^{n}$, $\left\vert f\left(
x\right)  \right\rangle \in\mathcal{H}_{2}^{1}$, $I\overset{\text{def}}%
{=}I_{\mathcal{H}_{2}^{1}}$ is the identity operator on the single-qubit
Hilbert space $\mathcal{H}_{2}^{1}$, and the symbol $\circ$ denotes the
ordinary composition of maps. In the remainder of the article, for the sake of
notational simplicity, we shall suppress use of the composition of maps
symbol. Quantum parallelism allows for a simultaneous evaluation of all
possible values of $f\left(  x\right)  $. However, measuring the superposition
state in Eq. (\ref{sup}) would yield only $f\left(  x\right)  $ for a single
value of $x$. In order to extract some global property of the function
$f\left(  x\right)  $ from the state in Eq. (\ref{sup}), one needs to exploit
the fact that multiple alternatives can interfere with one another. This leads
to the concept of quantum interference.

\subsection{Interference}

Assume that $\left\vert 0\right\rangle $ and $\left\vert 1\right\rangle $ are
the computational basis states while $\alpha$ and $\beta$ are \emph{complex}
numbers whose modulus is not greater than unity. Then, a qubit $\left\vert
q\right\rangle $ is a coherent quantum mechanical superposition that can be
written as,%
\begin{equation}
\left\vert q\right\rangle \overset{\text{def}}{=}\alpha\left\vert
0\right\rangle +\beta\left\vert 1\right\rangle \text{,}%
\end{equation}
where $\alpha$ and $\beta$ are known as probability amplitudes. We note that
while in classical information theory a bit can have only one out of two
alternative outcomes, $0$ or $1$, a qubit can exhibit an infinity of possible
outcomes (states). To a given probability amplitude there corresponds a
probability obtained by computing its modulus squared. Calculating
probabilities in this manner endows quantum computing with the nonclassical
feature of quantum interference. For example, note that to an amplitude
probability $\alpha+\beta$ there corresponds a probability given by,%
\begin{equation}
\left\vert \alpha+\beta\right\vert ^{2}=\left\vert \alpha\right\vert
^{2}+\left\vert \beta\right\vert ^{2}+\alpha^{\ast}\beta+\alpha\beta^{\ast
}\text{,} \label{amplitude}%
\end{equation}
where $\alpha^{\ast}$ denotes the \emph{complex} conjugate of $\alpha$. In the
language of probabilistic classical computation adapted to quantum mechanics,
from Eq. (\ref{amplitude}) one concludes that the probability of the quantum
computational path with amplitude $\alpha+\beta$ depends not only separately
on the paths with amplitude $\alpha$ and $\beta$ but also, and this is the key
aspect, on the interference of such paths \cite{ekert96}. Indeed, it was also
thanks to the constructive interference of quantum amplitudes that Grover was
capable of designing quantum operations enhancing the presence of the searched
target state \cite{galindo02}. A clever choice of the function to be evaluated
together with a final transformation that allows efficient determination of
useful global information about the function itself can be regarded as the
essence of quantum algorithm design \cite{nielsenbook}. For an experimental
realization of the quantum interference phenomenon in terms of photons on a
silicon chip produced from a single ring-resonator source, we refer to
\cite{preble15}.

\subsection{Entanglement}

Entanglement is a physical property of composite quantum systems and emerges
as non-local correlations that are absent in composite classical systems. From
a practical standpoint, consider a bipartite quantum system $S$ composed of
two subsystems $A$ and $B$ which may themselves be composite systems. Let
$\mathcal{H}_{A}$ and $\mathcal{H}_{B}$ be the $N_{A}$-dimensional and $N_{B}%
$-dimensional Hilbert spaces for $A$ and $B$, respectively. If $\left\{
\left\vert a_{i}\right\rangle \right\}  _{i=1,...,N_{A}}$ and $\left\{
\left\vert b_{j}\right\rangle \right\}  _{j=1,...,N_{B}}$ are orthonormal
bases for $\mathcal{H}_{A}$ and $\mathcal{H}_{B}$, respectively, then an
orthonormal basis for $S$ is $\left\{  \left\vert a_{i}b_{j}\right\rangle
\right\}  _{i=1,...,N_{A}\text{; }j=1,...,N_{B}}$. A normalized pure state
$\left\vert \psi\right\rangle \in\mathcal{H}_{S}$ of $S$,%
\begin{equation}
\left\vert \psi\right\rangle \overset{\text{def}}{=}\sum_{i,j}c_{ij}\left\vert
a_{i}b_{j}\right\rangle \text{,}%
\end{equation}
represents an entangled state if it cannot be expressed as a direct product of
states $\left\vert \psi_{A}\right\rangle \in\mathcal{H}_{A}$ and $\left\vert
\psi_{B}\right\rangle \in\mathcal{H}_{B}$,%
\begin{equation}
\left\vert \psi\right\rangle \neq\left\vert \psi_{A}\right\rangle \left\vert
\psi_{B}\right\rangle \text{.}%
\end{equation}
From a conceptual standpoint, entanglement can be understood by investigating
the various ways in which information, one of the most fundamental concepts in
quantum physics, can be distributed within a composite system. First, assume
that information contained in any system, be it composite or individual, is
not infinite. Second, assume that the information contained in a composite
system can be partitioned into the information encoded in the individual
subsystems and into the information carried by the correlations between
observations performed on the individual subsystems. The essence of quantum
entanglement can be explained as follows. For a classical composite system,
knowledge of all properties of each individual subsystem of the composite
system allows one to draw conclusions about how much information is contained
in the correlations. For a quantum entangled system, such a conclusion cannot
be reached any longer: entangled quantum states can carry more information in
joint properties than what may be concluded from knowledge of the individual
subsystems \cite{brukner01}. In recent years, entanglement has undergone
foundational investigations \cite{cafaroijqi}, more theoretically oriented
studies of relativistic nature \cite{alsing12, cafaroijtp}, and very practical
investigations \cite{fanto14}.\textbf{ }For an experimental realization of
quantum entanglement in terms of orbital angular momentum states of photons,
we refer to \cite{mair01}. For a detailed review on quantum entanglement, we
refer to \cite{horodecki09} and \cite{ziman11}.

\section{Grover's iterative procedure}

In what follows, we report what Grover's algorithm roughly does by following
very closely Grover's pedagogical description \cite{grover01}. Take into
consideration the following problem from a crossword puzzle (solution:
computer),%
\begin{equation}
-\text{ }-\text{ M }-\text{ }-\text{ T E }-\text{.}%
\end{equation}
If we limit ourselves to a classical computer and to an online dictionary with
$N=10^{6}$ words arranged alphabetically, we could write a classical software
that finds the solution after inspecting about $N/2=5\times10^{5}$ words.
However, if we use a quantum computer and the same online dictionary, we could
write a quantum software (Grover's quantum search algorithm) that solves the
puzzle in about $\sqrt{N}=10^{3}$ steps. In what follows, we describe the
iterative procedure that characterizes Grover's quantum search algorithm in
standard mathematical terms \cite{grover97, nielsenbook, grover01, kayebook}.

\subsection{Standard mathematical description}

\emph{Step-0}: Initialization of the quantum registers. Prepare the query and
target registers. Suppose we set the target register to $\left\vert
0\right\rangle \in\mathcal{H}_{2}^{1}$ and the query register to $\left\vert
0\right\rangle ^{\otimes n}\in\mathcal{H}_{2}^{n}$ with $N\overset{\text{def}%
}{=}2^{n}$. The composite state of the registers becomes,%
\begin{equation}
\left\vert \Psi_{\text{step-}0}\right\rangle =\left\vert 0\right\rangle
^{\otimes n}\left\vert 0\right\rangle \text{.}%
\end{equation}

\emph{Step-1}: Application of the Walsh-Hadamard transform, $H^{\otimes n}$.
Apply the Walsh-Hadamard transform $H^{\otimes n}$ to the query register and
the transformation $H\sigma_{x}$ to the target register where $\sigma_{x}$ is
the bit-flip operator. The composite state of the registers becomes,%
\begin{equation}
\left\vert \Psi_{\text{step-}1}\right\rangle =\frac{1}{\sqrt{N}}\sum
_{x=0}^{N-1}\left\vert x\right\rangle \left[  \frac{\left\vert 0\right\rangle
-\left\vert 1\right\rangle }{\sqrt{2}}\right]  =\frac{1}{\sqrt{N}}\sum
_{x=0}^{N-1}\left\vert x\right\rangle \left[  \frac{1}{\sqrt{2}}\sum_{y=0}%
^{1}\left(  -1\right)  ^{y}\left\vert y\right\rangle \right]  \text{.}%
\end{equation}
The effect of the Walsh-Hadamard transform is that of generating a uniform
superposition of basis states $\left\{  \left\vert x\right\rangle \right\}
_{x=0,1,...,N-1}$. The effect of $H\sigma_{x}$ on $\left\vert 0\right\rangle $
is that of producing a new target state $H\left\vert 1\right\rangle
=\left\vert -\right\rangle \overset{\text{def}}{=}\left(  \left\vert
0\right\rangle -\left\vert 1\right\rangle \right)  /\sqrt{2}$. The usefulness
of the latter effect will become clear in step-$2$.

\emph{Step-2}: Application of the oracle, $U_{f_{\bar{x}}}$. Assume that the
set of all query values $\mathcal{X}\overset{\text{def}}{=}\left\{
0,1...,N-1\right\}  $ can be partitioned in two sets, $\mathcal{X}%
_{\text{good}}$ and $\mathcal{X}_{\text{bad}}$ with%
\begin{equation}
\mathcal{X}_{\text{good}}\overset{\text{def}}{=}\left\{  \bar{x}\right\}
\text{, with }f_{\bar{x}}\left(  x\right)  =1\text{,}%
\end{equation}
and,%
\begin{equation}
\mathcal{X}_{\text{bad}}\overset{\text{def}}{=}\mathcal{X}\backslash\left\{
\bar{x}\right\}  \text{, with }f_{\bar{x}}\left(  x\right)  =0\text{.}%
\end{equation}
Applying the operator $U_{f_{\bar{x}}}$ to $\left\vert \Psi_{\text{step-}%
1}\right\rangle $, one obtains%
\begin{equation}
\left\vert \Psi_{\text{step-}2}\right\rangle =\frac{1}{\sqrt{2N}}\sum
_{x=0}^{N-1}\left(  -1\right)  ^{f_{\bar{x}}\left(  x\right)  }\left\vert
x\right\rangle \sum_{y=0}^{1}\left\vert y\right\rangle \text{.}%
\end{equation}
Note also that the probability amplitude of $\left\vert \bar{x}\right\rangle $
picks up a $-1$ phase shift after the application of $U_{f_{\bar{x}}}$. As a
consequence, the mean value of the amplitudes slightly decreases. Note that
the flip of the target state in step-$1$ is cleverly exploited in step-$2$
since the target register is in an eigenstate and can be therefore ignored.
This allows one to focus only on the query register. Furthermore, observe that
the action of $U_{f_{\bar{x}}}$ on $\left\vert \Psi_{\text{step-}%
1}\right\rangle $ can be rewritten as,%
\begin{equation}
\left\vert \Psi_{\text{step-}2}\right\rangle =U_{f_{\bar{x}}}\left\vert
\Psi_{\text{step-}1}\right\rangle =\left(  U_{\bar{x}}\otimes I_{\mathcal{H}%
_{2}^{1}}\right)  \left\vert \Psi_{\text{step-}1}\right\rangle \text{.}
\label{util}%
\end{equation}
The quantity $I_{\mathcal{H}_{2}^{1}}$ is the identity operator acting on the
target register while $U_{\bar{x}}$ is the operator that flips the amplitude
of the marked state in the query register and is defined as,%
\begin{equation}
U_{\bar{x}}\left\vert x\right\rangle \overset{\text{def}}{=}\left(
I_{\mathcal{H}_{2}^{n}}-2\left\vert \bar{x}\right\rangle \left\langle \bar
{x}\right\vert \right)  \left\vert x\right\rangle =\left\{
\begin{array}
[c]{c}%
\left\vert x\right\rangle \text{, }x\neq\bar{x}\\
\\
-\left\vert \bar{x}\right\rangle \text{, }x=\bar{x}%
\end{array}
\right.  \text{,}%
\end{equation}
where $I_{\mathcal{H}_{2}^{n}}$ denotes the identity operator acting on the
query register.

\emph{Step-3}: Application of the inversion about the mean operator,
$U_{\psi^{\perp}}$. Focusing on the query register state, consider the state%
\begin{equation}
\left\vert \psi\right\rangle \overset{\text{def}}{=}H^{\otimes n}\left\vert
0\right\rangle ^{\otimes n}=\frac{1}{\sqrt{N}}\sum_{x=0}^{N-1}\left\vert
x\right\rangle \text{,} \label{fidef}%
\end{equation}
together with the vector spaces $V_{\psi}\overset{\text{def}}{=}$%
\emph{Span}$\left\{  \left\vert \psi\right\rangle \right\}  $ and $V_{\psi
}^{\perp}\overset{\text{def}}{=}$\emph{Span}$\left\{  H\left\vert
x\right\rangle \right\}  $ with $x\neq0...0$. Note that $V_{\psi}^{\perp}$ is
orthogonal to $V_{\psi}$ since $\left\langle 0...0\left\vert H^{\dagger
}H\right\vert x\right\rangle =0$, $\forall x\neq0...0$ since $H^{\dagger
}=H^{-1}=H$. We wish to introduce an operator $U_{\psi^{\bot}}$ such that it
acts like the identity on $V_{\psi}\subset\mathcal{H}_{2}^{n}$ and like a
phase shift of $-1$ on states in $V_{\psi}^{\perp}\subset\mathcal{H}_{2}^{n}$.
The operator acting on the joint space of both registers that exhibits these
properties is the so-called inversion about the mean operator defined as,%
\begin{equation}
U_{\psi^{\bot}}\overset{\text{def}}{=}\left(  H^{\otimes n}\otimes
I_{\mathcal{H}_{2}^{1}}\right)  U_{f_{\bar{x}}=0}\left(  H^{\otimes n}\otimes
I_{\mathcal{H}_{2}^{1}}\right)  \text{,}%
\end{equation}
that is, using Eq. (\ref{util}) and the properties of tensor products,%
\begin{equation}
U_{\psi^{\bot}}=\left(  H^{\otimes n}U_{\bar{x}=0}H^{\otimes n}\right)
\otimes I_{\mathcal{H}_{2}^{1}}=\left(  2\left\vert \psi\right\rangle
\left\langle \psi\right\vert -I_{\mathcal{H}_{2}^{n}}\right)  \otimes
I_{\mathcal{H}_{2}^{1}}\text{.} \label{ut}%
\end{equation}
Applying $U_{\psi^{\bot}}$ in Eq. (\ref{ut}) to $\left\vert \Psi
_{\text{step-}2}\right\rangle $ yields,%
\begin{equation}
\left\vert \Psi_{\text{step-}3}\right\rangle =U_{\psi^{\bot}}\left\vert
\Psi_{\text{step-}2}\right\rangle \text{.}%
\end{equation}
We point out that the action of the operator $2\left\vert \psi\right\rangle
\left\langle \psi\right\vert -I_{\mathcal{H}_{2}^{n}}$ on an arbitrary vector
$\left\vert \phi\right\rangle $ in $\mathcal{H}_{2}^{n}$,%
\begin{equation}
\left\vert \phi\right\rangle =\sum_{x=0}^{N-1}\alpha_{x}\left\vert
x\right\rangle \text{,}%
\end{equation}
with $\alpha_{x}\in%
\mathbb{C}
$ for any $x\in\mathcal{X}$ is given by,%
\begin{equation}
\left(  2\left\vert \psi\right\rangle \left\langle \psi\right\vert
-I_{\mathcal{H}_{2}^{n}}\right)  \left\vert \phi\right\rangle =\sum
_{x=0}^{N-1}\left(  2\mu_{\alpha}-\alpha_{x}\right)  \left\vert x\right\rangle
\text{,} \label{iam}%
\end{equation}
where $\mu_{\alpha}$ is the mean of the amplitudes defined as,%
\begin{equation}
\mu_{\alpha}\overset{\text{def}}{=}\frac{1}{N}\sum_{x=0}^{N-1}\alpha
_{x}\text{.}%
\end{equation}
For a detailed derivation of Eq. (\ref{iam}), we refer to Appendix A. We point
out that this explicit derivation was useful since it allows to identify and
correct a typographical error that appears in \cite{kayebook}\textbf{.} Eq.
(\ref{iam}) justifies the terminology of inversion about the mean operator
associated to $2\left\vert \psi\right\rangle \left\langle \psi\right\vert
-I_{\mathcal{H}_{2}^{n}}$. The net effect of this operator is that of
enhancing the amplitude of the target state $\left\vert \bar{x}\right\rangle $
and slightly lowering the amplitudes of all the remaining basis states. At
this point, we can finally define the Grover iterate operator $G$ in terms of
$U_{f_{\bar{x}}}$ and $U_{\psi^{\bot}}$ as follows,%
\begin{equation}
G\overset{\text{def}}{=}U_{\psi^{\bot}}U_{f_{\bar{x}}}\text{.}%
\end{equation}

\emph{Step-4}: Iterations of the Grover iterate, $G^{k}$ with $G=U_{\psi
^{\perp}}U_{f_{\bar{x}}}$. This step requires the application of the quantum
search iterate a number of $k$-times. This number is approximately estimated
by requiring that for $N\gg1$ the probability of the amplitude of the target
state $\left\vert \bar{x}\right\rangle $ approaches unity,%
\begin{equation}
\left\vert \left\langle \bar{x}\left\vert G^{k}\right\vert \psi\right\rangle
\right\vert ^{2}\overset{N\rightarrow\infty}{\longrightarrow}1\text{.}%
\end{equation}
Grover uncovered that \cite{grover97, grover01},%
\begin{equation}
k_{\text{Grover}}\left(  N\right)  \overset{N\gg1}{\approx}\frac{\pi}{4}%
\sqrt{N}\text{.} \label{kgrover}%
\end{equation}
Eq. (\ref{kgrover}) ends our standard mathematical treatment. The explicit
derivation of Eq. (\ref{kgrover}) appears in the next subsection.

\subsection{The quadratic speedup}

We begin by applying the Walsh-Hadamard transform $H^{\otimes n}$ to the state
$\left\vert 0\right\rangle ^{\otimes n}$ in order to construct a uniform
amplitude initial state $\left\vert \psi\right\rangle $ given in Eq.
(\ref{fidef}),%
\begin{equation}
\left\vert \psi\right\rangle \overset{\text{def}}{=}\frac{1}{\sqrt{N}}%
\sum_{x=0}^{N-1}\left\vert x\right\rangle \text{.} \label{psi1}%
\end{equation}
We note that $\left\vert \psi\right\rangle $ can be decomposed as,%
\begin{equation}
\left\vert \psi\right\rangle =\frac{1}{\sqrt{N}}\sum_{x=0}^{N-1}\left\vert
x\right\rangle =\frac{1}{\sqrt{N}}\left\vert \bar{x}\right\rangle
+\sqrt{1-\frac{1}{N}}\sum_{x\in\mathcal{X}\backslash\left\{  \bar{x}\right\}
}\left\vert x\right\rangle \text{,}%
\end{equation}
that is,%
\begin{equation}
\left\vert \psi\right\rangle =\frac{1}{\sqrt{N}}\left\vert \bar{x}%
\right\rangle +\sqrt{\frac{N-1}{N}}\left\vert \psi_{\text{bad}}\right\rangle
\text{,} \label{fi1}%
\end{equation}
where $\left\vert \bar{x}\right\rangle $ is the target state, $\mathcal{X}$ is
the set previously defined, and
\begin{equation}
\left\vert \psi_{\text{bad}}\right\rangle \overset{\text{def}}{=}\sum
_{x\in\mathcal{X}\backslash\left\{  \bar{x}\right\}  }\left\vert
x\right\rangle \text{.}%
\end{equation}
Observe that the application of the Grover iterate $G$ onto $\left\vert
\psi\right\rangle $ yields states that belong to \emph{Span}$\left\{
\left\vert \bar{x}\right\rangle \text{, }\left\vert \psi_{\text{bad}%
}\right\rangle \right\}  $, a two-dimensional subspace of $\mathcal{H}_{2}%
^{n}$. Furthermore, it happens to be convenient to introduce an additional
basis of such a two-dimensional space in order to better understand the
iterative procedure proposed by Grover. Specifically, consider the orthogonal
states $\left\vert \psi\right\rangle $ in Eq. (\ref{psi1}) and $\left\vert
\bar{\psi}\right\rangle $ where,%
\begin{equation}
\left\vert \bar{\psi}\right\rangle \overset{\text{def}}{=}\sqrt{\frac{N-1}{N}%
}\left\vert \bar{x}\right\rangle -\frac{1}{\sqrt{N}}\left\vert \psi
_{\text{bad}}\right\rangle \text{.} \label{fi2}%
\end{equation}
To quantify the relation between the two orthogonal bases $\left\{  \left\vert
\bar{x}\right\rangle \text{, }\left\vert \psi_{\text{bad}}\right\rangle
\right\}  $ and $\left\{  \left\vert \psi\right\rangle \text{, }\left\vert
\bar{\psi}\right\rangle \right\}  $, introduce an angle $\theta$ such that%
\begin{equation}
\sin\left(  \theta\right)  \overset{\text{def}}{=}\frac{1}{\sqrt{N}}\text{,
and }\cos\left(  \theta\right)  \overset{\text{def}}{=}\sqrt{\frac{N-1}{N}%
}\text{.} \label{sin}%
\end{equation}
Using Eqs. (\ref{fi1}), (\ref{fi2}), and (\ref{sin}), it follows that%
\begin{equation}
\binom{\left\vert \psi\right\rangle }{\left\vert \bar{\psi}\right\rangle
}=\left(
\begin{array}
[c]{cc}%
\sin\left(  \theta\right)  & \cos\left(  \theta\right) \\
\cos\left(  \theta\right)  & -\sin\left(  \theta\right)
\end{array}
\right)  \binom{\left\vert \bar{x}\right\rangle }{\left\vert \psi_{\text{bad}%
}\right\rangle }\text{.} \label{la}%
\end{equation}
From Eq. (\ref{la}), it is straightforward to verify that%
\begin{equation}
\binom{\left\vert \bar{x}\right\rangle }{\left\vert \psi_{\text{bad}%
}\right\rangle }=\left(
\begin{array}
[c]{cc}%
\sin\left(  \theta\right)  & \cos\left(  \theta\right) \\
\cos\left(  \theta\right)  & -\sin\left(  \theta\right)
\end{array}
\right)  \binom{\left\vert \psi\right\rangle }{\left\vert \bar{\psi
}\right\rangle }\text{.} \label{la2}%
\end{equation}
The iterative procedure begins in the state $\left\vert \psi\right\rangle $,%
\begin{equation}
\left\vert \psi\right\rangle =\sin\left(  \theta\right)  \left\vert \bar
{x}\right\rangle +\cos\left(  \theta\right)  \left\vert \psi_{\text{bad}%
}\right\rangle \text{.} \label{fis}%
\end{equation}
Let us consider the first iteration by computing $G\left\vert \psi
\right\rangle =U_{\psi^{\perp}}U_{f_{\bar{x}}}\left\vert \psi\right\rangle $.
Using Eqs. (\ref{util}) and (\ref{fis}) together with the orthogonality of
states $\left\vert \bar{x}\right\rangle $ and $\left\vert \psi_{\text{bad}%
}\right\rangle $, we obtain%
\begin{equation}
U_{f_{\bar{x}}}\left\vert \psi\right\rangle =\cos\left(  2\theta\right)
\left\vert \psi\right\rangle -\sin\left(  2\theta\right)  \left\vert \bar
{\psi}\right\rangle \text{.} \label{uff1}%
\end{equation}
Applying $U_{\psi^{\perp}}$ in Eq. (\ref{ut}) to $U_{f_{\bar{x}}}\left\vert
\psi\right\rangle $ in Eq. (\ref{uff1}) together with standard trigonometric
relations,\textbf{ }we find%
\begin{equation}
G\left\vert \psi\right\rangle =U_{\psi^{\perp}}U_{f_{\bar{x}}}\left\vert
\psi\right\rangle =\sin\left(  3\theta\right)  \left\vert \bar{x}\right\rangle
+\cos\left(  3\theta\right)  \left\vert \psi_{\text{bad}}\right\rangle
\text{.}%
\end{equation}
Using some mathematical reasoning, it is straightforward to verify that after
$k$-iterations of $G$, the angle $\theta_{k}$ that specifies $G^{k}$ is such
that $\theta_{k+1}=\theta_{k}+2\theta$ and $\theta_{1}=3\theta$. Therefore, it
follows that $\theta_{k}=\left(  2k+1\right)  \theta$ and $G^{k}\left\vert
\psi\right\rangle $ becomes%
\begin{equation}
G^{k}\left\vert \psi\right\rangle =\sin\left[  \left(  2k+1\right)
\theta\right]  \left\vert \bar{x}\right\rangle +\cos\left[  \left(
2k+1\right)  \theta\right]  \left\vert \psi_{\text{bad}}\right\rangle \text{.}
\label{iter}%
\end{equation}
From Eq. (\ref{iter}), the probability of the amplitude of the target state
$\left\vert \bar{x}\right\rangle $ becomes,%
\begin{equation}
\left\vert \left\langle \bar{x}\left\vert G^{k}\right\vert \psi\right\rangle
\right\vert ^{2}=\sin^{2}\left[  \left(  2k+1\right)  \theta\right]  \text{.}
\label{proba}%
\end{equation}
Observe that the probability in Eq. (\ref{proba}) approaches unity when
$\left(  2k+1\right)  \theta$ approaches $\pi/2$. From the first relation in
Eq. (\ref{sin}), in the limit of $N\gg1$, we obtain%
\begin{equation}
\frac{\pi}{2}=\left(  2k+1\right)  \theta\overset{N\gg1}{\approx}\frac{\left(
2k+1\right)  }{\sqrt{N}}\overset{N\gg1}{\approx}\frac{2}{\sqrt{N}}k\text{,}%
\end{equation}
that is, finally,%
\begin{equation}
k_{\text{Grover}}\left(  N\right)  \overset{N\gg1}{\approx}\frac{\pi}{4}%
\sqrt{N}\text{.} \label{quadratic}%
\end{equation}
Eq. (\ref{quadratic}) concludes our formal reexamination of the quadratic
speedup relation. As a side remark, we point out that within the framework of
quantum amplitude amplification techniques \cite{brassard00}, quantum
operations allow to amplify the amplitudes of good output states. Since the
corresponding probabilities are defined as the squares of the amplitudes, the
amplification is quadratically faster than in the classical case. Having said
that, the quadratic improvement in the complexity of Grover's search algorithm
can be explained also by observing that Grover's search is a special case of
the amplitude amplification technique \cite{brassard00}.

\subsection{Matrix representation of the Grover iterate}

Observe that the action of the oracle $U_{f_{\bar{x}}}$ in Eq. (\ref{util}) on
the state $\left\vert \psi\right\rangle =\sin\left(  \theta\right)  \left\vert
\bar{x}\right\rangle +\cos\left(  \theta\right)  \left\vert \psi_{\text{bad}%
}\right\rangle $ in Eq. (\ref{fis}) is given by,%
\begin{equation}
U_{f_{\bar{x}}}\left\vert \psi\right\rangle =-\sin\left(  \theta\right)
\left\vert \bar{x}\right\rangle +\cos\left(  \theta\right)  \left\vert
\psi_{\text{bad}}\right\rangle \text{.}%
\end{equation}
Therefore, $U_{f_{\bar{x}}}$ describes a reflection about the vector
$\left\vert \psi_{\text{bad}}\right\rangle $ in the plane defined by
$\left\vert \bar{x}\right\rangle $ and $\left\vert \psi_{\text{bad}%
}\right\rangle $. Similarly, we note that the action of $U_{\psi^{\perp}}$ in
Eq. (\ref{ut}) on the state $\left\vert \bar{x}\right\rangle =\sin\left(
\theta\right)  \left\vert \psi\right\rangle +\cos\left(  \theta\right)
\left\vert \bar{\psi}\right\rangle $ in Eq. (\ref{la2}) is given by,%
\begin{equation}
U_{\psi^{\perp}}\left\vert \bar{x}\right\rangle =\sin\left(  \theta\right)
\left\vert \psi\right\rangle -\cos\left(  \theta\right)  \left\vert \bar{\psi
}\right\rangle \text{.}%
\end{equation}
Therefore, $U_{\psi^{\perp}}$ performs a reflection about the vector
$\left\vert \psi\right\rangle $ in the plane defined by $\left\vert \bar
{x}\right\rangle $ and $\left\vert \psi_{\text{bad}}\right\rangle $ because
\emph{Span}$\left\{  \left\vert \bar{x}\right\rangle \text{, }\left\vert
\psi_{\text{bad}}\right\rangle \right\}  =$\emph{Span}$\left\{  \left\vert
\psi\right\rangle \text{, }\left\vert \bar{\psi}\right\rangle \right\}  $.
Since the Grover iterate $G$ is the composition of two reflections, it is a
rotation. In particular, after some straightforward algebra, the matrix
representation $\left[  G\right]  $ of the operator $G$ restricted to the
two-dimensional space spanned by $\left\vert \psi_{\text{bad}}\right\rangle $
and $\left\vert \bar{x}\right\rangle $ becomes,%
\begin{equation}
\left[  G\right]  =\left(
\begin{array}
[c]{cc}%
\left\langle \psi_{\text{bad}}\left\vert G\right\vert \psi_{\text{bad}%
}\right\rangle  & \left\langle \psi_{\text{bad}}\left\vert G\right\vert
\bar{x}\right\rangle \\
\left\langle \bar{x}\left\vert G\right\vert \psi_{\text{bad}}\right\rangle  &
\left\langle \bar{x}\left\vert G\right\vert \bar{x}\right\rangle
\end{array}
\right)  =\left(
\begin{array}
[c]{cc}%
\cos\left(  2\theta\right)  & -\sin\left(  2\theta\right) \\
\sin\left(  2\theta\right)  & \cos\left(  2\theta\right)
\end{array}
\right)  \text{.} \label{mg}%
\end{equation}
Note that $\left[  G\right]  $ is an orthogonal matrix since $\det\left(
\left[  G\right]  \right)  =+1$ and $\left[  G\right]  ^{t}\left[  G\right]
=\left[  G\right]  \left[  G\right]  ^{t}=I_{2\times2}$ where $I_{2\times2}$
denotes the $2\times2$ identity matrix and the symbol $t$ denotes the
transposition operation. Therefore, $\left[  G\right]  $ does indeed represent
a rotation. In summary: i) the Grover iterate is the product of two
reflections; ii) the product of two reflections is a rotation; iii) the Grover
iterate is a rotation in the two-dimensional space spanned by $\left\vert
\psi_{\text{bad}}\right\rangle $ and $\left\vert \bar{x}\right\rangle $ that
rotates state vectors by $2\theta$ radians. For the sake of completeness, we
also point out that, following the same computational steps needed to compute
$\left[  G\right]  $ in Eq. (\ref{mg}), the matrix representations $\left[
U_{\psi^{\perp}}\right]  $ and $\left[  U_{f_{\bar{x}}}\right]  $ of the
operators $U_{\psi^{\perp}}$ and $U_{f_{\bar{x}}}$ restricted to the
two-dimensional space spanned by $\left\vert \psi_{\text{bad}}\right\rangle $
and $\left\vert \bar{x}\right\rangle $ are given by,%
\begin{equation}
\left[  U_{\psi^{\perp}}\right]  =\left(
\begin{array}
[c]{cc}%
\left\langle \psi_{\text{bad}}\left\vert U_{\psi^{\perp}}\right\vert
\psi_{\text{bad}}\right\rangle  & \left\langle \psi_{\text{bad}}\left\vert
U_{\psi^{\perp}}\right\vert \bar{x}\right\rangle \\
\left\langle \bar{x}\left\vert U_{\psi^{\perp}}\right\vert \psi_{\text{bad}%
}\right\rangle  & \left\langle \bar{x}\left\vert U_{\psi^{\perp}}\right\vert
\bar{x}\right\rangle
\end{array}
\right)  =\left(
\begin{array}
[c]{cc}%
\cos\left(  2\theta\right)  & \sin\left(  2\theta\right) \\
\sin\left(  2\theta\right)  & -\cos\left(  2\theta\right)
\end{array}
\right)  \label{uf}%
\end{equation}
and,%
\begin{equation}
\left[  U_{f_{\bar{x}}}\right]  =\left(
\begin{array}
[c]{cc}%
\left\langle \psi_{\text{bad}}\left\vert U_{f_{\bar{x}}}\right\vert
\psi_{\text{bad}}\right\rangle  & \left\langle \psi_{\text{bad}}\left\vert
U_{f_{\bar{x}}}\right\vert \bar{x}\right\rangle \\
\left\langle \bar{x}\left\vert U_{f_{\bar{x}}}\right\vert \psi_{\text{bad}%
}\right\rangle  & \left\langle \bar{x}\left\vert U_{f_{\bar{x}}}\right\vert
\bar{x}\right\rangle
\end{array}
\right)  =\left(
\begin{array}
[c]{cc}%
1 & 0\\
0 & -1
\end{array}
\right)  \text{,} \label{uff}%
\end{equation}
respectively. Observe that from Eqs. (\ref{mg}), (\ref{uf}), and (\ref{uff}),
one obtains, as expected, that%
\begin{equation}
\left[  G\right]  \overset{\text{def}}{=}\left[  U_{\psi^{\perp}}U_{f_{\bar
{x}}}\right]  =\left[  U_{\psi^{\perp}}\right]  \left[  U_{f_{\bar{x}}%
}\right]  \text{.}%
\end{equation}
Finally, notice that $\left[  U_{\psi^{\perp}}\right]  $ and $\left[
U_{f_{\bar{x}}}\right]  $ are orthogonal matrices with $\det\left(  \left[
U_{\psi^{\perp}}\right]  \right)  =\det\left(  \left[  U_{f_{\bar{x}}}\right]
\right)  =-1$. Therefore, $\left[  U_{\psi^{\perp}}\right]  $ and $\left[
U_{f_{\bar{x}}}\right]  $ do indeed represent reflections.

\section{Geometric algebra and the digital description}

From the standard mathematical description of Grover's algorithm, we realize
that some of the key concepts that need to be translated into the GA language
are those of reflection, rotation, quantum state, and quantum operators. In
what follows, we introduce the strictly necessary GA tools needed to provide a
GA description of the key aspects of Grover's search algorithm. For the sake
of completeness, some additional details appear in Appendix B.

A GA is a graded linear space whose elements are known as multivectors. For
instance, a scalar is a grade-$0$ multivector, a vector is a grade-$1$
multivector, a bivector is a grade-$2$ multivector, and so on. The so-called
pseudoscalar is the highest-grade element in the algebra. Multivectors can be
either homogeneous or non-homogeneous. In the former case, they only contain
elements of a single grade. In the latter case, they contain elements with
multiple grades. Furthermore, homogeneous vectors form a subspace which is
closed under scalar multiplication and addition. For further details, we shall
refer to our Appendix B. \ For more foundational purposes, we refer to more
extensive works on the use of GA in quantum information processing
\cite{doran, lasenby93, doran93, doran96, somaroo99, havel02}.

\subsection{Reflections}

Recall from Section III that the Grover iterate is a rotation defined in terms
of the product of two reflections. GA offers a powerful platform for handling
rotations and reflections by means of bivectors (a grade-$2$ multivector). Let
us consider two vectors $a$ and $b$ that belong to the GA of physical space
$\mathfrak{cl}(3)$ (for further details, see Appendix B). Let us further
assume that $a$ and $b$ diverge by an angle $\theta\in\left(  0,\pi\right)  $.
Using the GA formalism, the vector $a$ can be written as a superposition of
two vectors $a_{\parallel}$ and $a_{\perp}$ as follows,%
\begin{equation}
a=a_{\parallel}+a_{\perp}\text{,}%
\end{equation}
where,%
\begin{equation}
a_{\parallel}=\left(  a\cdot b\right)  \frac{b}{\left\vert b\right\vert ^{2}%
}=\left(  a\cdot b\right)  b^{-1}\text{,}%
\end{equation}
and,%
\begin{equation}
a_{\perp}=a-a_{\parallel}=a-\left(  a\cdot b\right)  b^{-1}=\left(  ab-a\cdot
b\right)  b^{-1}=\left(  a\wedge b\right)  b^{-1}\text{.}%
\end{equation}
Vectors $a_{\parallel}$ and $a_{\perp}$ denote vectors parallel and
perpendicular to $b$, respectively. The reflection of a vector $v$ across the
line $a$ is obtained by sending $v=v_{\parallel}+v_{\perp}$ to the mirror
image $v^{\prime}=v_{\parallel}-v_{\perp}$,%
\begin{align}
v^{\prime}  &  =v_{\parallel}-v_{\perp}=\left(  v\cdot a\right)
a^{-1}-\left(  v\wedge a\right)  a^{-1}\nonumber\\
&  =\left(  v\cdot a-v\wedge a\right)  a^{-1}\nonumber\\
&  =\left(  a\cdot v+a\wedge v\right)  a^{-1}\nonumber\\
&  =ava^{-1}\text{,}%
\end{align}
that is,%
\begin{equation}
v\rightarrow v^{\prime}=ava^{-1}:\text{single reflection.} \label{reflection}%
\end{equation}
Note that the composition of two reflections of $v$, first across $a$ and then
across $b$ is defined as,%
\begin{equation}
v\rightarrow v^{\prime}=ava^{-1}\rightarrow v^{\prime\prime}=bv^{\prime}%
b^{-1}=bava^{-1}b^{-1}=(ba)v(ba)^{-1}\text{,}%
\end{equation}
therefore,%
\begin{equation}
v\rightarrow v^{\prime\prime}=(ba)v(ba)^{-1}:\text{composition of two
reflections.} \label{2reflection}%
\end{equation}
For further details on reflections, we refer to \cite{doran}.

\subsection{Rotations}

To keep a smooth reading flow, we refer to Appendix B for further mathematical
and notational details. Using GA, the unit vector $e_{\theta}\in
\mathfrak{cl}(3)$ obtained from $e_{1}$ by a rotation by an angle $\theta$ in
the $e_{1}e_{2}$ plane is given by \cite{doran},%
\begin{align}
e_{\theta}  &  =e_{1}\cos\left(  \theta\right)  +e_{2}\sin\left(
\theta\right) \nonumber\\
&  =e_{1}\left[  \cos\left(  \theta\right)  +e_{1}e_{2}\sin\left(
\theta\right)  \right] \nonumber\\
&  =e_{1}\exp\left[  e_{1}e_{2}\theta\right] \nonumber\\
&  =\exp\left[  e_{2}e_{1}\theta\right]  e_{1}\text{,} \label{eteta}%
\end{align}
that is,%
\begin{equation}
e_{1}\rightarrow e_{\theta}\overset{\text{def}}{=}\mathcal{R}_{e_{1}e_{2}%
}\left(  \theta\right)  e_{1}=\exp\left[  e_{2}e_{1}\theta\right]
e_{1}\text{.} \label{rot1}%
\end{equation}
The quantity $\mathcal{R}_{e_{1}e_{2}}\left(  \theta\right)  $ in Eq.
(\ref{rot1}) denotes the bivector operator that describes the rotation by
$\theta$ of $e_{1}$ in the $e_{1}e_{2}$ plane. The rotation of a vector $v$
that does not necessarily belong to the rotation plane $e_{1}e_{2}$ can be
described as,%
\begin{equation}
v\rightarrow v^{\prime}=\mathcal{R}_{e_{1}e_{2}}\left(  \theta\right)
v=\exp\left[  e_{2}e_{1}\frac{\theta}{2}\right]  v\exp\left[  e_{1}e_{2}%
\frac{\theta}{2}\right]  :\text{rotation} \label{f1}%
\end{equation}
For the sake of completeness, we point out that $e_{1}$ belongs to the
rotation plane $e_{1}e_{2}$, and
\begin{equation}
\exp\left[  e_{2}e_{1}\theta\right]  \left(  e_{1}\right)  =\exp\left[
e_{2}e_{1}\frac{\theta}{2}\right]  \left(  e_{1}\right)  \exp\left[
e_{1}e_{2}\frac{\theta}{2}\right]  \text{.}%
\end{equation}
Let us now demonstrate the equivalence of a rotation to a pair of successive
reflections in intersecting planes. Observe that from $e_{\theta}$ in Eq.
(\ref{eteta}), we obtain%
\begin{equation}
e_{1}e_{\theta}=\exp\left[  e_{1}e_{2}\theta\right]  =\cos\left(
\theta\right)  +e_{1}e_{2}\sin\left(  \theta\right)  \text{.} \label{f2}%
\end{equation}
Therefore, using Eqs. (\ref{f1}) and (\ref{f2}), we find%
\begin{align}
v^{\prime}  &  =\exp\left[  e_{2}e_{1}\frac{\theta}{2}\right]  v\exp\left[
e_{1}e_{2}\frac{\theta}{2}\right] \nonumber\\
&  =\left(  e_{\frac{\theta}{2}}e_{1}\right)  v\left(  e_{1}e_{\frac{\theta
}{2}}\right) \nonumber\\
&  =\left(  e_{\frac{\theta}{2}}e_{3}\right)  \left(  e_{3}e_{1}\right)
v\left(  e_{1}e_{3}\right)  \left(  e_{3}e_{\frac{\theta}{2}}\right)  \text{,}%
\end{align}
that is, the rotated vector $v^{\prime}$ becomes
\begin{equation}
v^{\prime}=\left(  e_{\frac{\theta}{2}}e_{3}\right)  \left(  e_{3}%
e_{1}\right)  v\left(  e_{1}e_{3}\right)  \left(  e_{3}e_{\frac{\theta}{2}%
}\right)  \text{.} \label{dr}%
\end{equation}
From Eq. (\ref{dr}), we observe that%
\begin{equation}
v^{\prime}=\left[  \text{rotation}_{e_{1}e_{2}}\left(  \theta\right)  \right]
\left(  v\right)  =\text{reflection}_{e_{\frac{\theta}{2}}e_{3}}\left[
\text{reflection}_{e_{3}e_{1}}\left(  v\right)  \right]  \text{,}%
\end{equation}
where,%
\begin{equation}
\theta=2\cos^{-1}\left[  n_{e_{\frac{\theta}{2}}e_{3}}\cdot n_{e_{3}e_{1}%
}\right]  \text{,} \label{dr2}%
\end{equation}
with $n_{\pi}$ denoting the unit normal to the plane $\pi$. From Eqs.
(\ref{dr}) and (\ref{dr2}), it becomes transparent using GA that a rotation is
equivalent to a pair of successive reflections in intersecting planes where
the angle of rotation is twice the angular opening between the planes that
characterize the reflections. Observe that it is straightforward to verify
within GA\ that both rotations and reflections are orthogonal (linear)
transformations since they both preserve the inner product. Furthermore, while
rotations are proper orthogonal transformations with $\det\left(
\text{rotation}\right)  =+1$, reflections are improper orthogonal
transformations with $\det\left(  \text{reflection}\right)  =-1$. Within GA,
these last two statements can be explained as follows. First, we point out
that the mirror image of an arbitrary vector $x$ in the plane through the
origin with normal $n$ is called the reflection along $n$ and is given by
$x^{\prime}=-nxn$ \cite{hestenes86}. As a consequence, it so happens that the
product of three transformed vectors becomes%
\begin{equation}
xyz\rightarrow x^{\prime}y^{\prime}z^{\prime}=-n\left(  xyz\right)  n\text{,}%
\end{equation}
where,%
\begin{align}
xyz  &  =x\left(  yz\right)  =x\left[  y\cdot z+y\wedge z\right]  =x\left(
y\cdot z\right)  +x\left(  y\wedge z\right) \nonumber\\
&  =\left(  y\cdot z\right)  x+x\cdot\left(  y\wedge z\right)  +x\wedge\left(
y\wedge z\right) \nonumber\\
&  =\left(  y\cdot z\right)  x+\left(  x\cdot y\right)  z-\left(  x\cdot
z\right)  y+x\wedge y\wedge z\nonumber\\
&  =\left(  y\cdot z\right)  x-\left(  x\cdot z\right)  y+\left(  x\cdot
y\right)  z+x\wedge y\wedge z\text{,}%
\end{align}
that is,%
\begin{equation}
xyz=\left(  y\cdot z\right)  x-\left(  x\cdot z\right)  y+\left(  x\cdot
y\right)  z+x\wedge y\wedge z\text{.}%
\end{equation}
The trivector part of the multivector $xyz$ is given by,%
\begin{equation}
\left\langle xyz\right\rangle _{\text{grade-}3}=x\wedge y\wedge z\text{,}%
\end{equation}
where, under reflection,%
\begin{equation}
x\wedge y\wedge z\rightarrow-n\left(  x\wedge y\wedge z\right)  n=-x\wedge
y\wedge z\text{,} \label{triva}%
\end{equation}
since vectors in the GA of physical space $\mathfrak{cl}(3)$ commute with all
pseudoscalars. Second, an outermorphism of a linear transformation
$\mathcal{R}$ is a transformation on $\mathfrak{cl}(3)$ which is linear,
grade-preserving, and preserves the outer product \cite{hestenes86}. Since for
linear transformations $\mathcal{R}$ on $\mathfrak{cl}(3)$ the determinant is
specified in terms of the action of its outermorphism $\mathcal{\text{\b{R} }%
}$on trivectors (that is, the pseudoscalars),%
\begin{equation}
\mathcal{\text{\b{R}}}\left(  x\wedge y\wedge z\right)  =\det\left(
\mathcal{R}\right)  x\wedge y\wedge z\text{,}%
\end{equation}
from Eq. (\ref{triva}) we conclude that $\det\left(  \text{reflection}\right)
=-1$ (improper orthogonal transformation). Furthermore, since a rotation is a
composition of two reflections, $\det\left(  \text{reflection}\right)  =+1$
(proper orthogonal transformation). Observe that $x\wedge y\wedge z$ denotes
an oriented volume of a parallelepiped with edges $x$, $y$, $z$. Therefore,
$\det\left(  \mathcal{R}\right)  $ is a factor that represents an induced
change in scale of the volume. In what follows, we shall employ the relevant
GA considerations presented here to understand the structure of the Grover
iterate, namely a rotation defined in terms of the product of two reflections
as evident from Eqs. (\ref{mg}), (\ref{uf}), and (\ref{uff}).

\subsection{The Grover iterate and the quadratic speedup}

Let us briefly recall that the Grover iterate is given by $G=U_{\psi^{\bot}%
}U_{f_{\bar{x}}}$ where $U_{f_{\bar{x}}}$ describes a reflection about the
vector $\left\vert \psi_{\text{bad}}\right\rangle $ in the plane defined by
$\left\vert \bar{x}\right\rangle $ and $\left\vert \psi_{\text{bad}%
}\right\rangle $, $U_{\psi^{\bot}}$ represents a reflection about the vector
$\left\vert \psi\right\rangle $ in the plane defined by $\left\vert \bar
{x}\right\rangle $ and $\left\vert \psi_{\text{bad}}\right\rangle $, and,
finally, $G$ is a rotation by $2\theta$ in the two-dimensional space spanned
by $\left\vert \bar{x}\right\rangle $ and $\left\vert \psi_{\text{bad}%
}\right\rangle $.\textbf{ }In what follows, a quantum state in the Hilbert
space $\mathcal{H}_{2}^{n}$ will be replaced by a multivector belonging to the
reduced even subalgebra space\textbf{ }$\left[  \mathfrak{cl}^{+}(3)\right]
^{n}/E_{n}$\textbf{ }where $E_{n}$ denotes the $n$-particle correlator defined
in Eq. (\ref{en}) (for further details, see Appendix B)\textbf{.} Using the GA
language, the quantum state $\left\vert \psi\right\rangle $ in Eq.
(\ref{psi1}) can be recast as%
\begin{equation}
e_{\psi}=\frac{1}{\sqrt{N}}\sum_{i=0}^{N-1}e_{i}=\frac{1}{\sqrt{N}}e_{\bar{x}%
}+\sqrt{\frac{N-1}{N}}e_{\text{bad}}\text{.}%
\end{equation}
After the application of the first above-mentioned reflection and using Eq.
(\ref{reflection}), the multivector $e_{\psi}$ becomes%
\begin{equation}
e_{\psi}\rightarrow e_{\text{bad}}e_{\psi}e_{\text{bad}}^{-1}=e_{\text{bad}%
}e_{\psi}\frac{e_{\text{bad}}}{\left\vert e_{\text{bad}}\right\vert ^{2}%
}=e_{\text{bad}}e_{\psi}e_{\text{bad}}\text{,}%
\end{equation}
since $\left\vert e_{\text{bad}}\right\vert ^{2}=1$. After the application of
the second above-mentioned reflection and employing Eq. (\ref{2reflection}),
we obtain%
\begin{equation}
e_{\text{bad}}e_{\psi}e_{\text{bad}}\rightarrow e_{\psi}\left(  e_{\text{bad}%
}e_{\psi}e_{\text{bad}}\right)  e_{\psi}^{-1}=e_{\psi}\left(  e_{\text{bad}%
}e_{\psi}e_{\text{bad}}\right)  \frac{e_{\psi}}{\left\vert e_{\psi}\right\vert
^{2}}=e_{\psi}\left(  e_{\text{bad}}e_{\psi}e_{\text{bad}}\right)  e_{\psi
}\text{,} \label{okk}%
\end{equation}
where, this time, we have exploited the fact that $\left\vert e_{\psi
}\right\vert ^{2}=1$. Therefore, from Eq. (\ref{okk}), we uncover that the
action of the Grover iterate $G$ on $e_{\psi}$ yields%
\begin{equation}
Ge_{\psi}=e_{\psi}e_{\text{bad}}e_{\psi}e_{\text{bad}}e_{\psi}=\left(
e_{\psi}e_{\text{bad}}\right)  e_{\psi}\left(  e_{\text{bad}}e_{\psi}\right)
\text{,}%
\end{equation}
that is,%
\begin{equation}
Ge_{\psi}=ge_{\psi}g^{\dagger}\text{,} \label{gag}%
\end{equation}
where the bivector $g$ is defined as,%
\begin{equation}
g\overset{\text{def}}{=}e_{\psi}e_{\text{bad}}\text{.} \label{g}%
\end{equation}
The dagger symbol $\dagger$ in Eq. (\ref{gag}) denotes the reversion operation
in GA. From Eq. (\ref{fis}), it is clear that%
\begin{equation}
e_{\psi}=\sin\left(  \theta\right)  e_{\bar{x}}+\cos\left(  \theta\right)
e_{\text{bad}}\text{.} \label{efi}%
\end{equation}
From Eqs. (\ref{g}) and (\ref{efi}), we find%
\begin{align}
g  &  =e_{\psi}e_{\text{bad}}=\left[  \sin\left(  \theta\right)  e_{\bar{x}%
}+\cos\left(  \theta\right)  e_{\text{bad}}\right]  e_{\text{bad}}\nonumber\\
&  =\cos\left(  \theta\right)  +e_{\bar{x}}e_{\text{bad}}\sin\left(
\theta\right) \nonumber\\
&  =\exp\left[  e_{\bar{x}}e_{\text{bad}}\theta\right]  \text{,} \label{gs}%
\end{align}
and,%
\begin{align}
g^{\dagger}  &  =e_{\text{bad}}e_{\psi}=e_{\text{bad}}\left[  \sin\left(
\theta\right)  e_{\bar{x}}+\cos\left(  \theta\right)  e_{\text{bad}}\right]
\nonumber\\
&  =\cos\left(  \theta\right)  +e_{\text{bad}}e_{\bar{x}}\sin\left(
\theta\right) \nonumber\\
&  =\exp\left[  e_{\text{bad}}e_{\bar{x}}\theta\right]  \text{.} \label{gc}%
\end{align}
Therefore, using Eqs. (\ref{gs}) and (\ref{gc}), Eq. (\ref{gag}) yields%
\begin{equation}
Ge_{\psi}=\exp\left[  e_{\bar{x}}e_{\text{bad}}\theta\right]  e_{\psi}%
\exp\left[  e_{\text{bad}}e_{\bar{x}}\theta\right]  \text{.} \label{gag2}%
\end{equation}
From the structural analogy with Eq. (\ref{f1}), it becomes evident that the
multivector $ge_{\psi}g^{\dagger}$ describes a rotation by $2\theta$ of
$e_{\psi}$ in the plane $e_{\text{bad}}e_{\bar{x}}$. From Eq. (\ref{gag2}), we
note that the multivector $G$ can be written as,%
\begin{equation}
G=\exp\left[  e_{\bar{x}}e_{\text{bad}}\left(  2\theta\right)  \right]
=\cos\left(  2\theta\right)  +e_{\bar{x}}e_{\text{bad}}\sin\left(
2\theta\right)  \text{.} \label{gj1}%
\end{equation}
Using the two relations in Eq. (\ref{sin}), it is apparent that%
\begin{equation}
\cos\left(  2\theta\right)  =\frac{N-2}{N}\text{, and }\sin\left(
2\theta\right)  =\frac{2\sqrt{N-1}}{N}\text{.} \label{gj2}%
\end{equation}
Finally, from Eqs. (\ref{gj1}) and (\ref{gj2}), the multivector $G$ becomes%
\begin{equation}
G=\frac{N-2}{N}+\frac{2\sqrt{N-1}}{N}e_{\bar{x}}e_{\text{bad}}\text{.}
\label{gag3}%
\end{equation}
Eqs. (\ref{gag}), (\ref{gag2}), and (\ref{gag3}) are the main GA\ equations
that we use to characterize the Grover iterate. In GA terms, $k$-iterations of
$G$ can be described as follows. From Eq. (\ref{gag}), we obtain%
\begin{equation}
G^{k}e_{\psi}=g^{k}e_{\psi}\left(  g^{\dagger}\right)  ^{k}=g^{2k}e_{\psi
}\text{.} \label{gk}%
\end{equation}
Using Eqs. (\ref{efi}) and (\ref{gj1}), the multivector $G^{k}e_{\psi}$ in Eq.
(\ref{gk}) becomes%
\begin{equation}
G^{k}e_{\psi}=\sin\left[  \left(  2k+1\right)  \theta\right]  e_{\bar{x}}%
+\cos\left[  \left(  2k+1\right)  \theta\right]  e_{\text{bad}}\text{.}%
\end{equation}
We observe that $G^{k}e_{\psi}=e_{\bar{x}}$ if and only if,
\begin{equation}
\frac{\pi}{2}=\left(  2k+1\right)  \theta\overset{N\gg1}{\approx}\frac
{2k+1}{\sqrt{N}}\text{,}%
\end{equation}
that is, if the number of iterations $k=k\left(  N\right)  $ in the asymptotic
limit for $N$ approaching infinity equals%
\begin{equation}
k_{\text{Grover}}^{\left(  \text{GA}\right)  }\left(  N\right)  \overset
{N\gg1}{\approx}\frac{\pi}{4}\sqrt{N}\text{.} \label{gaquadratic}%
\end{equation}
Eq. (\ref{gaquadratic}) is the GA analog of Eq. (\ref{quadratic}) and exhibits
the known quadratic speedup relation obtained, this time, from purely GA
arguments. Observe that within the GA framework, both quantum states and
quantum operators are elements of the same\textbf{ }\emph{real} space $\left[
\mathfrak{cl}^{+}(3)\right]  ^{n}/E_{n}$. This is an important conceptual
unifying feature that might find some support also from an experimental
standpoint as we will briefly argue in what follows. For a detailed
GA\ formulation of logic operations in quantum computing, we refer to our own
work presented in \cite{cafaro2011}.

\subsection{Geometric algebra and experimental implementations}

It is not unusual to employ unphysical concepts in intermediate steps in
theoretical physics. However, it is always good to keep in mind that whatever
is being described here occurs in a real-world laboratory \cite{peres04}. For
instance, quantum phenomena do not happen in a \emph{complex} Hilbert space.
They actually occur in a laboratory where you can only see detectors and
emitters (lasers and ion guns, for instance) but no Hermitian operators.

Several experimental proposals for realizing Grover's algorithm have been
presented since its theoretical discovery. In \cite{chuang98, jones98},
Grover's quantum search algorithm was experimentally implemented on NMR
quantum computers. For details on nuclear magnetic resonance principles, we
refer to \cite{ernst87}. In \cite{kwiat99, scully01}, a quantum optical
implementation of Grover's algorithm appeared. For details on linear optical
quantum computing, we refer to \cite{kok07}. In \cite{loss01}, Grover's
algorithm was implemented using molecular magnets ($Fe_{8}$ and $Mn_{12}$)
embedded in a crystal. In \cite{loss03}, Grover's algorithm was implemented
using large nuclear spins in semiconductors ($^{27}Al$, $^{55}Mn$, and
$^{67}Mn$ with nuclear spin $I=\frac{5}{2}$, and $^{73}Ge$ and $^{113}In$ with
nuclear spin $I=\frac{9}{2}$). It is interesting to note that the
implementation of quantum computations using large nuclear spins in $GaAs$
(nuclear spin $I=\frac{3}{2}$) semiconductors is based on a \emph{unary}
representation \cite{poggio02}. This means that once the control over $2I$
magnetic fields is established and provided that there exists a sufficient
signal amplification due to the spin ensemble, quantum computation occurs with
a single pulse \cite{poggio02}. As a consequence, nuclear spins can be
essentially used either as qubits or as logical gate actions on qubits in
these types of experimental settings \cite{loss03}. This physical
implementation of Grover's algorithm seems to support the conceptual unifying
feature provided by GA and, in our view, deserves further investigation. For
the sake of completeness, we point out that the first application of GA to
quantum error correction in liquid NMR quantum computing appeared in
\cite{cory00}. We also emphasize that within the framework of quantum
computing with holograms \cite{miller11, alsing15}, photons are used to
realize both qubits and logic gates acting on them. For instance, the CNOT
gate can be constructed with a single linear momentum photon in a
four-dimensional state space while a qubit can be realized as the polarization
state of a photon. In particular, photons are the carriers of quantum
information and they interact thanks to the use of interferometers. To make
interferometers more stable against environmental noise, Alsing and
collaborators have used holograms of interferometers staked in a photo-thermal
refractive piece of glass \cite{alsing15}. Within this theoretical framework,
an holographic CNOT gate was realized and its functionality was verified by
means of a tomographic analysis in which, due to imperfections of materials,
probabilities do not sum to unity resulting in a violation of the unitarity
condition. To better understand the connection between physical
implementations based on high-dimensional spin systems and the GA language, it
may be worthwhile recasting the qudit search problem in GA terms. We leave
this line of investigation to future investigations.

\section{The digital-to-analog transition}

From a digital quantum computing viewpoint, Grover's algorithm can be regarded
as a definite (discrete-time) sequence of elementary unitary transformations
acting on qubits. Specifically, given an initial input state, the output of
the algorithm becomes the input state after the action of the sequence of
transformations\textbf{ }presented in Section III. In digital terms, the
length of the algorithm equals the number of unitary transformations that
compose the quantum computational software. In \cite{farhi98}, an analog
version of Grover's algorithm was proposed. The search problem was recast in
terms of finding the normalized eigenvector $\left\vert \bar{x}\right\rangle $
corresponding to the only nonvanishing eigenvalue $E$ of an Hamiltonian
$H_{\bar{x}}$ acting on a \emph{complex} $N$-dimensional Hilbert space. The
search ends when the system is in the state $\left\vert \bar{x}\right\rangle
$. More specifically, consider \ the (continuos-time) quantum mechanical
Schrodinger evolution under the time-independent Hamiltonian,%
\begin{equation}
H_{\text{Farhi-Gutmann}}=H_{\bar{x}}+H_{D}\text{,} \label{FG}%
\end{equation}
where $H_{\bar{x}}$ and the driving Hamiltonian $H_{D}$ are defined as
\cite{farhi98},%
\begin{equation}
H_{\bar{x}}\overset{\text{def}}{=}E\left\vert \bar{x}\right\rangle
\left\langle \bar{x}\right\vert \text{, and }H_{D}\overset{\text{def}}%
{=}E\left\vert \psi\right\rangle \left\langle \psi\right\vert \text{,}
\label{fari}%
\end{equation}
respectively. The state $\left\vert \psi\right\rangle $ in Eq. (\ref{fari})
denotes a normalized initial state that does not depend on $\left\vert \bar
{x}\right\rangle $ with $\left\langle \bar{x}|\psi\right\rangle =1/\sqrt
{N}\neq0$. Farhi and Gutmann argued that if $\left\vert \bar{x}\right\rangle $
is chosen from a fixed and known orthonormal basis (or, if $\left\vert \bar
{x}\right\rangle $ is chosen uniformly at random), $\left\vert \bar
{x}\right\rangle $ can be found in a time interval $\Delta t$ that is
proportional to $\sqrt{N}$ \cite{farhi98},%
\begin{equation}
\Delta t\propto\frac{1}{E}\sqrt{N}\text{.} \label{tii}%
\end{equation}
In particular, they showed that this time interval in Eq. (\ref{tii}) is
optimal even if $H_{D}=H_{D}\left(  t\right)  $ in Eq. (\ref{FG}). Despite the
same quadratic speedup behavior, Fenner observed that the Hamiltonian in Eq.
(\ref{FG}) does not generate the analog version of the discrete path that
characterizes Grover's algorithm in the digital setting \cite{fenner00}. To
accomplish this goal, it was proposed in \cite{fenner00} an alternative
time-independent search Hamiltonian defined as,%
\begin{equation}
H_{\text{Fenner}}\overset{\text{def}}{=}\frac{2i_{%
\mathbb{C}
}}{E}\left[  H_{\bar{x}}\text{, }H_{D}\right]  =\frac{2i_{%
\mathbb{C}
}E}{\sqrt{N}}\left[  \left\vert \bar{x}\right\rangle \left\langle
\psi\right\vert -\left\vert \psi\right\rangle \left\langle \bar{x}\right\vert
\right]  \text{,} \label{fenner}%
\end{equation}
where\textbf{ }$i_{%
\mathbb{C}
}$\textbf{ }denotes the \emph{complex} imaginary unit. In particular, it was
shown that the time-independent Grover iterate $G$ can be exactly matched on
the whole Hilbert space $\mathcal{H}_{2}^{n}$ with $N\overset{\text{def}}%
{=}2^{n}$ by the following time-independent unitary time-evolution operator
\cite{fenner00},%
\begin{equation}
G_{\text{Fenner}}\overset{\text{def}}{=}e^{-\frac{i_{%
\mathbb{C}
}}{\hbar}H_{\text{Fenner}}\bar{t}}\text{,}%
\end{equation}
where $\hbar\overset{\text{def}}{=}h/2\pi$, $h$ is the Planck constant
($h\approx6.63\times10^{-34}\left[  \text{MKSA}\right]  $), and $\bar{t}$
approaches $\pi/4\sqrt{N}$ as $N$ approaches infinity and is formally defined
as \cite{fenner00},%
\begin{equation}
\bar{t}\overset{\text{def}}{=}\frac{\pi-2\cos^{-1}\left(  1/\sqrt{N}\right)
}{2}\cdot\frac{N}{\sqrt{N-1}}\text{.}%
\end{equation}
For further details on\textbf{ }more general quantum search Hamiltonians, we
refer to \cite{bae02}. In what follows, we set $\hbar$ equal to one.

\subsection{Matrix algebra and the Fenner iterate}

In what follows, we analyze the Fenner iterate from a matrix algebra
viewpoint. Let us assume that $\left\vert \psi\right\rangle =\alpha\left\vert
\bar{x}\right\rangle +\beta\left\vert \psi_{\text{bad}}\right\rangle $ with
$\left\langle \bar{x}|\psi_{\text{bad}}\right\rangle =0$, $\alpha
\overset{\text{def}}{=}1/\sqrt{N}$, $\beta\overset{\text{def}}{=}\sqrt{\left(
N-1\right)  /N}\in%
\mathbb{R}
_{+}\backslash\left\{  0\right\}  $, and $\alpha^{2}+\beta^{2}=1$. The matrix
representation of $H_{\text{Fenner}}$ in Eq. (\ref{fenner}) on the
two-dimensional space spanned by $\left\vert \bar{x}\right\rangle $ and
$\left\vert \psi_{\text{bad}}\right\rangle $ is given by,%
\begin{align}
\left[  H_{\text{Fenner}}\right]   &  =\left(
\begin{array}
[c]{cc}%
\left\langle \bar{x}\left\vert H_{\text{Fenner}}\right\vert \bar
{x}\right\rangle  & \left\langle \bar{x}\left\vert H_{\text{Fenner}%
}\right\vert \psi_{\text{bad}}\right\rangle \\
\left\langle \psi_{\text{bad}}\left\vert H_{\text{Fenner}}\right\vert \bar
{x}\right\rangle  & \left\langle \psi_{\text{bad}}\left\vert H_{\text{Fenner}%
}\right\vert \psi_{\text{bad}}\right\rangle
\end{array}
\right) \nonumber\\
&  =\frac{2i_{%
\mathbb{C}
}\beta}{\sqrt{N}}\left(
\begin{array}
[c]{cc}%
0 & 1\\
-1 & 0
\end{array}
\right) \nonumber\\
&  =\frac{2i_{%
\mathbb{C}
}\beta}{\sqrt{N}}\sigma_{z}\sigma_{x}\text{,}%
\end{align}
where $\sigma_{x}$ and $\sigma_{z}$ are Pauli matrices. Observing that
$\left(  \sigma_{z}\sigma_{x}\right)  ^{2}=-I_{2\times2}$ and $\left\{
\sigma_{x}\text{, }\sigma_{z}\right\}  \overset{\text{def}}{=}\sigma_{x}%
\sigma_{z}+\sigma_{z}\sigma_{x}=0_{2\times2}$ (where $0_{2\times2}$ denotes
the two-by-two null matrix) , after some algebra (for further details, see
Appendix C), we obtain%
\begin{equation}
G\left(  t\right)  =e^{-i_{%
\mathbb{C}
}H_{\text{Fenner}}t}=e^{\frac{2\beta}{\sqrt{N}}\sigma_{z}\sigma_{x}t}%
=\cos\left(  \frac{2\beta}{\sqrt{N}}t\right)  I_{2\times2}+\sin\left(
\frac{2\beta}{\sqrt{N}}t\right)  \sigma_{z}\sigma_{x}\text{,} \label{derivare}%
\end{equation}
and the properly normalized time-dependent quantum state $\left\vert
\psi\left(  t\right)  \right\rangle $ becomes,%
\begin{align}
\left\vert \psi\left(  t\right)  \right\rangle  &  =G\left(  t\right)
\left\vert \psi\right\rangle \nonumber\\
&  =\left[  \cos\left(  \frac{2\beta}{\sqrt{N}}t\right)  -\frac{\alpha}{\beta
}\sin\left(  \frac{2\beta}{\sqrt{N}}t\right)  \right]  \left\vert
\psi\right\rangle +\frac{1}{\beta}\sin\left(  \frac{2\beta}{\sqrt{N}}t\right)
\left\vert \bar{x}\right\rangle \text{,} \label{fitt}%
\end{align}
From Eq. (\ref{fitt}), we note that the probability of the amplitude of the
target state $\left\vert \bar{x}\right\rangle $ is given by,%
\begin{equation}
\left\vert \left\langle \bar{x}\left\vert G\left(  t\right)  \right\vert
\psi\right\rangle \right\vert ^{2}=\left[  \alpha\cos\left(  \frac{2\beta
}{\sqrt{N}}t\right)  +\beta\sin\left(  \frac{2\beta}{\sqrt{N}}t\right)
\right]  ^{2}\text{.} \label{chichi}%
\end{equation}
From Eq. (\ref{chichi}), recalling that $\beta\overset{\text{def}}{=}%
\sqrt{\left(  N-1\right)  /N}$, we conclude that%
\begin{equation}
\left\vert \left\langle \bar{x}\left\vert G\left(  t\right)  \right\vert
\psi\right\rangle \right\vert ^{2}\overset{N\gg1}{\approx}\beta\sin\left(
\frac{2\beta}{\sqrt{N}}t\right)  =1
\end{equation}
requires%
\begin{equation}
t_{\text{Fenner}}\left(  N\right)  =\frac{N}{2\sqrt{N-1}}\sin^{-1}\left(
\sqrt{\frac{N}{N-1}}\right)  \overset{N\gg1}{\approx}\frac{\pi}{4}\sqrt
{N}\text{.} \label{tfenner}%
\end{equation}
Eq. (\ref{tfenner}) is the Hamiltonian search analog of Eq. (\ref{kgrover})
obtained in the original digital setting by Grover.

\subsection{Geometric algebra, the Fenner iterate, and the Grover iterate}

In what follows, we analyze the Fenner iterate and its connection to the
continuous-time generalization of the Grover iterate from a GA
perspective.\textbf{ }From Eq. (\ref{fitt}), recalling that $\left[
\mathfrak{cl}^{+}(3)\right]  ^{n}/E_{n}\ni e_{\psi}=\alpha e_{\bar{x}}+\beta
e_{\text{bad}}$ and setting $\theta_{\text{F}}\left(  t\right)  \overset
{\text{def}}{=}2\beta/\sqrt{N}t$, we obtain%
\begin{equation}
e_{\psi\left(  t\right)  }=\left\{  \beta\sin\left[  \theta_{\text{F}}\left(
t\right)  \right]  +\alpha\cos\left[  \theta_{\text{F}}\left(  t\right)
\right]  \right\}  e_{\bar{x}}+\left\{  -\alpha\sin\left[  \theta_{\text{F}%
}\left(  t\right)  \right]  +\beta\cos\left[  \theta_{\text{F}}\left(
t\right)  \right]  \right\}  e_{\text{bad}}\text{.} \label{lips}%
\end{equation}
From Eq. (\ref{lips}), we observe that%
\begin{equation}
\left[  e_{\psi\left(  t\right)  }\right]  _{\left\{  e_{\bar{x}}\text{,
}e_{\text{bad}}\right\}  }=\left(
\begin{array}
[c]{cc}%
\beta & \alpha\\
-\alpha & \beta
\end{array}
\right)  \binom{\sin\left[  \theta_{\text{F}}\left(  t\right)  \right]  }%
{\cos\left[  \theta_{\text{F}}\left(  t\right)  \right]  }\text{,} \label{ef}%
\end{equation}
that is, the multivector $e_{\psi\left(  t\right)  }$ expanded in the basis
$\left\{  e_{\bar{x}}\text{, }e_{\text{bad}}\right\}  $ has components that
can be obtained via a rotation of the components $\sin\left[  \theta
_{\text{F}}\left(  t\right)  \right]  $ and $\cos\left[  \theta_{\text{F}%
}\left(  t\right)  \right]  $. Indeed, the matrix%
\begin{equation}
A\overset{\text{def}}{=}\left(
\begin{array}
[c]{cc}%
\beta & \alpha\\
-\alpha & \beta
\end{array}
\right)  \text{,}%
\end{equation}
is such that $\det\left(  A\right)  =+1$ and $AA^{t}=A^{t}A=I_{2\times2}$. It
is therefore, a rotation. We also notice that,%
\begin{equation}
e_{\text{Grover}\left(  t\right)  }=\sin\left[  \theta_{\text{G}}\left(
t\right)  \right]  e_{\bar{x}}+\cos\left[  \theta_{\text{G}}\left(  t\right)
\right]  e_{\text{bad}}\text{,}%
\end{equation}
that is,%
\begin{equation}
\left[  e_{\text{Grover}\left(  t\right)  }\right]  _{\left\{  e_{\bar{x}%
}\text{, }e_{\text{bad}}\right\}  }=\binom{\sin\left[  \theta_{\text{G}%
}\left(  t\right)  \right]  }{\cos\left[  \theta_{\text{G}}\left(  t\right)
\right]  }\text{.} \label{eg}%
\end{equation}
To compare Eqs. (\ref{ef}) and (\ref{eg}), let $\theta_{\text{G}}\left(
t\right)  =\theta_{\text{F}}\left(  t\right)  =\theta\left(  t\right)  $.
Observe that the temporal linearity of $\theta_{\text{G}}\left(  t\right)  $
arises due to the fact\textbf{ }that the Hamiltonian being used for the
quantum search is assumed to be time-independent. We wish to uncover a new
basis of vectors $\left\{  e_{\bar{x}}^{\prime}\text{, }e_{\text{bad}}%
^{\prime}\right\}  $,%
\[
e_{\bar{x}}^{\prime}=f_{1}\left(  e_{\bar{x}}\text{, }e_{\text{bad}}\right)
\text{, and }e_{\text{bad}}^{\prime}=f_{2}\left(  e_{\bar{x}}\text{,
}e_{\text{bad}}\right)  \text{,}%
\]
such that in this new basis, one has%
\begin{equation}
\left[  e_{\text{Grover}\left(  t\right)  }\right]  _{\left\{  e_{\bar{x}%
}^{\prime}\text{, }e_{\text{bad}}^{\prime}\right\}  }=\left(
\begin{array}
[c]{cc}%
\beta & \alpha\\
-\alpha & \beta
\end{array}
\right)  \binom{\sin\left[  \theta_{\text{F}}\left(  t\right)  \right]  }%
{\cos\left[  \theta_{\text{F}}\left(  t\right)  \right]  }\text{.}%
\end{equation}
It is straightforward to check that the desired new basis is given by,%
\begin{equation}
\binom{e_{\bar{x}}^{\prime}}{e_{\text{bad}}^{\prime}}=\left(
\begin{array}
[c]{cc}%
\beta & \alpha\\
-\alpha & \beta
\end{array}
\right)  \binom{e_{\bar{x}}}{e_{\text{bad}}}\text{,}%
\end{equation}
that is,%
\begin{equation}
\binom{e_{\bar{x}}}{e_{\text{bad}}}=\left(
\begin{array}
[c]{cc}%
\beta & -\alpha\\
\alpha & \beta
\end{array}
\right)  \binom{e_{\bar{x}}^{\prime}}{e_{\text{bad}}^{\prime}}\text{.}%
\end{equation}
To identify $e_{\bar{x}}^{\prime}$ and $e_{\text{bad}}^{\prime}$, we impose,%
\begin{equation}
\left[  G_{\text{Grover}}\left(  t\right)  \right]  _{\text{old}}%
=e^{e_{\bar{x}}e_{\text{bad}}\theta_{G}\left(  t\right)  }\rightarrow\left[
G_{\text{Grover}}\left(  t\right)  \right]  _{\text{new}}=e^{e_{\bar{x}%
}^{\prime}e_{\text{bad}}^{\prime}\theta_{G}\left(  t\right)  }=e^{e_{3}%
e_{1}\theta_{F}\left(  t\right)  }=\left[  G_{\text{Fenner}}\left(  t\right)
\right]  _{\text{old}}\text{,} \label{dx1}%
\end{equation}
so that,%
\begin{equation}
e_{\bar{x}}^{\prime}=e_{3}\text{, and }e_{\text{bad}}^{\prime}=e_{1}\text{.}%
\end{equation}
We finally obtain,%
\begin{equation}
\binom{e_{3}}{e_{1}}=\left(
\begin{array}
[c]{cc}%
\beta & \alpha\\
-\alpha & \beta
\end{array}
\right)  \binom{e_{\bar{x}}}{e_{\text{bad}}}\text{ and, }\binom{e_{\bar{x}}%
}{e_{\text{bad}}}=\left(
\begin{array}
[c]{cc}%
\beta & -\alpha\\
\alpha & \beta
\end{array}
\right)  \binom{e_{3}}{e_{1}}\text{,}%
\end{equation}
with $e_{\bar{x}}^{\prime}e_{\text{bad}}^{\prime}=e_{\bar{x}}e_{\text{bad}}$
since,%
\begin{align}
e_{\bar{x}}^{\prime}e_{\text{bad}}^{\prime}  &  =e_{3}e_{1}\nonumber\\
&  =ie_{2}\nonumber\\
&  =\left(  \beta e_{\bar{x}}+\alpha e_{\text{bad}}\right)  \left(  -\alpha
e_{\bar{x}}+\beta e_{\text{bad}}\right) \nonumber\\
&  =-\alpha\beta+\beta^{2}e_{\bar{x}}e_{\text{bad}}-\alpha^{2}e_{\text{bad}%
}e_{\bar{x}}+\alpha\beta\nonumber\\
&  =\left(  \alpha^{2}+\beta^{2}\right)  e_{\bar{x}}e_{\text{bad}}\nonumber\\
&  =e_{\bar{x}}e_{\text{bad}}\text{.} \label{dx2}%
\end{align}
From Eqs. (\ref{dx1}) and (\ref{dx2}), we conclude that in GA terms the Fenner
iterate can be described in terms of the continuos-time Grover multivector
$e_{\text{Grover}\left(  t\right)  }$ which, in turn, is essentially a
rotation by $2\theta$ about the $e_{2}$ axis. We conclude that GA methods also
have utility in the digital-to-analog transition descriptions of quantum
computational software.

\section{Information geometry and the analog description}

Having performed the transition from the digital to the analog formulation of
Grover's algorithm, we are now ready to present an IG-based discussion
on\textbf{ }quantum computational software suitable for solving search
problems on a quantum hardware.

\subsection{Statistical distinguishability}

Before presenting the IG description of Grover's algorithm, several
considerations have to be carried out. Information geometry consists of the
application of differential geometrical methods to the study of families of
probabilities, both classical and quantum, either parametric or nonparametric
\cite{amari}.

In the classical information geometric setting, there is essentially a unique
statistical distance that quantifies the distinguishability between two
probability distributions $p_{\theta}\left(  x\right)  $ and $p_{\theta
+d\theta}\left(  x\right)  $. The quantity $x$ is in the domain of definition
$\mathcal{X}$ of the probability distribution while $\theta=\left(  \theta
^{1},...,\theta^{n}\right)  $ are the $n$-\emph{real} statistical parameters
that parametrize $p_{\theta}\left(  x\right)  $ that belongs to the
$n$-dimensional statistical model (or, statistical manifold) $\mathcal{M}_{s}%
$,%
\begin{equation}
\mathcal{M}_{s}\overset{\text{def}}{=}\left\{  p_{\theta}\left(  x\right)
=p\left(  x|\theta\right)  :\theta=\left(  \theta^{1},...,\theta^{n}\right)
\in\Theta\right\}  \text{,}%
\end{equation}
where $\Theta\subset%
\mathbb{R}
^{n}$ is the so-called parameter space. Except for an overall multiplicative
constant, there is a unique monotone Riemannian metric with the property of
having its line element reduced under stochastic maps known as Markov
morphisms \cite{cencov, campbell}. This metric is the so-called Fisher-Rao
information metric defined as,%
\begin{equation}
g_{ij}^{\left(  \text{FR}\right)  }\left(  \theta\right)  \overset{\text{def}%
}{=}\int_{\mathcal{X}}p_{\theta}\left(  x\right)  \partial_{i}\log\left[
p_{\theta}\left(  x\right)  \right]  \partial_{j}\log\left[  p_{\theta}\left(
x\right)  \right]  dx\text{,} \label{fr}%
\end{equation}
where $\partial_{i}\overset{\text{def}}{=}\frac{\partial}{\partial\theta^{i}}$
with $1\leq i\leq n$ and the infinitesimal line element $ds^{2}$ on
$\mathcal{M}_{s}$ becomes,%
\begin{equation}
ds^{2}\overset{\text{def}}{=}g_{ij}^{\left(  \text{FR}\right)  }\left(
\theta\right)  d\theta^{i}d\theta^{j}\text{.}%
\end{equation}
It is convenient to note that since,%
\begin{equation}
\sqrt{p}\frac{\partial\log p}{\partial\theta}=\frac{1}{\sqrt{p}}\frac{\partial
p}{\partial\theta}=2\frac{\partial\sqrt{p}}{\partial\theta}\text{,}%
\end{equation}
Eq. (\ref{fr}) can be rewritten as,%
\begin{equation}
g_{ij}^{\left(  \text{FR}\right)  }\left(  \theta\right)  =4\int_{\mathcal{X}%
}\partial_{i}\sqrt{p_{\theta}\left(  x\right)  }\partial_{j}\sqrt{p_{\theta
}\left(  x\right)  }dx\text{.} \label{fr2}%
\end{equation}
In the quantum information setting, probability distributions are replaced by
density matrices. Quantum statistical distinguishability between density
matrices requires that their distance on the space of density matrices must
decrease under coarse-graining (that is, stochastic maps). Unlike the
classical case, there are infinitely many monotone Riemannian metrics on the
space of density matrices that fulfill this requirement \cite{morozova, petz,
grasselli}. Even quantum generalizations of the very same classical Fisher-Rao
information metric are not unique. In general, two classically identical
expressions in Eqs. (\ref{fr}) and (\ref{fr2}) differ when they are extended
to a quantum setting. This difference is a manifestation of the
non-commutative nature of quantum mechanics and is reminiscent of the idea of
quantum discord \cite{luo}. Substituting the integral in Eq. (\ref{fr2}) with
the trace and the probability distributions $p_{\theta}$ with density
operators $\rho_{\theta}$, this heuristic quantum generalization of Eq.
(\ref{fr2}) becomes the so-called quantum Wigner-Yanase metric \cite{wigner,
gibilisco},%
\begin{equation}
g_{ij}^{\left(  \text{WY}\right)  }\left(  \theta\right)  \overset{\text{def}%
}{=}4\text{Tr}\left[  \left(  \partial_{i}\rho_{\theta}\right)  \left(
\partial_{j}\rho_{\theta}\right)  \right]  \text{.} \label{gwy}%
\end{equation}
For pure quantum states $\rho_{\theta}\overset{\text{def}}{=}\left\vert
\psi_{\theta}\right\rangle \left\langle \psi_{\theta}\right\vert $ with
normalized quantum states $\left\vert \psi_{\theta}\right\rangle $ such that
$\psi_{\theta}=\psi\left(  \theta\right)  $ and $\rho_{\theta}^{2}%
=\rho_{\theta}$, after some algebra (for further details, we refer to
\cite{cafaro2012B}), it follows that Eq. (\ref{gwy}) can be rewritten as%
\begin{equation}
g_{ij}^{\left(  \text{WY}\right)  }\left(  \theta\right)  =4\left[
\operatorname{Re}\left\langle \partial_{i}\psi_{\theta}|\partial_{j}%
\psi_{\theta}\right\rangle +\left\langle \partial_{i}\psi_{\theta}%
|\psi_{\theta}\right\rangle \left\langle \partial_{j}\psi_{\theta}%
|\psi_{\theta}\right\rangle \right]  \text{,} \label{gwy2}%
\end{equation}
where $\operatorname{Re}\left(  \cdot\right)  $ denotes the \emph{real} part
of a \emph{complex} number.

\subsection{The metric structure}

The metric in Eq. (\ref{gwy2}) is exactly four times the so-called
Fubini-Study metric employed in the standard geometric analysis of quantum
evolution \cite{anandan90}. The Fubini-Study metric is a gauge invariant
metric on the \emph{complex} projective Hilbert space, the manifold of
Hilbert-space rays \cite{provost}. If we assume that $\left\vert \psi_{\theta
}\right\rangle $ in Eq. (\ref{gwy2}) is given by,%
\begin{equation}
\left\vert \psi_{\theta}\right\rangle =\sum_{l=1}^{N}\sqrt{p_{l}\left(
\theta\right)  }e^{i_{%
\mathbb{C}
}\phi_{l}\left(  \theta\right)  }\left\vert l\right\rangle \text{,}%
\end{equation}
after some straightforward but tedious algebra (for details, we refer to
Appendix A in Ref. \cite{cafaro2012B}), we obtain that the infinitesimal
Wigner-Yanase line element becomes%
\begin{equation}
ds^{2}\overset{\text{def}}{=}g_{ij}^{\left(  \text{WY}\right)  }\left(
\theta\right)  d\theta^{i}d\theta^{j}=\left\{  \sum_{l=1}^{N}\frac{\dot{p}%
_{l}^{2}}{p_{l}}+4\left[  \sum_{l=1}^{N}p_{l}\dot{\phi}_{l}^{2}-\left(
\sum_{l=1}^{N}p_{l}\dot{\phi}_{l}\right)  ^{2}\right]  \right\}  d\theta
^{2}\text{,} \label{linea}%
\end{equation}
where,%
\begin{equation}
\dot{p}_{l}\overset{\text{def}}{=}\frac{dp_{l}\left(  \theta\right)  }%
{d\theta}\text{ and, }\dot{\phi}_{l}\overset{\text{def}}{=}\frac{d\phi
_{l}\left(  \theta\right)  }{d\theta}\text{.}%
\end{equation}
In what follows, we assume that the density operator $\rho_{\theta}$ is
parametrized by $n$-\emph{real} parameters $\theta=\left(  \theta
^{1},...,\theta^{n}\right)  $ and satisfies the standard von Neumann equation,%
\begin{equation}
\frac{\partial\rho_{\theta}}{\partial\theta}+\frac{i_{%
\mathbb{C}
}}{\hbar}\left[  T\text{, }\rho_{\theta}\right]  =0\text{,}%
\end{equation}
where $T$ is the generator of temporal shift such that,%
\begin{equation}
\rho_{\text{initial}}\overset{\text{def}}{=}\rho_{0}\rightarrow\rho
_{\text{final}}\overset{\text{def}}{=}\rho_{\theta}=e^{-\frac{i_{%
\mathbb{C}
}}{\hbar}\theta T}\rho_{0}e^{\frac{i_{%
\mathbb{C}
}}{\hbar}\theta T}\text{.} \label{roe}%
\end{equation}
The set of quantum states $\rho_{\theta}$ in Eq. (\ref{roe}) characterizes the
quantum evolution manifold of pure quantum states. Furthermore, the set of
parameters $\theta$ can be viewed as a local coordinate system on such
manifold endowed with a metric structure defined by the Wigner-Yanase metric
in Eq. (\ref{gwy2}). Finally, we point out that once the monotone Riemannian
metric in Eq. (\ref{gwy2}) is explicitly known, the other differential
geometric quantities such as, for instance, the Christoffel connection
coefficients, scalar and sectional curvatures, and the Riemannian curvature
tensor can be calculated from it, in principle. For example, assuming
$\theta^{m}=\theta^{m}\left(  \tau\right)  $, the geodesic equation is given
by%
\begin{equation}
\frac{d^{2}\theta^{m}\left(  \tau\right)  }{d\tau^{2}}+\Gamma_{ij}^{m}\left(
\theta\right)  \frac{d\theta^{i}}{d\tau}\frac{d\theta^{j}}{d\tau}=0\text{,}
\label{GE}%
\end{equation}
where the Christoffel connection coefficients $\Gamma_{ij}^{m}$ are defined as
\cite{amari},%
\begin{equation}
\Gamma_{ij}^{m}\overset{\text{def}}{=}\frac{1}{2}g^{ml}\left(  \partial
_{i}g_{lj}+\partial_{j}g_{il}-\partial_{l}g_{ij}\right)  \text{,}%
\end{equation}
where, from a classical standpoint, $g_{ij}$ equals $g_{ij}^{\left(
\text{FR}\right)  }$ in Eq. (\ref{fr}). For further technical details on both
classical and quantum aspects of information geometry, we refer to
\cite{amari}. For a classical derivation of Eq. (\ref{GE}), we refer to
Appendix F.

\subsection{Information geometric description: the dissipationless nature}

We are now ready to present an IG description of Grover's algorithm in terms
of optimal (dissipationless) geodesic paths. To better emphasize its
relevance, a comment on the dissipationless nature of quantum computing is required.

\subsubsection{On the dissipationless nature of quantum computing}

A quantum computer obeys the laws of quantum mechanics. Its unique feature is
that it can control a superposition of computation paths simultaneously and
produce a final state that depends on the interference of these paths. To
preserve quantum interferences that form the computation, quantum systems that
act as qubits must be sufficiently isolated from external influences.
Therefore, a quantum computer is non-dissipative and can operate at low
temperatures \cite{kane98}. As mentioned earlier, the dissipationless nature
of quantum computing is due to the fact that, in principle, quantum
computation can only occur in systems that are almost completely isolated from
the environment. Therefore, despite the extraordinarily difficult practical
conditions to satisfy, they must dissipate no energy during the process of
computation. In \cite{kane98}, the minimum operable temperature of a quantum
computer was assumed to be $T=100mK$. Observe that the dissipationless nature
of quantum computing and the low temperature operations of the computer imply
that, in principle, fluctuations (perturbations) are negligible since they can
be kept extremely small. In summary, isolation means unitarity. Unitarity
means conservation of probability and invertibility. Invertibility means
reversibility. Reversibility means no dissipation. Experimentally, there is a
number of factors that can go wrong and we can witness a departure from a
unitary and dissipationless scenario. From and experimental standpoint, the
fact that measured probabilities are not exactly zero or one is primarily due
to imperfect laser-cooling, imperfect state and detector preparation, and
decoherence effects. Several sources responsible for decoherence can be
identified when considering a physical realization of a quantum computer
\cite{cirac95, monroe95, divincenzo00a}: instabilities in the laser beam
power, instabilities in the relative position of the ion with respect to the
beams, coupling of internal vibrational modes due to trap imperfections,
fluctuating external magnetic fields which can modulate the qubit phases,
instabilities in the voltage amplitude, external heating and dissipation of
the ion motion \cite{itano89, itano95} and, finally, experimental
uncertainties in interaction times, laser detunings, positions of ions, and
phases of lasers. Interestingly, Grover pointed out in \cite{grover98} that
one of the limitations of his quantum search algorithm was the assumption of
the absolute absence of fluctuations in the operators $U$ and $U^{-1}$ during
the algorithmic\textbf{ }sequence: these operators stay the same at all time
steps. In what follows, we shall see the elegance and power of IG methods when
applied to the characterization of quantum computational software.

\subsubsection{Dissipationless information geometric evolution}

In its digital representation, Grover's algorithm is essentially a sequence of
unitary transformations on a pure state that evolves with discrete $k$. In the
limiting scenario of $N$ approaching infinity, we can replace $\theta
_{k}=\left(  2k+1\right)  \theta$ with $\theta\left(  \tau\right)  $. Then,
the output state in Eq. (\ref{iter}) can be approximately described by a
quantum mechanical wave-vector $\left\vert \psi_{\theta}\right\rangle $ given
by,%
\begin{equation}
\left\vert \psi_{\theta}\right\rangle \overset{\text{def}}{=}%
{\displaystyle\sum\limits_{l=0}^{N-1}}
\sqrt{p_{l}\left(  \theta\right)  }\left\vert l\right\rangle =\sqrt
{p_{0}\left(  \theta\right)  }\left\vert 0\right\rangle +%
{\displaystyle\sum\limits_{l=1}^{N-1}}
\sqrt{p_{l}\left(  \theta\right)  }\left\vert l\right\rangle \text{,}
\label{azz}%
\end{equation}
where $\left\vert \bar{x}\right\rangle $ and $\left\vert \psi_{\text{bad}%
}\right\rangle $ are being replaced with $\left\vert 0\right\rangle $ and the
second term in the RHS of Eq. (\ref{azz}), respectively. Furthermore
$p_{0}\left(  \theta\right)  $ and $p_{l}\left(  \theta\right)  $ with $1\leq
l\leq N-1$ are defined as,%
\begin{equation}
p_{0}\left(  \theta\right)  \overset{\text{def}}{=}\sin^{2}\left(
\theta\right)  \text{, and }p_{l}\left(  \theta\right)  \overset{\text{def}%
}{=}\frac{1}{N-1}\cos^{2}\left(  \theta\right)  \text{,} \label{azz1}%
\end{equation}
respectively. At this point, we introduce the so-called Fisher information
function $\mathcal{F}\left(  \theta\right)  $ defined as \cite{cover}%
\begin{equation}
\mathcal{F}\left(  \theta\right)  \overset{\text{def}}{=}\sum_{l=0}^{N-1}%
p_{l}\left(  \frac{\partial\log p_{l}}{\partial\theta}\right)  ^{2}\text{.}
\label{fif}%
\end{equation}
It can be shown that $\mathcal{F}\left(  \theta\right)  $ is invariant under
unitary transformations applied to arbitrary normalized quantum states
$\left\vert \psi_{\theta}\right\rangle $ (for further details, see
\cite{cafaro2012B}). For further details on the Fisher information function
from both information-theoretic and statistical mechanical standpoints, we
refer to Appendix D\textbf{.} For the specific case of normalized states in
Eq. (\ref{azz}), using Eqs. (\ref{linea}), (\ref{azz}), and (\ref{azz1}), the
function $\mathcal{F}\left(  \theta\right)  $ in Eq. (\ref{fif}) becomes%
\begin{equation}
\mathcal{F}\left(  \theta\right)  =\sum_{l=0}^{N-1}\frac{\dot{p}_{l}^{2}%
}{p_{l}}=4\sum_{l=0}^{N-1}\left(  \frac{\partial\sqrt{p_{l}}}{\partial\theta
}\right)  ^{2}=4\text{.} \label{FF}%
\end{equation}
In the rest of the manuscript, Eq. (\ref{FF}), a peculiarity of Grover's
information geometric evolution, shall be referred to as the
\emph{parametric-independence constraint} on the Fisher information function.
The relevance of the constancy of the Fisher information function from a
quantum information geometric viewpoint can be understood once we observe the
connection between $\mathcal{F}\left(  \theta\right)  $ and the generalized
mechanical kinetic energy $\mathcal{K}\left(  \theta\right)  $ (where $\theta$
is viewed as a temporal shift) \cite{luo33},%
\begin{equation}
\mathcal{K}\left(  \theta\right)  \overset{\text{def}}{=}\sum_{l=0}%
^{N-1}\left\vert \frac{\partial\psi_{\theta}\left(  l\right)  }{\partial
\theta}\right\vert ^{2}\text{,} \label{relation1}%
\end{equation}
where $\psi_{\theta}\left(  l\right)  =\left\langle l|\psi_{\theta
}\right\rangle $ and $p_{l}\left(  \theta\right)  =\left\vert \psi_{\theta
}\left(  l\right)  \right\vert ^{2}$. The quantity $\mathcal{K}\left(
\theta\right)  $ in Eq. (\ref{relation1}) can be rewritten as (for further
details, we refer to Appendix E),%
\begin{equation}
\mathcal{K}\left(  \theta\right)  =\frac{1}{4}\mathcal{F}\left(
\theta\right)  +\sum_{l=0}^{N-1}J_{\theta}^{2}\left(  l\right)  \left\vert
\psi_{\theta}\left(  l\right)  \right\vert ^{2}\text{,} \label{J3}%
\end{equation}
where $J_{\theta}\left(  l\right)  $ plays the role of the normalized quantum
mechanical current density and is defined as \cite{luo33},
\begin{equation}
J_{\theta}\left(  l\right)  \overset{\text{def}}{=}\frac{1}{2i_{%
\mathbb{C}
}\left\vert \psi_{\theta}\left(  l\right)  \right\vert ^{2}}\left(
\frac{\partial\psi_{\theta}\left(  l\right)  }{\partial\theta}\psi_{\theta
}^{\ast}\left(  l\right)  -\psi_{\theta}\left(  l\right)  \frac{\partial
\psi_{\theta}^{\ast}\left(  l\right)  }{\partial\theta}\right)  \text{,}
\label{jtetal}%
\end{equation}
while $\mathcal{F}\left(  \theta\right)  $ is the same as in\ Eq. (\ref{FF}).
The symbol $\ast$ in Eq. (\ref{jtetal}) denotes \emph{complex} conjugation.
Assuming,\textbf{\ }%
\begin{equation}
\psi_{\theta}\left(  l\right)  \overset{\text{def}}{=}\sqrt{p_{l}\left(
\theta\right)  }e^{i\phi_{l}\left(  \theta\right)  }\text{\textbf{,}}%
\end{equation}
after some algebra, we obtain%
\begin{equation}
\mathcal{K}\left(  \theta\right)  =\left\langle \dot{\psi}_{\theta}|\dot{\psi
}_{\theta}\right\rangle \text{, }\mathcal{F}\left(  \theta\right)  =\sum
_{l=0}^{N-1}\frac{\dot{p}_{l}^{2}}{p_{l}}\text{ and, }J_{\theta}\left(
l\right)  =\dot{\phi}_{l}\left(  \theta\right)  \text{,} \label{J4}%
\end{equation}
\textbf{\ }with\textbf{\ }$\dot{\psi}_{\theta}\overset{\text{def}}{=}%
\frac{d\psi\left(  \theta\right)  }{d\theta}$. Using Eq. (\ref{J4}), Eq.
(\ref{J3}) becomes%
\begin{equation}
\left\langle \dot{\psi}_{\theta}|\dot{\psi}_{\theta}\right\rangle =\frac{1}%
{4}\sum_{l=0}^{N-1}\frac{\dot{p}_{l}^{2}}{p_{l}}+\sum_{l=0}^{N-1}p_{l}%
\dot{\phi}_{l}^{2}\text{.} \label{relation2}%
\end{equation}
Since $\mathcal{F}\left(  \theta\right)  =4$ and $J_{\theta}\left(  l\right)
=0$\textbf{\ }for any\textbf{\ }$0\leq l\leq N-1$\textbf{, }from Eq.
(\ref{relation2}) we arrive at%
\begin{equation}
\mathcal{K}\left(  \theta\right)  =\left\langle \dot{\psi}_{\theta}|\dot{\psi
}_{\theta}\right\rangle =1\text{.} \label{KK}%
\end{equation}
Finally, we conclude that the constancy of the Fisher information function in
Eq. (\ref{FF}) implies that Grover's algorithm is characterized by a constant
mechanical kinetic energy in Eq. (\ref{KK}) with the absence of any
dissipation of statistical nature. \ A few considerations of note are as
follows. Our information geometric investigation can be linked to other lines
of research involving the concept of information, Riemannian geometry, and
thermodynamics. Recall that the analog (continuous-time) version of Grover's
algorithm as proposed by Farhi and Guttmann required the application of a
time-independent Hamiltonian for a time\textbf{ }interval $\Delta
t\propto\sqrt{N}$. In \cite{cerf01}, an analog version of Grover's algorithm
with a time-dependent Hamiltonian satisfying the adiabaticy condition at all
times (that is, locally) by varying the evolution rate of the driving
Hamiltonian was proposed. Such local adiabatic evolution of the analog version
of Grover's algorithm exhibits the same quadratic speedup as the original
version of the algorithm. Furthermore, exploiting fundamental limits\textbf{
}to the minimum time duration (maximum speed) for the dynamical evolution to
an orthogonal quantum state \cite{vaidman92, margolus98} (specifically,
$\Delta t\geq\frac{h}{4}\frac{1}{\Delta E}$ with $\Delta E$ denoting the
uncertainty in energy of the system and $h$ being the Planck constant) and
establishing a novel time-energy relation, a general upper bound for the
evolution speed for driving a quantum state to a target state where the
driving Hamiltonian is time-independent was estimated in \cite{luo04} and,
under special conditions, it was shown that this speed reduces to that
proposed by Farhi and Guttmann \cite{luo04}. The relevance of the concept of
Fisher information in statistical mechanics was discussed in \cite{crooks12}
where it was reported that the Fisher information of a thermodynamic system
characterizes the size of fluctuations about equilibrium. Specifically, it was
shown that the Fisher information of a probability distribution at thermal
equilibrium with respect to the inverse temperature $\beta\overset{\text{def}%
}{=}\left(  k_{B}T\right)  ^{-1}$ equals the energy fluctuations
\cite{crooks12},%
\begin{equation}
\mathcal{F}\left(  \beta\right)  =\sigma_{E}^{2}\overset{\text{def}}%
{=}\left\langle \left(  E-\left\langle E\right\rangle \right)  ^{2}%
\right\rangle \text{,} \label{statmech}%
\end{equation}
where $k_{B}$ is the Boltzmann constant with $k_{B}\approx1.38\times
10^{-23}\left[  \text{MKSA}\right]  $. For more details on the statistical
mechanical nature of the Fisher information function, we refer to Appendix D.
More recently, inspired by the study of Riemannian geometric aspects of
nonequilibrium thermodynamics \cite{ruppeiner79, ruppeiner95} applied to
nanoscale systems \cite{crooks07, crooksPRL12}, the concept of Fisher
information together with differential geometric methods have been employed to
uncover optimal (minimum dissipation, energy efficient) protocols for
designing nonequilibrium nanoscale (stochastic) machines, heat engines, and
magnetic refrigerators \cite{crooks15}. Within this Riemannian geometric
framework, geodesics represent nonequilibrium control protocols with the
lowest achievable dissipation. It would certainly be valuable to investigate
the IG description of quantum computational software in the presence of
possible energy fluctuations.\ We leave this topic for future efforts.

\subsubsection{Geodesic paths}

Some technical details on the minimization of the action functional in
differential geometric terms appear in Appendix F. Here,\textbf{ }the geodesic
path related to Grover's algorithm can be obtained by minimizing the action
functional $\mathcal{S}\left[  p_{l}\left(  \theta\right)  \right]  $ defined
as,%
\begin{equation}
\mathcal{S}\left[  p_{l}\left(  \theta\right)  \right]  \overset{\text{def}%
}{=}\int\sqrt{ds_{\text{WY}}^{2}}=\int\mathcal{L}\left(  \dot{p}_{l}\left(
\theta\right)  \text{, }p_{l}\left(  \theta\right)  \right)  d\theta\text{,}%
\end{equation}
with the Lagrangian quantity $\mathcal{L}\left(  \dot{p}_{l}\left(
\theta\right)  \text{, }p_{l}\left(  \theta\right)  \right)  $ given by,%
\begin{equation}
\mathcal{L}\left(  \dot{p}_{l}\left(  \theta\right)  \text{, }p_{l}\left(
\theta\right)  \right)  \overset{\text{def}}{=}\left[  \sum_{k=0}^{N-1}%
\frac{\dot{p}_{l}^{2}\left(  \theta\right)  }{p_{l}\left(  \theta\right)
}\right]  ^{\frac{1}{2}}\text{, } \label{lagrangian2}%
\end{equation}
where the parametric probabilities $p_{l}\left(  \theta\right)  $ satisfy the
normalization constraint,%
\begin{equation}
\sum_{l=0}^{N-1}p_{l}\left(  \theta\right)  =1\text{.} \label{magna}%
\end{equation}
Following \cite{wootters}, we can simplify our analysis by taking into
consideration the change of variable $p_{l}\left(  \theta\right)  \rightarrow
q_{l}^{2}\left(  \theta\right)  $. Employing the method of Lagrange
multipliers, the new action functional\textbf{ }$\mathcal{S}_{new}\left[
q_{l}\left(  \theta\right)  \right]  $ to be minimized becomes,%
\begin{equation}
\mathcal{S}_{new}\left[  q_{l}\left(  \theta\right)  \right]  =\int
\mathcal{L}_{new}\left(  \dot{q}_{l}\left(  \theta\right)  \text{, }%
q_{l}\left(  \theta\right)  \right)  d\theta=\int\left\{  \left[  4\sum
_{l=1}^{N}\dot{q}_{l}^{2}\left(  \theta\right)  \right]  ^{\frac{1}{2}%
}-\lambda\left(  \sum_{l=1}^{N}q_{l}^{2}\left(  \theta\right)  -1\right)
\right\}  d\theta\text{,}%
\end{equation}
where $\lambda$ is the Lagrange multiplier and $\mathcal{L}_{new}\left(
\dot{q}_{l}\left(  \theta\right)  \text{, }q_{l}\left(  \theta\right)
\right)  $ is the new Lagrangian quantity. The path that minimizes the
action\textbf{ }functional $\mathcal{S}_{new}\left[  q_{l}\left(
\theta\right)  \right]  $ satisfies the so-called \emph{actuality constraint},%
\begin{equation}
\frac{\delta\mathcal{S}_{new}\left[  q_{l}\left(  \theta\right)  \right]
}{\delta q_{l}\left(  \theta\right)  }=0\text{.} \label{actual}%
\end{equation}
After some algebra, it is found that the constraint in Eq. (\ref{actual})
leads to the Euler-Lagrange equation,%
\begin{equation}
\frac{d^{2}q_{l}\left(  \theta\right)  }{d\theta^{2}}-\frac{\mathcal{\dot{L}%
}\left(  \dot{q}_{l}\left(  \theta\right)  \text{, }q_{l}\left(
\theta\right)  \right)  }{\mathcal{L}\left(  \dot{q}_{l}\left(  \theta\right)
\text{, }q_{l}\left(  \theta\right)  \right)  }\frac{dq_{l}\left(
\theta\right)  }{d\theta}+\frac{\lambda}{2}\mathcal{L}\left(  \dot{q}%
_{l}\left(  \theta\right)  \text{, }q_{l}\left(  \theta\right)  \right)
q_{l}\left(  \theta\right)  =0\text{,} \label{GExx}%
\end{equation}
where $\mathcal{\dot{L}}=\frac{d\mathcal{L}}{d\theta}$ with $\mathcal{L}%
\left(  \dot{q}_{l}\left(  \theta\right)  \text{, }q_{l}\left(  \theta\right)
\right)  $ given in Eq. (\ref{lagrangian2}) and $q_{l}^{2}\left(
\theta\right)  =p_{l}\left(  \theta\right)  $. In the case of Grover's
dynamics, $\mathcal{F}\left(  \theta\right)  =4$ and, therefore,
$\mathcal{L}\left(  \dot{q}_{l}\left(  \theta\right)  \text{, }q_{l}\left(
\theta\right)  \right)  =2$ and $\mathcal{\dot{L}}\left(  \dot{q}_{l}\left(
\theta\right)  \text{, }q_{l}\left(  \theta\right)  \right)  =0$. Therefore,
setting the Lagrange multiplier $\lambda=1$ in order to satisfy Eq.
(\ref{magna}), the geodesic Eq. (\ref{GExx}) becomes%
\begin{equation}
\frac{d^{2}q_{l}\left(  \theta\right)  }{d\theta^{2}}+q_{l}\left(
\theta\right)  =0\text{.} \label{GE2}%
\end{equation}
The solution\textbf{ }$q\left(  \theta\right)  $\textbf{ }of Eq. (\ref{GE2})
is given by\textbf{,}%
\begin{equation}
q\left(  \theta\right)  =\left(  q_{0}\left(  \theta\right)  \text{, }%
q_{1}\left(  \theta\right)  \text{,..., }q_{N-1}\left(  \theta\right)
\right)  \text{,}%
\end{equation}
with\textbf{,}%
\begin{equation}
q_{0}\left(  \theta\right)  =\sin\left(  \theta\right)  \text{, and }%
q_{l}\left(  \theta\right)  =\frac{1}{\sqrt{N-1}}\cos\left(  \theta\right)
\text{,} \label{fuckyou}%
\end{equation}
for any\textbf{ }$1\leq l\leq N-1$\textbf{. }Recalling that\textbf{ }%
$q_{l}^{2}\left(  \theta\right)  =p_{l}\left(  \theta\right)  $\textbf{, }Eq.
(\ref{fuckyou}) leads to the the\textbf{ }$N$\textbf{-}dimensional probability
vector\textbf{ }$p\overset{\text{def}}{=}\left(  p_{0}\left(  \theta\right)
\text{, }p_{1}\left(  \theta\right)  \text{,..., }p_{N-1}\left(
\theta\right)  \right)  $\textbf{ }with\textbf{ }$p_{l}\left(  \theta\right)
$\textbf{ }defined in Eq. (\ref{azz1})\textbf{. }The quantity $p$
characterizes a geodesic path that satisfies both the parametric-independence
and actuality constraints in\ Eqs. (\ref{FF}) and (\ref{actual}),
respectively. It is the path for which the quantum\textbf{ }Fisher information
action functional $\mathcal{S}_{new}\left[  q_{l}\left(  \theta\right)
\right]  $ achieves an extremal value.

\subsection{Information geometric description: the quadratic speedup}

We investigate here the IG perspective of the quadratic speedup relation and
the superfluity of the Walsh-Hadamard operation in achieving this improvement
in a quantum search problem.

\subsubsection{The metric on the complex projective Hilbert space}

We recall that the necessity of introducing a metric on the manifold of
Hilbert space rays, the so-called \emph{complex} projective Hilbert space $%
\mathbb{C}
P$, is due to the non-observability of a phase of a vector in the Hilbert
space $\mathcal{H}$ \cite{provost}: $\left\vert \psi\left(  \theta\right)
\right\rangle $ and $e^{i_{%
\mathbb{C}
}\alpha\left(  \theta\right)  }\left\vert \psi\left(  \theta\right)
\right\rangle $ with $\theta$ denoting an $n$-dimensional parameter define the
same point on the manifold of rays. Wootters showed that the absolute
statistical distance between two different preparations of the same quantum
system equals the angle in Hilbert space between the corresponding rays
\cite{wootters},%
\begin{equation}
\cos^{2}\left(  \frac{1}{2}\theta\right)  =\left\vert \left\langle \psi
|\phi\right\rangle \right\vert ^{2}\text{,} \label{wootters}%
\end{equation}
where the RHS in Eq. (\ref{wootters}) denotes the probability of transition
from state $\left\vert \psi\right\rangle $ to state $\left\vert \phi
\right\rangle $. Up to a constant factor, the only Riemannian metric on the
set of rays which is invariant under all possible unitary time evolutions is
the angle in Hilbert space. For neighboring pure states $\left\vert
\psi\right\rangle $ and $\left\vert \phi\right\rangle $, let us set the
statistical distance $ds_{\text{PS}}=\theta$. Using Eq. (\ref{wootters}), we
obtain
\begin{equation}
ds_{\text{FS}}^{2}\overset{\text{def}}{=}\frac{1}{4}ds_{\text{PS}}%
^{2}=\left\{  \cos^{-1}\left[  \left\vert \left\langle \psi|\phi\right\rangle
\right\vert \right]  \right\}  ^{2}\text{,} \label{fsxx}%
\end{equation}
where $ds_{\text{FS}}^{2}$ denotes the Fubini-Study infinitesimal line element
\cite{anandan90, braunstein94}. In the working hypothesis that $ds_{\text{PS}%
}\ll1$, Eq. (\ref{wootters}) yields%
\begin{equation}
\cos^{2}\left(  \frac{1}{2}ds_{\text{PS}}\right)  \approx1-\frac{1}%
{4}ds_{\text{PS}}^{2}=\left\vert \left\langle \psi|\phi\right\rangle
\right\vert ^{2}\text{.} \label{fsxx2}%
\end{equation}
Combining Eqs. (\ref{fsxx}) and (\ref{fsxx2}), we finally obtain the
expression of the natural metric on the manifold of Hilbert space rays, the
Fubini-Study metric \cite{braunstein94},%
\begin{equation}
ds_{\text{FS}}^{2}=1-\left\vert \left\langle \psi|\phi\right\rangle
\right\vert ^{2}\text{.}%
\end{equation}
From an information geometric viewpoint, as mentioned earlier, we shall be
using the Wigner-Yanase metric which is essentially four times the
Fubini-Study metric.

\subsubsection{Information geometric steps-counting}

Within the information geometric setting, the number of steps needed to arrive
at the final target state $\left\vert \psi_{\theta_{f}}\right\rangle $ from
the initial state $\left\vert \psi_{\theta_{i}}\right\rangle $ can be
estimated as follows.

First, we note that the distance covered in a single step leading to
$\left\vert \psi_{\theta_{i+1}}\right\rangle \overset{\text{def}}%
{=}G\left\vert \psi_{\theta_{i}}\right\rangle $ from $\left\vert \psi
_{\theta_{i}}\right\rangle $ where $G$ is the Grover iterate can be computed
in an explicit manner using the Wigner-Yanase metric once we find $\left\vert
\psi_{\theta_{i+1}}\right\rangle $. Recall that the Grover iterate can be
written as $G\overset{\text{def}}{=}-I_{i}U^{-1}I_{f}U$ where $U$ is a unitary
operator while $I_{i}$ and $I_{f}$ are defined as,
\begin{equation}
I_{i}\overset{\text{def}}{=}1-2\left\vert \psi_{\theta_{i}}\right\rangle
\left\langle \psi_{\theta_{i}}\right\vert \text{, and }I_{f}\overset
{\text{def}}{=}1-2\left\vert \psi_{\theta_{f}}\right\rangle \left\langle
\psi_{\theta_{f}}\right\vert \text{,} \label{iif}%
\end{equation}
respectively. We emphasize that the unitary operator $U$ was originally
represented by the Walsh-Hadamard operation in Grover's quantum search
algorithm. However, Grover showed later that any unitary operation can be used
in his algorithm to arrive at the same quadratic speedup \cite{grover98}.
Using Eq. (\ref{iif}) and recalling that%
\begin{equation}
U_{if}\overset{\text{def}}{=}\left\langle \psi_{\theta_{i}}\left\vert
U\right\vert \psi_{\theta_{f}}\right\rangle \text{, }U_{if}^{\ast}%
\overset{\text{def}}{=}\left\langle \psi_{\theta_{f}}\left\vert U^{-1}%
\right\vert \psi_{\theta_{i}}\right\rangle \text{, and }\left\vert
U_{if}\right\vert ^{2}\overset{\text{def}}{=}U_{if}U_{if}^{\ast}\text{,}%
\end{equation}
after some algebra, we obtain%
\begin{align}
\left\vert \psi_{\theta_{i+1}}\right\rangle  &  =G\left\vert \psi_{\theta_{i}%
}\right\rangle \nonumber\\
&  =-I_{i}U^{-1}I_{f}U\left\vert \psi_{\theta_{i}}\right\rangle \nonumber\\
&  =\left(  1-4\left\vert U_{fi}\right\vert ^{2}\right)  \left\vert
\psi_{\theta_{i}}\right\rangle +2U_{fi}U^{-1}\left\vert \psi_{\theta_{f}%
}\right\rangle \text{,}%
\end{align}
that is,%
\begin{equation}
\left\vert \psi_{\theta_{i+1}}\right\rangle =\left(  1-4\left\vert
U_{fi}\right\vert ^{2}\right)  \left\vert \psi_{\theta_{i}}\right\rangle
+2U_{fi}U^{-1}\left\vert \psi_{\theta_{f}}\right\rangle \text{.} \label{step1}%
\end{equation}
The Wigner-Yanase line element between $\left\vert \psi_{\theta_{i}%
}\right\rangle $ and $\left\vert \psi_{\theta_{i+1}}\right\rangle $ is given
by,%
\begin{equation}
\left[  ds_{\text{WY}}^{2}\right]  _{i\rightarrow i+1}=4\left[  1-\left\vert
\left\langle \psi_{\theta_{i}}|\psi_{\theta_{i+1}}\right\rangle \right\vert
^{2}\right]  \text{.} \label{dstep1}%
\end{equation}
Substituting Eq. (\ref{step1}) into Eq. (\ref{dstep1}), we find%
\begin{equation}
\left[  ds_{\text{WY}}^{2}\right]  _{i\rightarrow i+1}=16\left\vert
U_{fi}\right\vert ^{2}\left[  1-\left\vert U_{fi}\right\vert ^{2}\right]
\text{.} \label{onestep}%
\end{equation}
Second, assume that\textbf{ }$\left\vert \psi_{\theta_{f}}\right\rangle
$\textbf{ }is reached through a succession of a countable set of intermediate
steps. Specifically,\textbf{ }$\left\vert \psi_{\theta_{f}}\right\rangle
$\textbf{ }can be reached as follows. First, we transition in\textbf{
}$\mathcal{N}_{s}$\textbf{ }steps from\textbf{ }$\left\vert \psi_{\theta_{i}%
}\right\rangle $\textbf{ }to\textbf{ }$\left\vert \tilde{\psi}_{\theta_{f}%
}\right\rangle $\textbf{,}%
\begin{equation}
\left\vert \tilde{\psi}_{\theta_{f}}\right\rangle \overset{\text{def}}%
{=}U^{-1}\left\vert \psi_{\theta_{f}}\right\rangle \text{.}%
\end{equation}
Then, with a single application of the unitary operator\textbf{ }$U$\textbf{,
}the target state\textbf{ }$\left\vert \psi_{\theta_{f}}\right\rangle
$\textbf{ }is reached. Observe that\textbf{ }$G$\textbf{ }preserves the
two-dimensional space spanned by\textbf{ }$\left\vert \psi_{\theta_{i}%
}\right\rangle $\textbf{ }and\textbf{ }$\left\vert \tilde{\psi}_{\theta_{f}%
}\right\rangle $\textbf{ }because of Eq. (\ref{step1}) and since%
\begin{equation}
G\left\vert \tilde{\psi}_{\theta_{f}}\right\rangle =\left\vert \tilde{\psi
}_{\theta_{f}}\right\rangle -2\left\langle \psi_{\theta_{i}}|\tilde{\psi
}_{\theta_{f}}\right\rangle \left\vert \psi_{\theta_{i}}\right\rangle \text{.}%
\end{equation}
We point out that in the original digital description of Grover's work, the
evolution of the algorithm occurs with a discrete number of iterations. In the
analog description, the evolution becomes approximately continuous in the
limiting case in which the dimensionality $N\overset{\text{def}}{=}2^{n}$ of
the Hilbert space approaches infinity. For this reason, it is reasonable to
assume that the number of iterations remains countable also in the analog
case. A simple calculation shows that the Wigner-Yanase line element between
$\left\vert \psi_{\theta_{i}}\right\rangle $ and $\left\vert \tilde{\psi
}_{\theta_{f}}\right\rangle \overset{\text{def}}{=}U^{-1}\left\vert
\psi_{\theta_{f}}\right\rangle $ is given by,
\begin{equation}
\left[  ds_{\text{WY}}^{2}\right]  _{i\rightarrow f}=4\left[  1-\left\vert
\left\langle \psi_{\theta_{i}}|\tilde{\psi}_{\theta_{f}}\right\rangle
\right\vert ^{2}\right]  =4\left[  1-\left\vert U_{fi}\right\vert ^{2}\right]
\text{.} \label{k-1-steps}%
\end{equation}
Assuming that\textbf{ }$\left\vert U_{fi}\right\vert \ll1$ (for further
details, see Appendix G)\textbf{, }it happens that%
\begin{equation}
\left[  ds_{\text{WY}}^{2}\right]  _{i\rightarrow i+1}=\left[  ds_{\text{WY}%
}^{2}\right]  _{i+1\rightarrow i+2}=...=\left[  ds_{\text{WY}}^{2}\right]
_{i+l-2\rightarrow i+l-1}=\left[  ds_{\text{WY}}^{2}\right]
_{i+l-1\rightarrow i+l}\text{,}%
\end{equation}
where\textbf{ }$l\in%
\mathbb{R}
$ and\textbf{ }$\left\vert \psi_{\theta_{i+l}}\right\rangle \overset
{\text{def}}{=}G^{l}\left\vert \psi_{\theta_{i}}\right\rangle $\textbf{ }is
equal to%
\begin{equation}
\left\vert \psi_{\theta_{i+l}}\right\rangle =\left(  1-4\left\vert
U_{fi}\right\vert ^{2}\right)  \left\vert \psi_{\theta_{i+l-1}}\right\rangle
+2U_{fi}G^{l-1}U^{-1}\left\vert \psi_{\theta_{f}}\right\rangle \text{.}%
\end{equation}
At this point, the number of equal-length steps $\mathcal{N}_{s}$ needed to
navigate a distance $\left[  ds_{\text{WY}}\right]  _{i\rightarrow f}$ becomes%
\begin{equation}
\mathcal{N}_{s}\overset{\text{def}}{=}\left(  \frac{\left[  ds_{\text{WY}}%
^{2}\right]  _{i\rightarrow f}}{\left[  ds_{\text{WY}}^{2}\right]
_{i\rightarrow i+1}}\right)  ^{\frac{1}{2}}\text{.}%
\end{equation}
Using Eqs. (\ref{onestep}) and (\ref{k-1-steps}), in the working hypothesis
that $\left\vert U_{fi}\right\vert \ll1$, we obtain%
\begin{equation}
\mathcal{N}_{s}\propto\frac{1}{\sqrt{\left\vert U_{fi}\right\vert ^{2}}%
}\text{.} \label{finaln}%
\end{equation}
Observe that when\textbf{ }$\left\vert U_{fi}\right\vert \ll1$\textbf{, }the
quantum states\textbf{ }$\left\vert \psi_{\theta_{i}}\right\rangle $\textbf{
}and $\left\vert \tilde{\psi}_{\theta_{f}}\right\rangle $\textbf{ }are
approximately orthogonal. Eq. (\ref{finaln}) implies that the number of steps
needed to arrive at the final target state from the initial state is inversely
proportional to the square root of the probability of transition from the
$i$-th state to the $f$-th state under the unitary evolution operator $U$. Eq.
(\ref{finaln}) is the quantum information geometric analog of what Grover and
Pati obtained by means of matrix algebra arguments and geometric quantum
mechanical methods in Refs. \cite{pati98} and \cite{grover98}, respectively.
Note that $U_{fi}$ denotes the amplitude of arriving at state $\left\vert
\psi_{\theta_{f}}\right\rangle $ by applying $U$ to $\left\vert \psi
_{\theta_{i}}\right\rangle $. At this point, if we perform an experiment and
observe the system, $\left\vert U_{fi}\right\vert ^{2}$ becomes the
probability of arriving at the target state $\left\vert \psi_{\theta_{f}%
}\right\rangle $ by starting from the initial basis state $\left\vert
\psi_{\theta_{i}}\right\rangle $. According to this line of reasoning, in
order to achieve a single success one needs to perform this experiment at
least $\mathcal{O}\left(  1/\left\vert U_{fi}\right\vert ^{2}\right)  $ number
of times. If $\left\vert U_{fi}\right\vert \ll1$, from Eq. (\ref{finaln}) we
conclude that the number of steps (repetition of this experiment) reduces to
$\mathcal{O}\left(  1/\left\vert U_{fi}\right\vert \right)  $. This is a
remarkable improvement achievable with any unitary $U$. As a final side
remark, we point out that assuming $N\overset{\text{def}}{=}2^{n}$ is needed
if $U$ equals the Walsh-Hadamard operation since this is the only case in
which this operator is well-defined. However, since the Grover iterate $G$ can
be constructed with arbitrary unitary operations $U$, this assumption can be
removed \cite{boyer98}. Our analysis based on quantum IG confirms this fact.

\section{On Grover's fixed-point quantum search algorithm}

In this Section, we present a qualitative discussion concerning Grover's fixed
point phase-$\frac{\pi}{3}$ quantum search algorithm \cite{grover05A,
grover05B} from both a geometric algebra and information geometry perspectives.

\subsection{Preliminaries}

Grover's original quantum search is characterized by an iterative procedure
that, in the asymptotic limit of a very large number $N$ of unsorted items,
allows us to identify with certainty one of the $N_{m}$ marked items in
approximately $k$-iterations with $k\approx\frac{\pi}{4}\sqrt{N/N_{m}\text{.}%
}$ We recall that Grover's original quantum search algorithm is constructed by
iteratively applying to $\left\vert 0\right\rangle ^{\otimes n}\in
\mathcal{H}_{2}^{n}$ the operator\textbf{ }$G$\textbf{,}%
\begin{equation}
G\overset{\text{def}}{=}-R_{\psi}^{\left(  \pi\right)  }U^{\dagger}R_{\bar{x}%
}^{\left(  \pi\right)  }U\text{,}%
\end{equation}
where the unitary $U$ is the Walsh-Hadamard transformation while $R_{\psi
}^{\left(  \pi\right)  }$ and $R_{\bar{x}}^{\left(  \pi\right)  }$ are the
selective inversion operators of the states $\left\vert \psi\right\rangle $
and $\left\vert \bar{x}\right\rangle $, respectively. They are given by,%
\begin{equation}
R_{\psi}^{\left(  \pi\right)  }\overset{\text{def}}{=}I-2\left\vert
\psi\right\rangle \left\langle \psi\right\vert \text{ and }R_{\bar{x}%
}^{\left(  \pi\right)  }\overset{\text{def}}{=}I-2\left\vert \bar
{x}\right\rangle \left\langle \bar{x}\right\vert \text{,}%
\end{equation}
respectively. The failure probability after $k$-iterations of Grover's
original search algorithm is given by,%
\begin{equation}
P_{\text{failure}}^{\left(  k\text{-iterations}\right)  }\left(  k\right)
=1-\left\vert \left\langle \bar{x}|G^{k}|\psi\right\rangle \right\vert
^{2}=1-\sin^{2}\left[  \left(  2k+1\right)  \theta\right]  \text{,}%
\end{equation}
where $P_{\text{failure}}^{\left(  k\text{-iterations}\right)  }\left(
k\right)  =P_{\text{failure}}^{\left(  k\text{-iterations}\right)  }\left(
k+T\right)  $ with $T\overset{\text{def}}{=}\pi/\left(  2\theta\right)  $. A
limitation of this search scheme is that success can be achieved only when the
number $N_{m}$ of marked items is known beforehand so that the number $k$ of
iterations landing the initial state closest to the marked state can be
predicted \cite{brassard97}. For instance, since the iterative procedure is
essentially a rotation, once the target is reached, further iterations will
drive the system away from the target state. As a consequence, the success
probability rapidly decreases and drops to zero. In summary, to know when to
terminate the algorithm, the fraction $N/N_{m}$ of marked items must be known
precisely. To undertake a quantum search when the number $N_{m}$ of marked
items is not known ahead of time, Grover proposed a different kind of search
method based upon a recursive procedure known as fixed-point quantum search
\cite{grover05A, grover05B}. In this different kind of quantum search, the
system is always moved towards the target state with the consequence that the
success probability is continuously amplified in a monotonic fashion. In other
towards, there is a monotonic convergence toward the solution.

Since its introduction, Grover's fixed-point quantum search algorithm has been
implemented under a variety of different schemes. In \cite{xiao05}, it was
implemented on a two-qubit NMR quantum computer. The two qubits were described
in terms of the $^{1}H$ (protium) and the $^{13}C$ (carbon-$13$) nuclei. The
quantum search was performed in the case of $N=4$ unsorted items and either
$N_{m}=1$ or $N_{m}=2$ marked items. In \cite{bhole16}, an improved version of
Grover's fixed-point quantum search algorithm exhibiting quadratic speedup
proposed by Yoder and collaborators in \cite{yoder14} was implemented on a
three-qubit quantum computer by using the bang-bang control technique. The
three qubits were described in terms of the $^{1}H$ (protium), the $^{13}C$
(carbon-$13$), and the $^{19}F$ (fluorine-$19$) nuclei. The quantum search was
performed in the case of $N=4$ unsorted items and $N_{m}=1$ marked items.

Grover noticed that the achievement of the monotonic convergence toward the
solution (the fixed point) could not have been achieved via iterations of the
same unitary transformation since unitary transformations are characterized by
eigenvalues of magnitude one. Therefore, any iteration is inherently periodic.
To solve this issue, Grover proposed a clever recursive scheme defined in
terms of suitably designed distinct unitary operations performed at successive
iterations.\ Specifically, the recursive scheme proposed by Grover is given
by,%
\begin{equation}
U_{k+1}\overset{\text{def}}{=}U_{k}R_{\psi}^{\left(  \pi/3\right)  }%
U_{k}^{\dagger}R_{\bar{x}}^{\left(  \pi/3\right)  }U_{k}\text{,}%
\end{equation}
where $U_{0}=U$, $U$ is some unitary operator while $R_{\psi}^{\left(
\pi/3\right)  }$ and $R_{\bar{x}}^{\left(  \pi/3\right)  }$ are the selective
$\pi/3$-phase shift operators of the initial (source) and final (target)
states, respectively. They are defined as,
\begin{equation}
R_{\psi}^{\left(  \pi/3\right)  }\overset{\text{def}}{=}I-\left[  1-e^{i_{%
\mathbb{C}
}\frac{\pi}{3}}\right]  \left\vert \psi\right\rangle \left\langle
\psi\right\vert \text{ and }R_{\bar{x}}^{\left(  \pi/3\right)  }%
\overset{\text{def}}{=}I-\left[  1-e^{i_{%
\mathbb{C}
}\frac{\pi}{3}}\right]  \left\vert \bar{x}\right\rangle \left\langle \bar
{x}\right\vert \text{,}%
\end{equation}
respectively. The failure probability after $k$-recursive steps of Grover's
fixed-point search algorithm is given by,%
\begin{equation}
P_{\text{failure}}^{\left(  k\text{-recursive steps}\right)  }\left(
k\right)  =1-\left\vert \left\langle \bar{x}|U_{k}|\psi\right\rangle
\right\vert ^{2}=\epsilon^{3^{k}}\text{,}%
\end{equation}
where $\epsilon\overset{\text{def}}{=}1-\left\vert \left\langle \bar{x}%
|U_{0}|\psi\right\rangle \right\vert ^{2}$ $\ll1$ denotes the probability of
failure after no-recursive step. For the sake of later convenience, let us
introduce the generalized Grover iterate $G\left(  \alpha\text{, }%
\beta\right)  \overset{\text{def}}{=}-R_{\psi}^{\left(  \alpha\right)
}R_{\bar{x}}^{\left(  \beta\right)  }$ defined as the product of two
generalized reflection operators with $\alpha$, $\beta\in%
\mathbb{R}
$. The two dimensional matrix representation of $G\left(  \alpha\text{, }%
\beta\right)  $ with respect to $\left\{  \left\vert \psi_{\text{bad}%
}\right\rangle \text{, }\left\vert \bar{x}\right\rangle \right\}  $ is given
by,%
\begin{align}
\left[  G\left(  \alpha\text{, }\beta\right)  \right]   &  =\left(
\begin{array}
[c]{cc}%
\left\langle \psi_{\text{bad}}|G\left(  \alpha\text{, }\beta\right)
|\psi_{\text{bad}}\right\rangle  & \left\langle \psi_{\text{bad}}|G\left(
\alpha\text{, }\beta\right)  |\bar{x}\right\rangle \\
\left\langle \bar{x}|G\left(  \alpha\text{, }\beta\right)  |\psi_{\text{bad}%
}\right\rangle  & \left\langle \bar{x}|G\left(  \alpha\text{, }\beta\right)
|\bar{x}\right\rangle
\end{array}
\right)  \nonumber\\
& \nonumber\\
&  =\left(
\begin{array}
[c]{cc}%
\left(  1-e^{i_{%
\mathbb{C}
}\alpha}\right)  \cos^{2}\left(  \theta\right)  -1 & e^{i_{%
\mathbb{C}
}\beta}\left(  1-e^{i_{%
\mathbb{C}
}\alpha}\right)  \sin\left(  \theta\right)  \cos\left(  \theta\right)  \\
\left(  1-e^{i_{%
\mathbb{C}
}\alpha}\right)  \sin\left(  \theta\right)  \cos\left(  \theta\right)   &
e^{i_{%
\mathbb{C}
}\beta}\left[  \left(  1-e^{i_{%
\mathbb{C}
}\alpha}\right)  \sin^{2}\left(  \theta\right)  -1\right]
\end{array}
\right)  \text{,}%
\end{align}
with $\sin\left(  \theta\right)  \overset{\text{def}}{=}\left\langle \bar
{x}|\psi\right\rangle $ and $\cos\left(  \theta\right)  \overset{\text{def}%
}{=}\left\langle \psi_{\text{bad}}|\psi\right\rangle $. We observe that for
$\alpha=\beta=\pi$ we recover the matrix representation of the original Grover
iterate (see Eq. (\ref{mg}) in Section III).

\subsection{Insights from geometric algebra}

We present a few remarks in order to make a parallel comparison between the
original Grover algorithm and the fixed-point search algorithm using GA. The
original Grover algorithm generates a sequence of quantum states $\left\{
\left\vert \psi_{k}\right\rangle \right\}  $ such that,%
\begin{equation}
\left\vert \psi_{k+1}\right\rangle =G\left\vert \psi_{k}\right\rangle
=G^{2}\left\vert \psi_{k-1}\right\rangle =...=G^{k-1}\left\vert \psi
_{2}\right\rangle =G^{k}\left\vert \psi_{1}\right\rangle \equiv G^{k}%
\left\vert \psi\right\rangle \text{,}\label{itersec}%
\end{equation}
where we assume that the operator $G$ in Eq. (\ref{itersec}) is given by
$G\overset{\text{def}}{=}-R_{\psi}^{\left(  \pi\right)  }\circ R_{\bar{x}%
}^{\left(  \pi\right)  }$. Using the GA language introduced in Section III,
the quantum state $\left\vert \psi_{k+1}\right\rangle $ expressed as the
action of $G^{k}$ onto $\left\vert \psi\right\rangle $ becomes the
nonhomogeneous multivector expressed as%
\begin{equation}
\left\vert \psi_{k+1}\right\rangle =G^{k}\left\vert \psi\right\rangle
\overset{\text{GA}}{\longrightarrow}g^{k}e_{\psi}\left(  g^{\dagger}\right)
^{k}\text{,}%
\end{equation}
where the bivector $g\overset{\text{def}}{=}e_{\psi}e_{\text{bad}}=\exp\left(
e_{\bar{x}}e_{\text{bad}}\theta\right)  $ with $e_{\psi}$, $e_{\text{bad}}$,
and\textbf{ }$e_{\bar{x}}\in\left[  \emph{cl}^{+}\left(  3\right)  \right]
^{n}/E_{n}$. The fixed-search quantum search algorithm generates a sequence of
quantum states $\left\{  \left\vert \psi_{k}\right\rangle \right\}  $ such
that,%
\begin{equation}
\left\vert \psi_{k+1}\right\rangle =\mathcal{G}_{k}\left\vert \psi
_{k}\right\rangle =\mathcal{G}_{k}\mathcal{G}_{k-1}\left\vert \psi
_{k-1}\right\rangle =...=\mathcal{G}_{k}\mathcal{G}_{k-1}...\mathcal{G}%
_{2}\mathcal{G}_{1}\left\vert \psi_{1}\right\rangle \equiv\mathcal{G}%
_{k}\mathcal{G}_{k-1}...\mathcal{G}_{2}\mathcal{G}_{1}\left\vert
\psi\right\rangle \text{,}\label{nesting}%
\end{equation}
where we consider that the operator $\mathcal{G}_{k}$ in Eq. (\ref{nesting})
is defined as $\mathcal{G}_{k}\overset{\text{def}}{=}R_{\psi_{k}}^{\left(
\pi/3\right)  }\circ R_{\bar{x}}^{\left(  \pi/3\right)  }$. More specifically,
following the analysis in Ref. \cite{iwai}, we obtain%
\begin{equation}
\left\vert \psi_{k}\right\rangle =\sqrt{1-\left\vert \left\langle \bar{x}%
|\psi_{k}\right\rangle \right\vert ^{2}}\left\vert \psi_{\text{bad, }%
k}\right\rangle +\left\langle \bar{x}|\psi_{k}\right\rangle \left\vert \bar
{x}\right\rangle \text{,}\label{151}%
\end{equation}
where,%
\begin{equation}
c_{k+1}=\left\langle \bar{x}|\psi_{k+1}\right\rangle \overset{\text{def}}%
{=}e^{i_{%
\mathbb{C}
}\frac{\pi}{3}}\left(  e^{i_{%
\mathbb{C}
}\frac{\pi}{3}}+\epsilon_{k}\right)  c_{k}\text{, }%
\end{equation}
$\epsilon_{k+1}=\epsilon_{k}^{3}$, $\epsilon_{k}\overset{\text{def}}%
{=}1-\left\vert c_{k}\right\vert ^{2}=\epsilon^{3^{k-1}}$, and%
\begin{equation}
\left\vert \psi_{\text{bad, }k+1}\right\rangle \overset{\text{def}}{=}e^{i_{%
\mathbb{C}
}\frac{\pi}{3}}\left\vert \psi_{\text{bad, }k}\right\rangle \text{.}%
\label{152}%
\end{equation}
Using Eqs. (\ref{151}) and (\ref{152}), $\left\vert \psi_{k}\right\rangle
=\mathcal{G}_{k-1}\left\vert \psi_{k-1}\right\rangle $ becomes%
\begin{equation}
\left\vert \psi_{k}\right\rangle =c_{k}\left\vert \bar{x}\right\rangle
+\sqrt{\epsilon_{k}}e^{i_{%
\mathbb{C}
}\frac{\pi}{3}\left(  k-1\right)  }\left\vert \psi_{\text{bad, }%
1}\right\rangle \text{.}\label{fikappa}%
\end{equation}
Using the GA language introduced in Section III, the quantum state $\left\vert
\psi_{k+1}\right\rangle $ expressed as the action of $\mathcal{G}%
_{k}\mathcal{G}_{k-1}...\mathcal{G}_{2}\mathcal{G}_{1}$ onto\textbf{
}$\left\vert \psi\right\rangle $\textbf{ }becomes the nonhomogeneous
multivector expressed as%
\begin{equation}
\left\vert \psi_{k+1}\right\rangle =\mathcal{G}_{k}\mathcal{G}_{k-1}%
...\mathcal{G}_{2}\mathcal{G}_{1}\left\vert \psi\right\rangle \overset
{\text{GA}}{\longrightarrow}\left(  g_{k}g_{k-1}...g_{2}g_{1}\right)  e_{\psi
}\left(  g_{k}g_{k-1}...g_{2}g_{1}\right)  ^{\dagger}\text{,}\label{gkk2}%
\end{equation}
where the bivector $g_{m}\overset{\text{def}}{=}e_{\psi}e_{\text{bad, }m}%
$\textbf{. }Eq. (\ref{gkk2}) is the analog of Eq. (\ref{gk}) in Section IV.
Observe that the sequence of nested rotations\textbf{ }$\mathcal{G}$\textbf{
}is expressed in the GA\ language in terms of a so-called sandwiching product
\cite{mann07}.

\subsection{Insights from information geometry}

We recall that the convergence toward a fixed point is achieved by means of
unitary transformations in Grover's fixed-point quantum algorithm. However,
such a convergence can also be obtained by introducing irreversible damping by
projective measurement operations in the quantum algorithm \cite{tulsi}.
Damping can be produced by coupling the system to a bath or, more
artificially, by employing a single ancilla spin that is measured after
undergoing a unitary evolution with the quantum system \cite{ari09}.

\subsubsection{Monotonic decrease of the Fisher information}

From Eq. (\ref{fikappa}), noticing that $\sqrt{\epsilon_{k}}=\exp\left[
\frac{1}{2}3^{k-1}\ln\left(  \epsilon\right)  \right]  $ with $\ln\left(
\epsilon\right)  \leq0$ since $0\leq\epsilon\ll1$, the output state of
Grover's fixed-point quantum search algorithm can be formally rewritten in
terms of a quantum mechanical wave-vector $\left\vert \psi_{\theta
}\right\rangle $ given by,%
\begin{equation}
\left\vert \psi_{\theta}\right\rangle \overset{\text{def}}{=}%
{\displaystyle\sum\limits_{l=0}^{N-1}}
\sqrt{p_{l}\left(  \theta\right)  }e^{i_{%
\mathbb{C}
}\phi_{l}\left(  \theta\right)  }\left\vert l\right\rangle =\sqrt{p_{0}\left(
\theta\right)  }e^{i_{%
\mathbb{C}
}\phi_{0}\left(  \theta\right)  }\left\vert 0\right\rangle +%
{\displaystyle\sum\limits_{l=1}^{N-1}}
\sqrt{p_{l}\left(  \theta\right)  }e^{i_{%
\mathbb{C}
}\phi_{l}\left(  \theta\right)  }\left\vert l\right\rangle \text{,}%
\label{azzz}%
\end{equation}
where $\left\vert \bar{x}\right\rangle $ and $\left\vert \psi_{\text{bad}%
}\right\rangle $ are being replaced with $\left\vert 0\right\rangle $ and the
second term in the RHS of Eq. (\ref{azzz}), respectively. For the sake of
clarity, we limit our discussion to $N=2$. In this case, probabilities
$p_{0}\left(  \theta\right)  $ and $p_{1}\left(  \theta\right)  $ in\ Eq.
(\ref{azzz}) can be parametrized as follows,%
\begin{equation}
p_{0}\left(  \theta\right)  \overset{\text{def}}{=}1-\xi\left(  \theta\right)
e^{-\theta}\text{, and }p_{1}\left(  \theta\right)  \overset{\text{def}}{=}%
\xi\left(  \theta\right)  e^{-\theta}\text{,}\label{azz12}%
\end{equation}
respectively. The function $\xi\left(  \theta\right)  $ is assumed to be some
differentiable function with values in the interval $\left]  0,1\right]  $ so
that the leading asymptotic behavior of the Fisher information $\mathcal{F}%
\left(  \theta\right)  $ is determined by the exponentially decaying term
$e^{-\theta}$ with $\theta$ being the temporal shift. Indeed\textbf{,}
$\mathcal{F}\left(  \theta\right)  $\textbf{ }is defined in terms of
$p_{0}\left(  \theta\right)  $ and $p_{1}\left(  \theta\right)  $ in Eq.
(\ref{azz12}) and becomes%
\begin{equation}
\mathcal{F}\left(  \theta\right)  =\left\{  \frac{\left[  \dot{\xi}\left(
\theta\right)  -\xi\left(  \theta\right)  \right]  ^{2}}{\xi\left(
\theta\right)  \left[  1-\xi\left(  \theta\right)  e^{-\theta}\right]
}\right\}  e^{-\theta}\text{,}\label{fifa}%
\end{equation}
where $\dot{\xi}=\frac{d\xi}{d\theta}$. Note that unlike the case of Grover's
original quantum search scheme, the Fisher information corresponding to
Grover's fixed-point quantum algorithm is not constant. Furthermore, although
the Fisher information does not depend on the presence of complex phase
factors in quantum states, such phase information becomes important when
linking the Fisher information to the generalized mechanical kinetic energy
$\mathcal{K}\left(  \theta\right)  $. In particular, the presence of
non-constant phases $\phi_{l}=\phi_{l}\left(  \theta\right)  $ with $\dot
{\phi}_{l}\left(  \theta\right)  \neq0$ implies the emergence of normalized
quantum mechanical current densities $J_{\theta}\left(  l\right)  =\dot{\phi
}_{l}\left(  \theta\right)  $. In what follows, as a simplifying working
assumption, we consider that $\phi_{l}\left(  \theta\right)  =\phi_{l}$ are
nonzero but constant phases. We leave a more rigorous investigation of this
nontrivial issue concerning a quantum information geometric characterization
of the effect of time-varying phases to future scientific efforts
\cite{carluccio, nico16}. Within our working hypothesis, the generalized
mechanical kinetic energy $\mathcal{K}\left(  \theta\right)  $ becomes%
\begin{equation}
\mathcal{K}\left(  \theta\right)  =\frac{1}{4}\left\{  \frac{\left[  \dot{\xi
}\left(  \theta\right)  -\xi\left(  \theta\right)  \right]  ^{2}}{\xi\left(
\theta\right)  \left[  1-\xi\left(  \theta\right)  e^{-\theta}\right]
}\right\}  e^{-\theta}\text{.}\label{KAKKA}%
\end{equation}
We remark that due to the polynomial nature of\textbf{ }$\xi\left(
\theta\right)  $, $\mathcal{K}\left(  \theta\right)  $ in Eq. (\ref{KAKKA})
exhibits a non-increasing behavior in the asymptotic limit of very large
$\theta$ values. It is interesting to observe that the monotonic decreasing
behavior of the Fisher information appears naturally in dissipative kinetic
models where energy is non-increasing. For instance, the energy in the
Boltzmann equation is decreasing when the binary collisions are dissipative
and the Fisher information decreases along the solutions of either the
Boltzmann \cite{toscani} or the Landau \cite{villani} equation of Maxwellian
molecules. In the language of thermodynamics, our approximate information
geometric analysis of Grover's fixed-point quantum search algorithm seems to
lead to the conclusion that the size of the energy fluctuations about
equilibrium represented by the Fisher information decreases during the
geometric evolution of the system towards the target item.

\subsubsection{Quantum search with damping}

Following the line of reasoning outlined in the case of Grover's original
quantum search algorithm, the geodesic path related to Grover's fixed-point
quantum search algorithm can be obtained by minimizing the action functional
$\mathcal{S}\left[  p_{l}\left(  \theta\right)  \right]  $ defined as,%
\begin{equation}
\mathcal{S}\left[  p_{l}\left(  \theta\right)  \right]  \overset{\text{def}%
}{=}\int\sqrt{ds_{\text{WY}}^{2}}\text{.}%
\end{equation}
It follows that the geodesic paths\textbf{ }$q_{l}=q_{l}\left(  \theta\right)
$\textbf{ }with\textbf{ }$p_{l}\left(  \theta\right)  \overset{\text{def}}%
{=}q_{l}^{2}\left(  \theta\right)  $ satisfy the relation,%
\begin{equation}
\frac{d^{2}q_{l}\left(  \theta\right)  }{d\theta^{2}}-\frac{\mathcal{\dot{L}%
}\left(  \dot{q}_{l}\left(  \theta\right)  \text{, }q_{l}\left(
\theta\right)  \right)  }{\mathcal{L}\left(  \dot{q}_{l}\left(  \theta\right)
\text{, }q_{l}\left(  \theta\right)  \right)  }\frac{dq_{l}\left(
\theta\right)  }{d\theta}+\frac{1}{2}\mathcal{L}\left(  \dot{q}_{l}\left(
\theta\right)  \text{, }q_{l}\left(  \theta\right)  \right)  q_{l}\left(
\theta\right)  =0\text{,}\label{UE}%
\end{equation}
where $\mathcal{L}\left(  \theta\right)  \overset{\text{def}}{=}%
\mathcal{L}\left(  \dot{q}_{l}\left(  \theta\right)  \text{, }q_{l}\left(
\theta\right)  \right)  $ denotes the Lagrangian of the system under
investigation with $\mathcal{L}\left(  \dot{p}_{l}\left(  \theta\right)
\text{, }p_{l}\left(  \theta\right)  \right)  \overset{\text{def}}{=}%
\sqrt{\mathcal{F}\left(  \theta\right)  }$. We observe that while Eq.
(\ref{UE}) reduces to a simple harmonic oscillator equation in the case of
Grover's original algorithm (see Eq. (\ref{GE2}) in Section VI), it now
becomes a slightly more complicated version of the standard damped harmonic
oscillator equation,%
\begin{equation}
\ddot{x}+b\dot{x}+kx=0\text{,}\label{dao}%
\end{equation}
where\textbf{ }$b$ denotes the damping factor, $k$ is the spring constant, and
$\dot{x}=\frac{dx}{dt}$. The general solution of Eq. (\ref{dao}) is given by,%
\begin{equation}
x\left(  t\right)  =Ae^{-\frac{b}{2}t}\sin\left(  \frac{\sqrt{4k-b^{2}}}%
{2}t+\varphi\right)  \text{,}\label{critical}%
\end{equation}
where $A$ and $\varphi$ are two \emph{real} integration constants. We shall
omit here the detailed investigation of geodesic paths in the case of the
fixed-point quantum algorithm. However, to better grasp the complications that
can arise when departing from the standard damped harmonic oscillator motion
and to get a bit closer to the actual quantum scenario, in what follows we
take into consideration the case in which $\mathcal{\dot{L}}/\mathcal{L}$ is
kept constant while $\mathcal{L}$ is not constant. Specifically, we assume
$\mathcal{L}\left(  \theta\right)  =\mathcal{L}_{0}e^{-\gamma\theta}$ with
$\mathcal{L}_{0}$ and $\gamma$ constant quantities in $%
\mathbb{R}
_{+}\backslash\left\{  0\right\}  $. In this case, omitting the index\textbf{
}$l$\textbf{ }in\textbf{ }$q_{l}\left(  \theta\right)  $\textbf{,} Eq.
(\ref{UE}) becomes%
\begin{equation}
\frac{d^{2}q\left(  \theta\right)  }{d\theta^{2}}+\gamma\frac{dq\left(
\theta\right)  }{d\theta}+\frac{1}{2}\mathcal{L}_{0}e^{-\gamma\theta}q\left(
\theta\right)  =0\text{.}\label{man}%
\end{equation}
The general solution of Eq. (\ref{man}) can be formally written in closed form
as,%
\begin{equation}
q\left(  \theta\right)  =\sqrt{\frac{\mathcal{L}_{0}}{2\gamma^{2}}}%
e^{-\frac{\gamma}{2}\theta}\left[  A\mathcal{J}_{1}\left(  \sqrt
{\frac{2\mathcal{L}_{0}}{\gamma^{2}}}e^{-\frac{\gamma}{2}\theta}\right)
+B\mathcal{Y}_{1}\left(  \sqrt{\frac{2\mathcal{L}_{0}}{\gamma^{2}}}%
e^{-\frac{\gamma}{2}\theta}\right)  \right]  \text{,}\label{Ax}%
\end{equation}
where $A$ and $B$ are suitable integration constants while $\mathcal{J}_{\nu
}\left(  x\right)  $ and $\mathcal{Y}_{\nu}\left(  x\right)  $ with integer
order $\nu=+1$ are Bessel functions of the first and the second kind
\cite{stegun}, respectively. We recall that the limiting form of the Bessel
function $\mathcal{J}_{\nu}\left(  x\right)  $ for small arguments in the case
in which $\nu$ is fixed and $x$ approaches zero is given by \cite{stegun},%
\begin{equation}
\mathcal{J}_{\nu}\left(  x\right)  \approx\left(  \frac{1}{2}x\right)  ^{\nu
}\frac{1}{\Gamma\left(  \nu+1\right)  }\text{,}\label{Axx}%
\end{equation}
where $\Gamma$ is the Euler Gamma function with $\Gamma\left(  \nu+1\right)
\overset{\text{def}}{=}\nu!$. If we set $\mathcal{L}_{0}=2$ and $\gamma=1$,
recalling that\textbf{ }$p_{l}\left(  \theta\right)  \overset{\text{def}}%
{=}q_{l}^{2}\left(  \theta\right)  $ and assuming $N=2$, we obtain from Eqs.
(\ref{Ax}) and (\ref{Axx}) the following asymptotic expansions of the
probabilities $p_{0}\left(  \theta\right)  $ and $p_{1}\left(  \theta\right)
$,
\begin{equation}
p_{0}\left(  \theta\right)  \approx1-\xi_{\mathcal{A}}\left(  \theta\right)
e^{-\theta}\text{, and }p_{1}\left(  \theta\right)  \approx\xi_{\mathcal{A}%
}\left(  \theta\right)  e^{-\theta}\label{dampedp}%
\end{equation}
respectively, for some constant\textbf{ }$\mathcal{A}\in\left]  0,1\right]
$\textbf{ }and\textbf{ }$\xi_{\mathcal{A}}\left(  \theta\right)
\approx\mathcal{A}e^{-\theta}$\textbf{. }Note that the functional form of the
probabilities in Eq. (\ref{dampedp}) obtained from our approximate information
geometric analysis based upon the integration of the geodesic equation is
consistent with that proposed in Eq. (\ref{azz12}). Such probabilities seem to
suggest a kind of critically damped quantum search and, in terms of the
standard damped harmonic oscillator solution in Eq. (\ref{critical}), they
would be linked to the case $b^{2}=4k$ (critically damped motion). This is a
quite intriguing finding that would deserve further investigation. Indeed,
observe that it is known that dissipation can be used in a constructive manner
in quantum search problems \cite{robert07, ari09}. For instance, in Ref.
\cite{ari09} it was shown that introducing dissipation into Grover's original
quantum search algorithm has positive effects because it leads to a more
robust search where the results oscillating between target and non-target
items can be damped out. This is exactly what our approximate analysis seems
to suggest. At this juncture, we remark in earnest that there are several
aspects of our investigation in need of deeper understanding. For instance,
take into consideration the phase factor $e^{i_{%
\mathbb{C}
}\phi\left(  x\right)  }$ that characterizes the wave function $\psi\left(
x\right)  \overset{\text{def}}{=}\sqrt{p\left(  x\right)  }e^{i_{%
\mathbb{C}
}\phi\left(  x\right)  }$ of a quantum state with $p\left(  x\right)
\overset{\text{def}}{=}\left\vert \psi\left(  x\right)  \right\vert ^{2}$
being a well-defined probability distribution.\textbf{ }The role played by
this factor in the differential geometric formulation of the quantum
mechanical evolution of pure states needs to be further clarified since, for
instance, the phase appears in the Fubini-Study metric. A full understanding
based upon a more rigorous analysis of the role of the phase factor together
with the monotonic decreasing behavior of dissipative quantum mechanical
systems within their information geometric characterization requires further
study and we leave it to future investigations. Once this analysis is
completed, it will be interesting to investigate the manner in which different
types of search algorithms manifest themselves from a fully quantum
information geometric standpoint. This last consideration seems to be
especially relevant due to the fact that, within the adiabatic quantum search
framework \cite{chuang16}, there exist algorithms that are fixed-point but
lack the Grover-like scaling, some are not fixed-point but exhibit the
Grover-like scaling, and some possess both features.

\section{Final Remarks}

In this article, we used methods of geometric algebra and information geometry
to enhance the algebraic efficiency and the geometrical significance of the
digital and analog representations of Grover's algorithm, respectively. A
summary of our main findings is presented below.

\subsection{Summary of results}

\begin{itemize}
\item Our analysis seems to confirm that the mathematical methods of GA and IG
cannot replace the physical intuition of the quantum algorithm developer.
However, GA is an excellent tool for assisting in the comprehension of the
algorithmic steps by mimicking them once these have been already uncovered
somehow (this is the real piece of art) by the developer. IG, in turn, seems
to yield very interesting insights. It provides fresh perspectives (its action
is not limited to mimicking) and does provide alternative intermediate
algorithmic steps that can lead to improvements with respect to the original
algorithm. Finally, its statistical nature, leads naturally to considerations
of thermodynamical nature which are essential in actual physical
implementations of any quantum computational software running on quantum hardware.

\item In reexamining the standard mathematical steps that specify Grover's
algorithm, we uncovered and fixed a typographical error that appears in
\cite{kayebook} that concerns the inversion about the mean operator. This
result appears in Eq. (\ref{iam}) and it is derived in detail in Eq.
(\ref{iamiam}) located in Appendix A.

\item Using the GA language, we presented an elegant characterization of the
Grover iterate in terms of a multivector in the reduced even subalgebra
$\left[  \mathfrak{cl}^{+}(3)\right]  ^{n}/E_{n}$. In particular, the power of
GA in handling reflections and rotations was emphasized. These results are
reported in Eqs. (\ref{gag}), (\ref{gag2}), and (\ref{gag3}). In addition, the
quadratic speedup relation was recovered in simple GA terms and appears in Eq.
(\ref{gaquadratic}). Details on the GA\ of physical space, spacetime, quantum
states and quantum operators are featured in Appendix B.

\item Quantum phenomena are generally observed only in carefully prepared
experiments in a very refined laboratory, not in a \emph{complex} Hilbert
space. Within GA, quantum states and quantum operators become united being
elements of the same real space $\left[  \mathfrak{cl}^{+}(3)\right]
^{n}/E_{n}$. This is an important conceptual unifying feature of GA that, we
argued, finds some support also in some experimental settings: \emph{unary
}representation of the implementation of Grover's algorithm with large nuclear
spins in semiconductors \cite{loss03, poggio02}, and quantum computing with
holograms \cite{miller11, alsing15}.

\item We showed that GA helps to clarify the digital-to-analog transition
descriptions of quantum computational software. Specifically, we demonstrated
that from a GA perspective, the Fenner iterate can be described in terms of
the continuos-time Grover multivector in $\left[  \mathfrak{cl}^{+}(3)\right]
^{n}/E_{n}$ which, in turn, is essentially a rotation by $2\theta$ about the
$e_{2}$ axis. This finding is reported in Eqs. (\ref{dx1}) and (\ref{dx2}).
For the sake of clarity, we recall that the angle $\theta$ equals $\sin
^{-1}\left[  \left\langle \bar{x}|\psi\right\rangle \right]  =\sin^{-1}\left[
N^{-1/2}\right]  $, while $e_{2}$ is a grade-$1$ multivector (that is, a
vector) that belongs to the Pauli algebra $\mathfrak{cl}(3)$ (for further
details, see Appendix B).

\item Considering the continuous-time description of Grover's algorithm, we
showed that the quantum search problem reduces to finding optimal\textbf{
}geodesic paths that minimize lengths on a manifold of pure density matrices
with a metric structure defined by the Wigner-Yanase metric tensor in Eq.
(\ref{gwy}). This result is reported in Eq. (\ref{GExx}).

\item We discussed the dissipationless nature of Grover's quantum search
algorithm from an IG perspective. Specifically, we computed the Fisher
information function and uncovered that it is constant (see Eq. (\ref{FF})).
Then, discussing the role of the Fisher information function in statistical
mechanics (see Eq. (\ref{statmech})) and showing its connection to a
generalized version of mechanical kinetic energy (see Eq. (\ref{J3})), we
concluded that the transport along selected geodesic paths from the initial
state toward the target state occurs without heat-like generation. Further
details on the Fisher information function and the mechanical kinetic energy
were confined to Appendix D and Appendix E.

\item The quadratic speedup behavior was recovered in pure information
geometric terms and with an arbitrary unitary operation replacing the
Walsh-Hadamard operation that appears in the original digital version of the
quantum search problem as discussed by Grover in \cite{grover97}. This finding
is reported in Eq. (\ref{finaln}).

\item A GA\ description of the sequence of nested rotations defined in terms
of generalized reflection operators that appear in Grover's fixed-point
quantum search algorithm was presented. This finding appears in Eq.
(\ref{gkk2}). We also provided an approximate IG characterization of the
dissipative nature of the fixed-point quantum search. In particular, we have
shown that the algorithm seems to be characterized by probabilities evolving
according to a damped-like harmonic oscillator type of motion. Finally, the
monotonic convergence behavior toward the target item as a consequence of the
monotonic decrease of the Fisher information is briefly addressed. Eqs.
(\ref{man}) and (\ref{dampedp}) summarize such insights in a more quantitative fashion.
\end{itemize}

\subsection{Limitations, improvements, and future investigations}

It is certainly a nontrivial task to cover in a satisfactory manner so many
different fields (quantum computing, geometric algebra, information geometry),
provide a unifying picture of a specific problem (quantum computational
software), and try to maintain some minimum link with physical implementations
occurring in laboratories. This article has been written so as to reach as
many readers (applied mathematicians, computer scientists, geometers, quantum
information theorists, statistical physicists) as possible. However, despite
the massive effort on the author's part, this work still remains preliminary
in nature. In any case, we think the relevance of this analysis is twofold.
First, there is a tutorial component. We presented a novel joint application
of special mathematical methods in the field of quantum computing. This may
inspire experts in geometric algebra and information geometry to seek new
applications for their theoretical methods within the field of quantum
information science. Second, there is research component. The effort of
recasting and analyzing in detail a known algorithm using mathematical methods
that emphasize its interpretation (geometry) and power (algebra) will
hopefully assist scientists in searching for new quantum algorithms. We think
that our hybrid geometrization of Grover's search algorithm does enhance our
intuition and visualization of the\textbf{ }art of quantum\textbf{ }algorithm
design. This, in turn, might help mitigating the non-triviality of designing
quantum algorithms. After all, this was also one of the hopes expressed by
Nielsen and collaborators \cite{nielsen06}. Possible future investigations can
be outlined as follows.

\begin{itemize}
\item First, it might be interesting to use GA and IG to explore departures
from the original set of working hypotheses that characterize Grover's quantum
search algorithm: from a unique to multiple targets, from a uniform to an
arbitrary initial amplitude distribution, from a flat superposition to an
entangled superposition for the initial state, from a power of two to an
arbitrary dimensionality of the search space. Given the enhanced algebraic
efficiency and the strengthened geometrical intuition provided by GA and IG,
this type of work would be\textbf{ }useful for checking in a faster manner the
consequences of various modifications\textbf{ }to the algorithm. Furthermore,
since GA and IG provide different mathematical structures,\textbf{ }they may
suggest different types of variations that may lead unexpectedly to searching
for new and yet unknown quantum algorithms. We leave such a psychological
excitement to future investigations.

\item Second, inspired in part by our works on quantum error correction with
both qubits and qudits \cite{carlo10, carlo12, carlo14}, we hope that the work
presented here paves the way to a new application of GA extended to quantum
search with qudits (higher-dimensional spin systems) \cite{ivanov12}. In
addition to pure theoretical motivations, we expect this line of research to
be especially relevant for enhancing our comprehension of the connection
between theory and experiment and, more specifically, the connection between
different experimental implementations of quantum search algorithms
with\ ordinary two-dimensional spin systems and higher-dimensional spin
systems \cite{loss03}.

\item Third, motivated by our thermodynamical analysis of information transfer
in complex systems \cite{cafaropre} and by our works on entropic
investigations of quantum error correction in the presence of imperfections
\cite{peter13, peter14}, we hope that the work presented here paves the way to
a thorough thermodynamic analysis of information geometric nature applied to
computational software used for search problems in quantum computing. We
expect this type of work to be especially important for finding optimal
algorithms where the thermodynamically-based concept of optimality has a
physical meaning directly related to the actual experimental realization of
the algorithm itself. As mentioned earlier, we are confident that this line of
research will further advance our understanding of the nontrivial connection
between experiment and theory in quantum computing.

\item Fourth, given the highly debated role of entanglement in quantum search
\cite{ekert98, lloyd99, braunstein02, jozsa03, vedral10} and following the
previously proposed (experimentally testable) measure of entanglement between
two quantum states described by density matrices from a statistical point of
view \cite{vedral97A, vedral97B}, it would be intriguing to investigate the
role of entanglement in the quantum search problem recast into information
geometric terms.
\end{itemize}

In conclusion, we hope that our work will have the attention of other
scientists with cross-disciplinary interests and will lead to both
technologically feasible and experimentally testable new significant advances
in quantum computing guided by physical intuition combined with powerful
mathematical tools.

\begin{acknowledgments}
The author is grateful to the\emph{ United States Air Force Research
Laboratory} (AFRL) Summer Faculty Fellowship Program for providing support for
this work. Any opinions, findings and conclusions or recommendations expressed
in this manuscript are those of the author and do not necessarily reflect the
views of AFRL. The author thanks Dr. Sean A. Ali for careful reading of the
manuscript. Furthermore, constructive criticisms from an anonymous referee
leading to an improved version of this manuscript are sincerely acknowledged
by the author.\textbf{ }Finally, the author thanks his lab advisor, Dr. Paul
M.\ Alsing, for insightful discussions and hospitality while at the AFRL in
Rome (New York).
\end{acknowledgments}

\bigskip\bigskip\pagebreak

\appendix

\section{Inversion about the mean operator}

In this Appendix, we wish to show that the action of the operator $2\left\vert
\psi\right\rangle \left\langle \psi\right\vert -I_{\mathcal{H}_{2}^{n}}$ that
appears in Eq. (\ref{ut}) on an arbitrary vector $\left\vert \phi\right\rangle
$ in $\mathcal{H}_{2}^{n}$,%
\begin{equation}
\left\vert \phi\right\rangle =\sum_{x=0}^{N-1}\alpha_{x}\left\vert
x\right\rangle \text{,}%
\end{equation}
with $\alpha_{x}\in%
\mathbb{C}
$ for any $x\in\mathcal{X}$ is given by,%
\begin{equation}
\left(  2\left\vert \psi\right\rangle \left\langle \psi\right\vert
-I_{\mathcal{H}_{2}^{n}}\right)  \left\vert \phi\right\rangle =\sum
_{x=0}^{N-1}\left(  2\mu_{\alpha}-\alpha_{x}\right)  \left\vert x\right\rangle
\text{,}%
\end{equation}
where $\mu_{\alpha}$ is the mean of the amplitudes defined as,%
\begin{equation}
\mu_{\alpha}\overset{\text{def}}{=}\frac{1}{N}\sum_{x=0}^{N-1}\alpha
_{x}\text{.}%
\end{equation}
Recall that $2\left\vert \psi\right\rangle \left\langle \psi\right\vert
-I_{\mathcal{H}_{2}^{n}}=H^{\otimes n}U_{\bar{x}=0}H^{\otimes n}$, therefore%
\begin{align}
\left(  2\left\vert \psi\right\rangle \left\langle \psi\right\vert
-I_{\mathcal{H}_{2}^{n}}\right)  \left\vert \phi\right\rangle  &  =\left(
H^{\otimes n}U_{\bar{x}=0}H^{\otimes n}\right)  \left(  \sum_{x=0}^{N-1}%
\alpha_{x}\left\vert x\right\rangle \right) \nonumber\\
&  =H^{\otimes n}U_{\bar{x}=0}\left(  \frac{1}{\sqrt{N}}\sum_{x=0}^{N-1}%
\alpha_{x}\sum_{y=0}^{N-1}\left(  -1\right)  ^{xy}\left\vert y\right\rangle
\right) \nonumber\\
&  =\frac{1}{\sqrt{N}}H^{\otimes n}\left(  \sum_{x,y=0}^{N-1}\left(
-1\right)  ^{xy}\alpha_{x}U_{\bar{x}=0}\left\vert y\right\rangle \right)
\nonumber\\
&  =\frac{1}{\sqrt{N}}H^{\otimes n}\left[  \sum_{x=0}^{N-1}\alpha_{x}%
\sum_{y=0}\left(  -1\right)  ^{xy}U_{\bar{x}=0}\left\vert y\right\rangle
+\sum_{x=0}^{N-1}\alpha_{x}\sum_{y=1}^{N-1}\left(  -1\right)  ^{xy}U_{\bar
{x}=0}\left\vert y\right\rangle \right] \nonumber\\
&  =\frac{1}{\sqrt{N}}H^{\otimes n}\left[  \sum_{x=0}^{N-1}\alpha
_{x}\left\vert 0\right\rangle -\sum_{x=0}^{N-1}\sum_{y=1}^{N-1}\left(
-1\right)  ^{xy}\alpha_{x}\left\vert y\right\rangle \right] \nonumber\\
&  =\frac{1}{\sqrt{N}}\sum_{x=0}^{N-1}\alpha_{x}H^{\otimes n}\left\vert
0\right\rangle -\frac{1}{\sqrt{N}}\sum_{x=0}^{N-1}\sum_{y=1}^{N-1}\left(
-1\right)  ^{xy}\alpha_{x}H^{\otimes n}\left\vert y\right\rangle \nonumber\\
&  =\frac{1}{\sqrt{N}}\sum_{x=0}^{N-1}\alpha_{x}\left[  \frac{1}{\sqrt{N}}%
\sum_{x=0}^{N-1}\left\vert x\right\rangle \right]  -\frac{1}{\sqrt{N}}\left\{
\sum_{x=0}^{N-1}\alpha_{x}\left[  \sqrt{N}\left\vert x\right\rangle
-H\left\vert 0\right\rangle \right]  \right\} \nonumber\\
&  =\sum_{x=0}^{N-1}\mu_{\alpha}\left\vert x\right\rangle -\frac{1}{\sqrt{N}%
}\left[  \sqrt{N}\sum_{x=0}^{N-1}\alpha_{x}\left\vert x\right\rangle -\left(
\sum_{x=0}^{N-1}\alpha_{x}\right)  \left(  H\left\vert 0\right\rangle \right)
\right] \nonumber\\
&  =\sum_{x=0}^{N-1}\left[  \mu_{\alpha}\left\vert x\right\rangle +\frac
{1}{\sqrt{N}}\alpha_{x}H\left\vert 0\right\rangle \right]  -\sum_{x=0}%
^{N-1}\alpha_{x}\left\vert x\right\rangle \nonumber\\
&  =\sum_{x=0}^{N-1}\mu_{\alpha}\left\vert x\right\rangle +\frac{1}{\sqrt{N}%
}\sum_{x=0}^{N-1}\alpha_{x}H\left\vert 0\right\rangle -\sum_{x=0}^{N-1}%
\alpha_{x}\left\vert x\right\rangle \nonumber\\
&  =\sum_{x=0}^{N-1}\mu_{\alpha}\left\vert x\right\rangle +\sum_{x=0}^{N-1}%
\mu_{\alpha}\left\vert x\right\rangle -\sum_{x=0}^{N-1}\alpha_{x}\left\vert
x\right\rangle \nonumber\\
&  =\sum_{x=0}^{N-1}\left(  2\mu_{\alpha}-\alpha_{x}\right)  \left\vert
x\right\rangle \text{,} \label{a4}%
\end{align}
that is,%
\begin{equation}
\left(  2\left\vert \psi\right\rangle \left\langle \psi\right\vert
-I_{\mathcal{H}_{2}^{n}}\right)  \left\vert \phi\right\rangle =\sum
_{x=0}^{N-1}\left(  2\mu_{\alpha}-\alpha_{x}\right)  \left\vert x\right\rangle
\text{.} \label{iamiam}%
\end{equation}
Note that the expression $xy$ in Eq. (\ref{a4}) with $x$ and $y$ in $\left\{
0,1\right\}  ^{n}$ denotes the bitwise inner product of $x$ and $y$, modulo
$2$. The possibility of reducing to modulo $2$ is due to the fact that
$\left(  -1\right)  ^{2}=1$. Furthermore, for the sake of clarity, we point
out that for $n=2$ and $N=4$ we have that $\mathcal{X}\overset{\text{def}}%
{=}\left\{  0,1,2,3\right\}  $ with $x=0\leftrightarrow00$,
$x=1\leftrightarrow01$, $x=2\leftrightarrow10$, and $x=3\leftrightarrow11$.

\section{Elements of geometric algebra}

In this Appendix, we introduce basic elements of GA of physical space and
spacetime \cite{doran}. These two geometric algebras shall be denoted as
$\mathfrak{cl}(3)$ and $\mathfrak{cl}\left(  1,3\right)  $, respectively.
Finally, we briefly present the GA formalism for quantum states and quantum operators.

\subsection{Physical space}

The key idea in GA is that of uniting the inner and outer products into a
single product, namely the \textit{geometric product. }This product is
associative and invertible. Consider two three-dimensional vectors $a$ and
$b$. The geometric product between them is defined as,%
\begin{equation}
ab\overset{\text{def}}{=}a\cdot b+a\wedge b\text{,}%
\end{equation}
where $a\cdot b$ is a scalar (a $0$-grade multivector), while $a\wedge
b=i(a\times b)$ is a bivector (a grade-$2$ multivector). The quantity $i$ is
not the imaginary unit $i_{%
\mathbb{C}
}$ usually employed in physics and is known as the unit \textit{pseudoscalar}.
Classical physics takes places in the three-dimensional Euclidean space $%
\mathbb{R}
^{3}$. Adding and multiplying vectors generate a geometric algebra denoted
$\mathfrak{cl}(3)$. The whole algebra can be generated by a right-handed set
of orthonormal vectors $\left\{  e_{1}\text{, }e_{2}\text{, }e_{3}\right\}  $
fulfilling the relation,%
\begin{equation}
e_{l}e_{m}=e_{l}\cdot e_{m}+e_{l}\wedge e_{m}=\delta_{lm}+\varepsilon
_{lmk}ie_{k}\text{,} \label{relation}%
\end{equation}
where $i$ is the above-mentioned unit three-dimensional pseudoscalar,%
\begin{equation}
i\equiv i_{\mathfrak{cl}(3)}\overset{\text{def}}{=}e_{1}e_{2}e_{3}\text{.}%
\end{equation}
We point out that Eq. (\ref{relation}) exhibits the same algebraic relations
as Pauli's $\sigma$-matrices. Indeed, the Pauli matrices constitute a
representation of the Clifford algebra $\mathfrak{cl}(3)$, also called the
Pauli algebra. The GA of physical space $\mathfrak{cl}(3)$ is generated by a
scalar, three vectors, three bivectors (area elements) and a trivector (volume
element). Therefore, $\mathfrak{cl}(3)$ is a eight-dimensional linear space,%
\begin{equation}
\dim_{%
\mathbb{R}
}\left[  \mathfrak{cl}(3)\right]  =%
{\displaystyle\sum\limits_{k=0}^{3}}
\dim\left[  \mathfrak{cl}^{\left(  k\right)  }(3)\right]  =%
{\displaystyle\sum\limits_{k=0}^{3}}
\binom{3}{k}=2^{3}=8
\end{equation}
where $\mathfrak{cl}^{\left(  k\right)  }(3)$ are the $\binom{3}{k}%
$-dimensional subspaces of $\mathfrak{cl}(3)$ spanned by the grade-$k$
multivectors in the algebra. A basis set $\mathcal{B}_{\mathfrak{cl}(3)}$ of
$\mathfrak{cl}(3)$ is,%
\begin{equation}
\mathcal{B}_{\mathfrak{cl}(3)}\overset{\text{def}}{=}\left\{  1\text{; }%
e_{1}\text{, }e_{2}\text{, }e_{3}\text{; }e_{1}e_{2}\text{, }e_{2}e_{3}\text{,
}e_{3}e_{1}\text{; }e_{1}e_{2}e_{3}\right\}  \text{.}%
\end{equation}
An arbitrary multivector $M\in$ $\mathfrak{cl}(3)$ can be decomposed as,%
\begin{align}
M  &  =%
{\displaystyle\sum\limits_{k=0}^{3}}
\left\langle M\right\rangle _{k}=\left\langle M\right\rangle _{0}+\left\langle
M\right\rangle _{1}+\left\langle M\right\rangle _{2}+\left\langle
M\right\rangle _{3}\nonumber\\
&  =\alpha+a+ib+i\beta=\text{scalar}+\text{ vector}+\text{ bivector}+\text{
trivector.} \label{multi}%
\end{align}
The quantities $\alpha$ and $\beta$ in\ Eq. (\ref{multi}) are \emph{real}
scalars while $a=a\cdot e^{k}e_{k}$ and $b=$ $b\cdot e^{k}e_{k}$ are vectors.
The quantity $\left\langle M\right\rangle _{k}$ is the grade-$k$
multivectorial part of the nonhomogeneous multivector $M\in$ $\mathfrak{cl}%
(3)$. The unit pseudoscalar $i$ is uniquely generated by the frame $\left\{
e_{1}\text{, }e_{2}\text{, }e_{3}\right\}  $ and it represents an oriented
unit volume. Furthermore, it is such that%
\begin{equation}
i^{2}=-1\text{, }iM=Mi\text{ \ }\forall M\in\mathfrak{cl}(3)\text{, and
}i^{\dagger}=-i\text{.} \label{reversion}%
\end{equation}
The symbol $\dagger$ in Eq. (\ref{reversion}) is called \textit{reversion or
Hermitian adjoint}. Within the Pauli algebra the operation of reversion plays
the role of complex conjugation\textit{ }in\textit{\ }$%
\mathbb{C}
$. For further details, we refer to \cite{doran}.

\subsection{Spacetime}

The spacetime algebra $\mathfrak{cl}(1,3)$ is the GA of Minkowski spacetime.
It is generated by four orthogonal basis vectors $\left\{  \gamma_{\mu
}\right\}  _{\mu=0,1,2,3}$ satisfying the relations%
\begin{equation}
\gamma_{\mu}\cdot\gamma_{\nu}=\frac{1}{2}\left(  \gamma_{\mu}\gamma_{\nu
}+\gamma_{\nu}\gamma_{\mu}\right)  =\text{diag}(+---)\text{, } \label{dirac1}%
\end{equation}
and,%
\begin{equation}
\gamma_{\mu}\wedge\gamma_{\nu}=\frac{1}{2}\left(  \gamma_{\mu}\gamma_{\nu
}-\gamma_{\nu}\gamma_{\mu}\right)  \text{,} \label{dirac2}%
\end{equation}
with $\mu$, $\nu\in\left\{  0,1,2,3\right\}  $. We note that Eqs.
(\ref{dirac1}) and (\ref{dirac2}) exhibit the same algebraic relations as
Dirac's $\gamma$-matrices. Indeed, the Dirac matrices constitute a
representation of the spacetime algebra $\mathfrak{cl}(1,3)$. From Eqs.
(\ref{dirac1}) and (\ref{dirac2}), it is straightforward to verify that%
\begin{equation}
\gamma_{0}^{2}=1\text{, }\gamma_{0}\cdot\gamma_{i}=0\text{, and }\gamma
_{i}\cdot\gamma_{j}=-\delta_{ij}\text{,}%
\end{equation}
with $i$, $j\in\left\{  1,2,3\right\}  $. The spacetime Clifford algebra
$\mathfrak{cl}(1,3)$ is a sixteen-dimensional linear space and a suitable
basis $\mathcal{B}_{\mathfrak{cl}(1,3)}$ is given by,%
\begin{equation}
\mathcal{B}_{\mathfrak{cl}(1,3)}\overset{\text{def}}{=}\left\{  1,\gamma_{\mu
},\gamma_{\mu}\wedge\gamma_{\nu},i\gamma_{\mu},i\right\}  \text{,}%
\end{equation}
whose elements represent scalars, vectors, bivectors, trivectors and
pseudoscalars, respectively. A general multivector $M\in\mathfrak{cl}(1,3)$
can be decomposed as%
\begin{equation}
M=%
{\displaystyle\sum\limits_{k=0}^{4}}
\left\langle M\right\rangle _{k}=\alpha+a+B+ib+i\beta\text{,}%
\end{equation}
where $\alpha$ and $\beta$ are \emph{real} scalars, $a$ and $b$\ are
\emph{real} spacetime vectors and $B$ is a bivector. A general spacetime
vector $a\in\mathfrak{cl}(1,3)$ can be written as%
\begin{equation}
a=a^{\mu}\gamma_{\mu}\text{,}%
\end{equation}
with $\mu\in\left\{  0,1,2,3\right\}  $. By selecting $\gamma_{0}$ as the
future-pointing timelike unit vector, the $\gamma_{0}$-vector determines a map
between spacetime vectors $a$ and the even subalgebra $\mathfrak{cl}%
^{+}\left(  1,3\right)  $ of the full spacetime algebra $\mathfrak{cl}(1,3)$
via the relation%
\begin{equation}
a\gamma_{0}=a_{0}+\mathbf{a}\text{,}%
\end{equation}
where $a_{0}\overset{\text{def}}{=}a\cdot\gamma_{0}$ and $\mathbf{a}%
\overset{\text{def}}{=}a\wedge\gamma_{0}$. Observe that the ordinary
three-dimensional vector $\mathbf{a}\in\mathfrak{cl}(3)$ becomes a spacetime
bivector $a\wedge\gamma_{0}$ in $\mathfrak{cl}(1,3)$. \ We remark that the
metric structure of the space whose GA is being built reflects the properties
of the unit pseudoscalar of the algebra. Indeed, the existence of a
pseudoscalar is equivalent to the existence of a metric. In $\mathfrak{cl}%
(1,3)$ the unit pseudoscalar is the highest-grade element,%
\begin{equation}
i\equiv i_{\mathfrak{cl}(1,3)}\overset{\text{def}}{=}\gamma_{0}\gamma
_{1}\gamma_{2}\gamma_{3}\text{,} \label{ispacetime}%
\end{equation}
and it represents an oriented unit four-dimensional volume element. Since $i$
in Eq. (\ref{ispacetime}) can be constructed from a right-handed vector basis
by the oriented product $\gamma_{0}\gamma_{1}\gamma_{2}\gamma_{3}$, the
corresponding volume element is said to be right-handed. The volume element
$i$ has magnitude $\left\vert i\right\vert =\left\langle i^{\dagger
}i\right\rangle _{0}^{\frac{1}{2}}=1$, where $\left\langle M\right\rangle
_{0}$ denotes the $0$-grade component of the multivector $M\in\mathfrak{cl}%
(1,3)$. The dagger $\dagger$ is the \textit{reverse} or the \textit{Hermitian
adjoint}.\textit{\ }For example, given a multivector $M=\gamma_{2}\gamma_{3}$,
$M^{\dagger}$ is obtained by reversing the order of vectors in the product.
That is, $M^{\dagger}=$\textit{\ }$\gamma_{3}\gamma_{2}=-\gamma_{2}\gamma_{3}$
. It is commonly said that $i$ specifies an orientation for spacetime. The
pseudoscalar in Eq. (\ref{ispacetime}) satisfies $i^{2}=\pm1$ with the sign
depending on the dimension and the signature of the space whose GA is being
taken into consideration. For instance, in spaces of positive definite metric,
the pseudoscalar has magnitude $\left\vert i\right\vert =1$ while the value of
$i^{2}$ depends only on the dimension $n$ of space as $i^{2}=\left(
-1\right)  ^{n\left(  n-1\right)  /2}$. Therefore in a space of even dimension
(like the Minkowski spacetime considered here, $n=4$), $i$ anticommutes with
odd-grade multivectors and commutes with even-grade elements of the algebra,%
\begin{equation}
iM=\pm Mi
\end{equation}
where the multivector $M$ is even for ($+$) and odd for ($-$). For further
details, we refer to \cite{doran}.

\subsection{Quantum states and quantum operators}

\subsubsection{Conceptual ideas}

In what follows, we describe the minimum amount of tools necessary to follow
our investigation in Sections III, IV, and V. The multiparticle spacetime
algebra (MSTA) is the GA of a relativistic configuration space and it is used
in geometric algebra to describe states and operators acting upon them. Within
this language, \emph{complex} Hilbert spaces and the imaginary unit $i_{%
\mathbb{C}
}$ in ordinary quantum mechanics are no longer fundamental. In addition, the
ordinary quantum mechanical tensor product which is used to construct both
multiparticle states and operators acting upon them and, in addition, to
isolate the Hilbert space of different particles, is replaced by the geometric
product within the GA language. The unique feature of the MSTA\ is that it
implies a separate copy of the three spatial dimensions for each particle, as
well as the time dimension. Specifically, MSTA is the GA of $n$-particle
configuration space which, for relativistic systems, consists of $n$ copies
(each copy is a $1$-particle space) of Minkowski spacetime. A convenient basis
for the MSTA is given by the set $\left\{  \gamma_{\mu}^{a}\right\}  $, where
$\mu\in\left\{  0,1,2,3\right\}  $ labels the spacetime vector and
$a\in\left\{  1,...,n\right\}  $ labels the particle space. These basis
vectors $\gamma_{\mu}^{a}$ fulfill the orthogonality conditions $\gamma_{\mu
}^{a}\cdot\gamma_{\nu}^{b}=\delta^{ab}\eta_{\mu\nu}$ with $\eta_{\mu\nu
}\overset{\text{def}}{=}$ diag$\left(  +\text{, }-\text{, }-\text{, }-\right)
$. Because of their orthogonality, vectors from different particle spaces
anticommute. Recall that,%
\begin{equation}
\dim_{%
\mathbb{R}
}\left[  \mathfrak{cl}\left(  1\text{, }3\right)  \right]  =16\text{, }\dim_{%
\mathbb{R}
}\left[  \mathfrak{cl}\left(  3\right)  \right]  =8\text{, and }\dim_{%
\mathbb{R}
}\left[  \mathfrak{cl}^{+}\left(  3\right)  \right]  =4\text{,}%
\end{equation}
where $\mathfrak{cl}^{+}\left(  3\right)  $ is the even subalgebra of
$\mathfrak{cl}\left(  3\right)  $. A basis for the entire MSTA has $2^{4n}$
degrees of freedom, that is%
\begin{equation}
\dim_{%
\mathbb{R}
}\left[  \mathfrak{cl}\left(  1\text{, }3\right)  \right]  ^{n}=2^{4n}\text{.}%
\end{equation}
In the GA version of nonrelativistic quantum mechanics, a single absolute time
identifies all of the individual time coordinates. For each $a$, this vector
is chosen to be $\gamma_{0}^{a}$. In such a spacetime split, bivectors are
used for modelling spatial vectors relative to these timelike vectors. A basis
set of relative vectors is then defined by $e_{k}^{a}\overset{\text{def}}%
{=}\gamma_{k}^{a}\gamma_{0}^{a}$, with $k\in\left\{  1,2,3\right\}  $ and
$a\in\left\{  1,...,n\right\}  $. The set $\left\{  \sigma_{k}^{a}\right\}  $
spans the GA of relative space $\mathfrak{cl}(3)\cong\mathfrak{cl}^{+}(1$,
$3)$ for each particle space. Therefore, each particle space has a basis given
by,%
\begin{equation}
1\text{, }\left\{  e_{k}\right\}  \text{, }\left\{  ie_{k}\right\}  \text{,
}i\text{,} \label{bf}%
\end{equation}
where, suppressing the particle space indices, the pseudoscalar $i$ is defined
as $i\overset{\text{def}}{=}e_{1}e_{2}e_{3}$. Observe that the basis in Eq.
(\ref{bf}) defines the Pauli algebra, that is to say the GA of the
three-dimensional physical (Euclidean) space. However, the ordinary three
Pauli $\sigma_{k}$ are no longer regarded in GA as three matrix-valued
components of a single isospace vector, but as three independent basis vectors
$e_{k}$ for real space. Unlike spacetime basis vectors, relative vectors
$\left\{  e_{k}^{a}\right\}  $ from separate particle spaces commute,
\begin{equation}
e_{k}^{a}e_{j}^{b}=e_{j}^{b}e_{k}^{a}\text{, }\forall a\neq b\text{.}%
\end{equation}
The direct product space of $n$ copies of $\mathfrak{cl}(3)$ denoted as,%
\begin{equation}
\left[  \mathfrak{cl}(3)\right]  ^{n}\overset{\text{def}}{=}\mathfrak{cl}%
(3)\otimes...\otimes\mathfrak{cl}(3)\text{,}%
\end{equation}
is generated by $\left\{  e_{k}^{a}\right\}  $ with $k\in\left\{
1,2,3\right\}  $ and $a\in\left\{  1,...,n\right\}  $. Within the MSTA
setting, $1$-particle Pauli spinors are represented as elements of
$\mathfrak{cl}^{+}(3)$, the even subalgebra of the Pauli algebra
$\mathfrak{cl}(3)$ spanned by $\left\{  1\text{, }ie_{k}\right\}  $. We have,%
\begin{equation}
\dim_{%
\mathbb{R}
}\left[  \mathcal{H}_{2}^{1}\right]  =4=\dim_{%
\mathbb{R}
}\left[  \mathfrak{cl}^{+}(3)\right]  \text{.}%
\end{equation}
We point out that within MSTA, the right multiplication by $ie_{3}$ plays the
role of the \emph{complex} imaginary unit $i_{%
\mathbb{C}
}$ in conventional quantum theory. However, in the $n$-particle algebra there
are $n$-copies of $ie_{3}$, namely $ie_{3}^{a}$ with $a\in\left\{
1,...,n\right\}  $. Therefore, in order to faithfully reproduce ordinary
quantum mechanics, the right-multiplication by all of these $ie_{3}^{a}$ must
lead to the same result. This constraint requires that the $n$-particle
multivector $\psi$ must satisfy,
\begin{equation}
\psi ie_{3}^{1}=\psi ie_{3}^{2}=\text{...}=\psi ie_{3}^{n-1}=\psi ie_{3}%
^{n}\text{.} \label{ccn}%
\end{equation}
The constraint conditions in Eq. (\ref{ccn}) are obtained by introducing the
$n$-particle correlator $E_{n}$ defined as,%
\begin{equation}
E_{n}\overset{\text{def}}{=}%
{\displaystyle\prod\limits_{b=2}^{n}}
\frac{1}{2}\left(  1-ie_{3}^{1}ie_{3}^{b}\right)  \text{,} \label{en}%
\end{equation}
such that,%
\begin{equation}
E_{n}ie_{3}^{a}=E_{n}ie_{3}^{b}=J_{n}\text{, }\forall a,b\in\left\{
1,...,n\right\}  \text{.} \label{eiax}%
\end{equation}
We remark that despite the fact that $E_{n}$ in Eq. (\ref{en}) has been
defined by correlating all the spaces\textbf{ }to the space with $a=1$, the
value of $E_{n}$ does not depend on this choice. The quantity $J_{n}$ in Eq.
(\ref{eiax}) defines the \emph{complex} structure and is such that $J_{n}%
^{2}=-E_{n}$. The number of \emph{real} degrees of freedom is reduced from
$4^{n}=\dim_{%
\mathbb{R}
}\left[  \mathfrak{cl}^{+}(3)\right]  ^{n}$ to the expected $2^{n+1}=\dim_{%
\mathbb{R}
}\mathcal{H}_{2}^{n}$ via right-multiplication by the quantum correlator
$E_{n}$ which can be regarded as a projection operation. From a physical
standpoint, this projection can be interpreted as locking the phases of the
various particles together. The \emph{reduced} even subalgebra space will be
denoted by $\left[  \mathfrak{cl}^{+}(3)\right]  ^{n}/E_{n}$ and is such that,%
\begin{equation}
\dim_{%
\mathbb{R}
}\left[  \mathcal{H}_{2}^{n}\right]  =2^{n+1}=\dim_{%
\mathbb{R}
}\left\{  \left[  \mathfrak{cl}^{+}(3)\right]  ^{n}/E_{n}\right\}  \text{.}%
\end{equation}
As elements of $\mathfrak{cl}^{+}(3)$ represent $1$-particle spinors (or,
single-qubit states in $\mathcal{H}_{2}^{1}$), multivectors belonging to
$\left[  \mathfrak{cl}^{+}(3)\right]  ^{n}/E_{n}$ can be regarded as
$n$-particle spinors (or, $n$-qubit states in $\mathcal{H}_{2}^{n}$). For
further details, we refer to \cite{doran, somaroo99}.

\subsubsection{Practical rules}

In what follows, we outline the minimum amount of translation rules from
elements of \emph{complex} Hilbert spaces to elements of \emph{real} geometric
Clifford algebras needed to follow our investigation. \ First, the GA analog
of a qubit $\left\vert \psi\right\rangle \in\mathcal{H}_{2}^{1}$ is given by,%
\begin{equation}
\left\vert \psi\right\rangle =\binom{a^{0}+i_{%
\mathbb{C}
}a^{3}}{-a^{2}+i_{%
\mathbb{C}
}a^{1}}\leftrightarrow\psi=a^{0}+a^{1}ie_{1}+a^{2}ie_{2}+a^{3}ie_{3}\text{,}%
\end{equation}
where $a^{0}$, $a^{1}$, $a^{2}$, and $a^{3}$ are \emph{real} constants and the
multivector $\psi\in\mathfrak{cl}^{+}(3)$. For example, $\left\vert
\psi\right\rangle =\binom{1}{-1}$ becomes $1+ie_{2}$. For any $k=1,2,3$, the
action of a Pauli operator $\sigma_{k}$ on a qubit $\left\vert \psi
\right\rangle $ becomes,%
\begin{equation}
\sigma_{k}\left\vert \psi\right\rangle \leftrightarrow e_{k}\psi e_{3}\text{.}%
\end{equation}
The multiplication of a qubit $\left\vert \psi\right\rangle $ by the
\emph{complex} imaginary unit $i_{%
\mathbb{C}
}$ is given by,%
\begin{equation}
i_{%
\mathbb{C}
}\left\vert \psi\right\rangle \leftrightarrow\psi ie_{3}\text{.}%
\end{equation}
The complex conjugation $\left\vert \psi\right\rangle ^{\ast}$ of a qubit
$\left\vert \psi\right\rangle $ becomes,%
\begin{equation}
\left\vert \psi\right\rangle ^{\ast}\leftrightarrow e_{2}\psi e_{2}\text{.}%
\end{equation}
The inner product between a qubit $\left\vert \psi\right\rangle $ and itself
is given by,%
\begin{equation}
\left\langle \psi|\psi\right\rangle \leftrightarrow\left\langle \psi^{\dagger
}\psi\right\rangle _{0\text{-grade}}=\left(  a^{0}\right)  ^{2}+\left(
a^{k}\right)  ^{2}\text{.}%
\end{equation}
The inner product $\left\langle \psi|\phi\right\rangle $ between two qubits
$\left\vert \phi\right\rangle $ and $\left\vert \psi\right\rangle $ becomes,%
\begin{equation}
\left\langle \psi|\phi\right\rangle \leftrightarrow\left\langle \psi^{\dagger
}\phi\right\rangle _{0\text{-grade}}-\left\langle \psi^{\dagger}\phi
ie_{3}\right\rangle _{0\text{-grade}}ie_{3}\text{.}%
\end{equation}
For example, if $\left\vert \psi\right\rangle =\binom{1}{i}$ and $\left\vert
\phi\right\rangle =\binom{1}{1}$, then $\left\langle \psi|\phi\right\rangle
=1-i_{%
\mathbb{C}
}$ becomes $1-ie_{3}$. The GA analog of a pure-state density matrix
$\rho_{\text{pure, single-qubit}}=\left\vert \psi\right\rangle \left\langle
\psi\right\vert $ with $\left\vert \psi\right\rangle \in\mathcal{H}_{2}^{1}$
is given by,%
\begin{align}
\left\vert \psi\right\rangle \left\langle \psi\right\vert  &  \leftrightarrow
\rho_{\text{pure, single-qubit}}^{\left(  \text{GA}\right)  }=\psi E_{+}%
^{1}\psi^{\dagger}\nonumber\\
&  =\psi\frac{1}{2}\left(  1+e_{3}^{1}\right)  \psi^{\dagger}\nonumber\\
&  =\frac{1}{2}\left(  1+\psi e_{3}^{1}\psi^{\dagger}\right) \nonumber\\
&  =\frac{1}{2}\left(  1+s\right)  \text{,}%
\end{align}
where multivectors $E_{+}^{1}$ and $s$ are defined as,
\begin{equation}
E_{+}^{1}\overset{\text{def}}{=}\frac{1}{2}\left(  1+e_{3}^{1}\right)  \text{
and, }s\overset{\text{def}}{=}\psi e_{3}^{1}\psi^{\dagger}\text{,}%
\end{equation}
respectively. For example, if $\left\vert \psi\right\rangle =\binom{1}{-1}$,
$\left\vert \psi\right\rangle \left\langle \psi\right\vert $ becomes $1-e_{1}%
$. The GA analog of a mixed-state density matrix,%
\begin{equation}
\rho_{\text{mixed, single-qubit}}=\sum_{i=1}^{n}p_{i}\left\vert \psi
_{i}\right\rangle \left\langle \psi_{i}\right\vert \text{,}%
\end{equation}
where $\left\vert \psi_{i}\right\rangle $ is a normalized single-qubit state,
is given by%
\begin{equation}
\sum_{i=1}^{n}p_{i}\left\vert \psi_{i}\right\rangle \left\langle \psi
_{i}\right\vert \leftrightarrow\rho_{\text{mixed, single-qubit}}^{\left(
\text{GA}\right)  }=\sum_{i=1}^{n}p_{i}\psi_{i}E_{+}^{1}\psi_{i}^{\dagger
}\text{,}%
\end{equation}
that is,%
\begin{align}
\rho_{\text{mixed, single-qubit}}^{\left(  \text{GA}\right)  }  &  =\sum
_{i=1}^{n}p_{i}\psi_{i}E_{+}^{1}\psi_{i}^{\dagger}\nonumber\\
&  =\sum_{i=1}^{n}p_{i}\psi_{i}\left[  \frac{1}{2}\left(  1+e_{3}^{1}\right)
\right]  \psi_{i}^{\dagger}\nonumber\\
&  =\frac{1}{2}\sum_{i=1}^{n}\left[  \left(  p_{i}+p_{i}\psi_{i}e_{3}^{1}%
\psi_{i}^{\dagger}\right)  \right] \nonumber\\
&  =\frac{1}{2}\left(  1+\sum_{i=1}^{n}p_{i}s_{i}\right) \nonumber\\
&  =\frac{1}{2}\left(  1+S\right)  \text{,}%
\end{align}
where multivectors $S$ and $s_{i}$ are defined as,%
\begin{equation}
S\overset{\text{def}}{=}\sum_{i=1}^{n}p_{i}s_{i}\text{ and, }s_{i}%
\overset{\text{def}}{=}\psi_{i}e_{3}^{1}\psi_{i}^{\dagger}\text{,}%
\end{equation}
respectively. For example, if $p_{1}=\frac{1}{4}$, $p_{2}=\frac{3}{4}$,
$\left\vert \psi_{1}\right\rangle =\binom{1}{0}$, and $\left\vert \psi
_{2}\right\rangle =\binom{0}{1}$, then $p_{1}\left\vert \psi_{1}\right\rangle
\left\langle \psi_{1}\right\vert +p_{2}\left\vert \psi_{2}\right\rangle
\left\langle \psi_{2}\right\vert $ becomes $\frac{1}{2}\left(  1-\frac{1}%
{2}e_{3}\right)  $. The GA analog of a mixed-state density matrix,%
\begin{equation}
\rho_{\text{mixed, multi-qubit}}=\sum_{i=1}^{n}p_{i}\left\vert \psi
_{i}\right\rangle \left\langle \psi_{i}\right\vert \text{,}%
\end{equation}
where $\left\vert \psi_{i}\right\rangle $ is a normalized multi-qubit state,
is given by%
\begin{equation}
\sum_{i=1}^{n}p_{i}\left\vert \psi_{i}\right\rangle \left\langle \psi
_{i}\right\vert \leftrightarrow\rho_{\text{mixed, multi-qubit}}^{\left(
\text{GA}\right)  }=\overline{\left(  \psi E_{n}\right)  E_{+}\left(  \psi
E_{n}\right)  ^{\sim}}\text{,}%
\end{equation}
that is,%
\begin{equation}
\rho_{\text{mixed, multi-qubit}}^{\left(  \text{GA}\right)  }=\overline
{\left(  \psi E_{n}\right)  E_{+}\left(  \psi E_{n}\right)  ^{\sim}}%
=\sum_{i=1}^{n}p_{i}\left(  \psi_{i}E_{n}\right)  E_{+}\left(  \psi_{i}%
E_{n}\right)  ^{\sim}\text{,} \label{joe}%
\end{equation}
where,%
\begin{equation}
E_{n}\overset{\text{def}}{=}%
{\displaystyle\prod\limits_{k=2}^{n}}
\frac{1}{2}\left(  1-ie_{3}^{1}ie_{3}^{k}\right)  \text{ and, }E_{+}%
\overset{\text{def}}{=}E_{+}^{1}...E_{+}^{n}\text{.} \label{cort}%
\end{equation}
For any $l\in\left\{  1,...,n\right\}  $,%
\begin{equation}
E_{+}^{l}\overset{\text{def}}{=}\frac{1}{2}\left(  1+e_{3}^{l}\right)
\text{.}%
\end{equation}
In Eq. (\ref{joe}), the over-line and tilde symbols denote the
ensemble-average and the space-time reversion, respectively. In addition,
$E_{n}$ is the $n$-particle correlator while $E_{+}$ is the geometric product
of $n$-idempotents. Finally, consider a quantum state given by,%
\begin{equation}
\left\vert \psi\right\rangle \overset{\text{def}}{=}\left\vert \psi_{i_{1}%
}\text{,..., }\psi_{i_{n}}\right\rangle \in\mathcal{H}_{2}^{n}\text{.}%
\end{equation}
The corresponding GA\ element is given by,%
\begin{equation}
\psi\overset{\text{def}}{=}\psi_{i_{1}}^{1}...\psi_{i_{n}}^{n}E_{n}\in\left[
\mathfrak{cl}^{+}(3)\right]  ^{n}/E_{n}\text{,}%
\end{equation}
where $E_{n}$ is the $n$-particle correlator in Eq. (\ref{cort}). The
superscripts denote which space the multivector belongs to. Observe that
vectors from different spaces anticommute since they are orthogonal.
Furthermore, bivectors from different spaces commute. We also point out that
$i_{%
\mathbb{C}
}\left\vert \psi\right\rangle \leftrightarrow\psi J_{n}$, where $J_{n}%
\overset{\text{def}}{=}E_{n}ie_{3}^{a}$. The index $a$ can be any index in
$\left\{  1,...,n\right\}  $ with $J_{n}^{2}=-E_{n}$, $J_{n}=J_{n}E_{n}%
=E_{n}J_{n}$, and $E_{n}^{2}=E_{n}$. The (non-simple) bivector $J_{n}$ defines
the \emph{complex} structure in the MSTA. For further technical details, we
refer to \cite{doran, lasenby93, doran93, doran96, somaroo99, havel02}.

\section{The unitary time-evolution operator}

In this Appendix, we show the computational steps needed to recover Eqs.
(\ref{derivare}) and (\ref{fitt}). In what follows, we set $\hbar=1$. Observe
that,%
\begin{align}
G\left(  t\right)   &  =e^{-i_{%
\mathbb{C}
}H_{\text{Fenner}}t}\nonumber\\
&  =e^{\frac{2\beta}{\sqrt{N}}\sigma_{z}\sigma_{x}t}\nonumber\\
&  =\sum_{k=0}^{+\infty}\frac{1}{k!}\left(  \frac{2\beta}{\sqrt{N}}\sigma
_{z}\sigma_{x}t\right)  ^{k}\nonumber\\
&  =I+\frac{2\beta}{\sqrt{N}}\sigma_{z}\sigma_{x}t+\left(  \frac{2\beta}%
{\sqrt{N}}\sigma_{z}\sigma_{x}\right)  ^{2}\frac{t^{2}}{2!}+\left(
\frac{2\beta}{\sqrt{N}}\sigma_{z}\sigma_{x}\right)  ^{3}\frac{t^{3}}%
{3!}\nonumber\\
&  +\left(  \frac{2\beta}{\sqrt{N}}\sigma_{z}\sigma_{x}\right)  ^{4}%
\frac{t^{4}}{4!}+\left(  \frac{2\beta}{\sqrt{N}}\sigma_{z}\sigma_{x}\right)
^{5}\frac{t^{5}}{5!}+...\nonumber\\
&  =I+\frac{2\beta}{\sqrt{N}}\left(  \sigma_{z}\sigma_{x}\right)  t+\left(
\frac{2\beta}{\sqrt{N}}\right)  ^{2}\left(  \sigma_{z}\sigma_{x}\right)
^{2}\frac{t^{2}}{2!}+\left(  \frac{2\beta}{\sqrt{N}}\right)  ^{3}\left(
\sigma_{z}\sigma_{x}\right)  ^{3}\frac{t^{3}}{3!}\nonumber\\
&  +\left(  \frac{2\beta}{\sqrt{N}}\right)  ^{4}\left(  \sigma_{z}\sigma
_{x}\right)  ^{4}\frac{t^{4}}{4!}+\left(  \frac{2\beta}{\sqrt{N}}\right)
^{5}\left(  \sigma_{z}\sigma_{x}\right)  ^{5}\frac{t^{5}}{5!}+...\nonumber\\
&  =I+\frac{2\beta}{\sqrt{N}}\left(  \sigma_{z}\sigma_{x}\right)  t-\left(
\frac{2\beta}{\sqrt{N}}\right)  ^{2}\frac{t^{2}}{2!}-\left(  \frac{2\beta
}{\sqrt{N}}\right)  ^{3}\left(  \sigma_{z}\sigma_{x}\right)  \frac{t^{3}}%
{3!}\nonumber\\
&  +\left(  \frac{2\beta}{\sqrt{N}}\right)  ^{4}\frac{t^{4}}{4!}+\left(
\frac{2\beta}{\sqrt{N}}\right)  ^{5}\left(  \sigma_{z}\sigma_{x}\right)
\frac{t^{5}}{5!}+...\nonumber\\
&  =\left[  1-\left(  \frac{2\beta}{\sqrt{N}}\right)  ^{2}\frac{t^{2}}%
{2!}+\left(  \frac{2\beta}{\sqrt{N}}\right)  ^{4}\frac{t^{4}}{4!}+...\right]
I_{2\times2}\nonumber\\
&  +\left[  \frac{2\beta}{\sqrt{N}}t-\left(  \frac{2\beta}{\sqrt{N}}\right)
^{3}\frac{t^{3}}{3!}+\left(  \frac{2\beta}{\sqrt{N}}\right)  ^{5}\frac{t^{5}%
}{5!}+...\right]  \sigma_{z}\sigma_{x}\nonumber\\
&  =\cos\left(  \frac{2\beta}{\sqrt{N}}t\right)  I_{2\times2}+\sin\left(
\frac{2\beta}{\sqrt{N}}t\right)  \sigma_{z}\sigma_{x}\text{,}%
\end{align}
that is, finally%
\begin{equation}
G\left(  t\right)  =e^{-i_{%
\mathbb{C}
}H_{\text{Fenner}}t}=\cos\left(  \frac{2\beta}{\sqrt{N}}t\right)  I_{2\times
2}+\sin\left(  \frac{2\beta}{\sqrt{N}}t\right)  \sigma_{z}\sigma_{x}\text{.}
\label{derivarea}%
\end{equation}
Eq. (\ref{derivarea}) is equivalent to Eq. (\ref{derivare}). Furthermore, the
time-dependent quantum state $\left\vert \psi\left(  t\right)  \right\rangle
\overset{\text{def}}{=}G\left(  t\right)  \left\vert \psi\right\rangle $
becomes,%
\begin{align}
\left\vert \psi\left(  t\right)  \right\rangle  &  =G\left(  t\right)
\left\vert \psi\right\rangle \nonumber\\
&  =\cos\left(  \frac{2\beta}{\sqrt{N}}t\right)  \left\vert \psi\right\rangle
+\sin\left(  \frac{2\beta}{\sqrt{N}}t\right)  \sigma_{z}\sigma_{x}\left\vert
\psi\right\rangle \nonumber\\
&  =\cos\left(  \frac{2\beta}{\sqrt{N}}t\right)  \left\vert \psi\right\rangle
+\sin\left(  \frac{2\beta}{\sqrt{N}}t\right)  \sigma_{z}\sigma_{x}\left(
\alpha\left\vert \bar{x}\right\rangle +\beta\left\vert \psi_{\text{bad}%
}\right\rangle \right) \nonumber\\
&  =\cos\left(  \frac{2\beta}{\sqrt{N}}t\right)  \left\vert \psi\right\rangle
+\sin\left(  \frac{2\beta}{\sqrt{N}}t\right)  \left(  \beta\left\vert \bar
{x}\right\rangle -\alpha\left\vert \psi_{\text{bad}}\right\rangle \right)
\nonumber\\
&  =\cos\left(  \frac{2\beta}{\sqrt{N}}t\right)  \left\vert \psi\right\rangle
+\sin\left(  \frac{2\beta}{\sqrt{N}}t\right)  \left\{  \beta\left\vert \bar
{x}\right\rangle -\alpha\left[  \frac{1}{\beta}\left(  \left\vert
\psi\right\rangle -\alpha\left\vert \bar{x}\right\rangle \right)  \right]
\right\} \nonumber\\
&  =\cos\left(  \frac{2\beta}{\sqrt{N}}t\right)  \left\vert \psi\right\rangle
+\beta\sin\left(  \frac{2\beta}{\sqrt{N}}t\right)  \left\vert \bar
{x}\right\rangle -\frac{\alpha}{\beta}\sin\left(  \frac{2\beta}{\sqrt{N}%
}t\right)  \left\vert \psi\right\rangle +\frac{\alpha^{2}}{\beta}\sin\left(
\frac{2\beta}{\sqrt{N}}t\right)  \left\vert \bar{x}\right\rangle \nonumber\\
&  =\left[  \cos\left(  \frac{2\beta}{\sqrt{N}}t\right)  -\frac{\alpha}{\beta
}\sin\left(  \frac{2\beta}{\sqrt{N}}t\right)  \right]  \left\vert
\psi\right\rangle +\frac{1}{\beta}\sin\left(  \frac{2\beta}{\sqrt{N}}t\right)
\left\vert \bar{x}\right\rangle \text{,}%
\end{align}
that is,%
\begin{equation}
\left\vert \psi\left(  t\right)  \right\rangle =\left[  \cos\left(
\frac{2\beta}{\sqrt{N}}t\right)  -\frac{\alpha}{\beta}\sin\left(  \frac
{2\beta}{\sqrt{N}}t\right)  \right]  \left\vert \psi\right\rangle +\frac
{1}{\beta}\sin\left(  \frac{2\beta}{\sqrt{N}}t\right)  \left\vert \bar
{x}\right\rangle \text{.} \label{fitta}%
\end{equation}
Eq. (\ref{fitta}) is equivalent to Eq. (\ref{fitt}). As a consistency check,
note that%
\begin{equation}
\left\langle \psi\left(  t\right)  |\psi\left(  t\right)  \right\rangle
=\left[  \cos\left(  x\right)  -\frac{\alpha}{\beta}\sin\left(  x\right)
\right]  ^{2}+\left[  \frac{1}{\beta}\sin\left(  x\right)  \right]  ^{2}%
+\frac{2\alpha}{\beta}\sin\left(  x\right)  \left[  \cos\left(  x\right)
-\frac{\alpha}{\beta}\sin\left(  x\right)  \right]  =1\text{,}%
\end{equation}
where $x\overset{\text{def}}{=}\frac{2\beta}{\sqrt{N}}t$, and $\alpha
=\left\langle \bar{x}|\psi\right\rangle =\left\langle \psi|\bar{x}%
\right\rangle $.

\section{On the Fisher information}

In this Appendix, we discuss the meaning of the Fisher information function
from both an information-theoretical and statistical mechanical standpoints.

\subsection{Information Theory}

Let us recall that the Fisher information\textbf{ }$\mathcal{F}\left(
\theta\right)  $ is a measure of the amount of information that an observable
random variable\textbf{ }$X$\textbf{ }carries about an unknown
parameter\textbf{ }$\theta$\textbf{ }upon which the probability\textbf{
}$p\left(  x|\theta\right)  =p_{\theta}\left(  x\right)  $\textbf{ }depends
and is defined as,%
\begin{equation}
\mathcal{F}\left(  \theta\right)  \overset{\text{def}}{=}\left\langle \left(
\frac{\partial\log p\left(  x|\theta\right)  }{\partial\theta}\right)
^{2}\right\rangle =\int p\left(  x|\theta\right)  \left(  \frac{\partial\log
p\left(  x|\theta\right)  }{\partial\theta}\right)  ^{2}dx\text{.}
\label{djuly}%
\end{equation}
In what follows, we assume to consider continuous random variables. The
probability\textbf{ }$p\left(  x|\theta\right)  $\textbf{ }is known as the
likelihood function while the quantity\textbf{ }$\partial_{\theta}\log
p\left(  x|\theta\right)  $\textbf{ }with\textbf{ }$\partial_{\theta}%
\overset{\text{def}}{=}\frac{\partial}{\partial\theta}$\textbf{ }is known as
the score. Observe that the expectation value of the score is zero. Indeed,
from the normalization condition%
\begin{equation}
\int p\left(  x|\theta\right)  dx=1\text{,}%
\end{equation}
it follows that,%
\begin{equation}
\frac{\partial}{\partial\theta}\left(  \int p\left(  x|\theta\right)
dx\right)  =0\text{.} \label{july1}%
\end{equation}
After some simple algebra, we find%
\begin{equation}
\frac{\partial\log p\left(  x|\theta\right)  }{\partial\theta}=\frac
{1}{p\left(  x|\theta\right)  }\frac{\partial p\left(  x|\theta\right)
}{\partial\theta}\text{.} \label{july2}%
\end{equation}
Using Eqs. (\ref{july1}) and (\ref{july2}), in the working hypothesis that
differentiation and integration can be interchanged, we obtain%
\begin{equation}
\left\langle \partial_{\theta}\log p\left(  x|\theta\right)  \right\rangle
=0\text{.} \label{score}%
\end{equation}
Therefore, from Eqs. (\ref{djuly}) and (\ref{score}), we conclude that the
Fisher information\textbf{ }$\mathcal{F}\left(  \theta\right)  $\textbf{ }can
be regarded as the variance of the score function. From an
information-theoretic standpoint it is convenient to observe that the Fisher
information can also be viewed as the negative of the expectation value of the
second derivative with respect to\textbf{ }$\theta$\textbf{ }of\textbf{ }$\log
p\left(  x|\theta\right)  $\textbf{,}%
\begin{equation}
\mathcal{F}\left(  \theta\right)  =-\left\langle \frac{\partial^{2}\log
p\left(  x|\theta\right)  }{\partial\theta^{2}}\right\rangle \text{.}%
\end{equation}
Indeed, note that%
\begin{align}
-\left\langle \frac{\partial^{2}\log p\left(  x|\theta\right)  }%
{\partial\theta^{2}}\right\rangle  &  =-\int p\left(  x|\theta\right)
\frac{\partial^{2}\log p\left(  x|\theta\right)  }{\partial\theta^{2}%
}dx\nonumber\\
&  =-\int p\left(  x|\theta\right)  \frac{\partial}{\partial\theta}\left[
\frac{\partial\log p\left(  x|\theta\right)  }{\partial\theta}\right]
dx\nonumber\\
&  =-\int p\left(  x|\theta\right)  \frac{\partial}{\partial\theta}\left[
\frac{1}{p\left(  x|\theta\right)  }\frac{\partial p\left(  x|\theta\right)
}{\partial\theta}\right]  dx\nonumber\\
&  =-\int p\left(  x|\theta\right)  \left[  \frac{\partial}{\partial\theta
}\left(  \frac{1}{p\left(  x|\theta\right)  }\right)  \frac{\partial p\left(
x|\theta\right)  }{\partial\theta}+\frac{1}{p\left(  x|\theta\right)  }%
\frac{\partial^{2}p\left(  x|\theta\right)  }{\partial\theta^{2}}\right]
dx\nonumber\\
&  =-\int p\left(  x|\theta\right)  \left[  -\frac{1}{p^{2}\left(
x|\theta\right)  }\left(  \frac{\partial p\left(  x|\theta\right)  }%
{\partial\theta}\right)  ^{2}+\frac{1}{p\left(  x|\theta\right)  }%
\frac{\partial^{2}p\left(  x|\theta\right)  }{\partial\theta^{2}}\right]
dx\nonumber\\
&  =\int\frac{1}{p\left(  x|\theta\right)  }\left(  \frac{\partial p\left(
x|\theta\right)  }{\partial\theta}\right)  ^{2}dx-\int\frac{\partial
^{2}p\left(  x|\theta\right)  }{\partial\theta^{2}}dx\nonumber\\
&  =\int\frac{1}{p\left(  x|\theta\right)  }p^{2}\left(  x|\theta\right)
\left(  \frac{\partial\log p\left(  x|\theta\right)  }{\partial\theta}\right)
^{2}dx-\frac{\partial^{2}}{\partial\theta^{2}}\int p\left(  x|\theta\right)
dx\nonumber\\
&  =\int p\left(  x|\theta\right)  \left(  \frac{\partial\log p\left(
x|\theta\right)  }{\partial\theta}\right)  ^{2}dx\nonumber\\
&  =\mathcal{F}\left(  \theta\right)  \text{.}%
\end{align}
This concludes the explicit verification\textbf{.}

\subsection{Statistical Mechanics}

Let us evaluate the Fisher information\textbf{ }$\mathcal{F}\left(
\theta\right)  $\textbf{ }for a thermodynamical system at equilibrium with
probability\textbf{ }$p\left(  x|\theta\right)  $\textbf{ }given by\textbf{,}%
\begin{equation}
p\left(  x|\theta\right)  =\frac{e^{-\beta E\left(  x,\theta\right)  }%
}{\mathcal{Z}\left(  \theta\right)  }\text{,} \label{jul11}%
\end{equation}
where\textbf{ }$\beta\overset{\text{def}}{=}\left(  k_{B}T\right)  ^{-1}%
$\textbf{ }and\textbf{ }$\mathcal{Z}\left(  \theta\right)  $\textbf{ }denotes
the partition function of the system defined as,%
\begin{equation}
\mathcal{Z}\left(  \theta\right)  \overset{\text{def}}{=}\int e^{-\beta
E\left(  x,\theta\right)  }dx\text{.} \label{july13}%
\end{equation}
In terms of the free energy\textbf{ }$F\left(  \theta\right)  $\textbf{, }the
partition function\textbf{ }$\mathcal{Z}\left(  \theta\right)  $ becomes%
\begin{equation}
\mathcal{Z}\left(  \theta\right)  =e^{-\beta F\left(  \theta\right)  }\text{,}%
\end{equation}
that is,%
\begin{equation}
F\left(  \theta\right)  \overset{\text{def}}{=}-k_{B}T\log\left[
\mathcal{Z}\left(  \theta\right)  \right]  \label{jul12}%
\end{equation}
Using Eqs. (\ref{jul11}) and (\ref{jul12}), the probability\textbf{ }$p\left(
x|\theta\right)  $\textbf{ }can be rewritten as%
\begin{equation}
p\left(  x|\theta\right)  =e^{\beta F\left(  \theta\right)  -\beta E\left(
x,\theta\right)  }\text{.} \label{pjuly}%
\end{equation}
Substituting Eq. \ (\ref{pjuly}) into Eq. (\ref{djuly}), after some algebra,
the Fisher information function becomes%
\begin{equation}
\mathcal{F}\left(  \theta\right)  =\beta^{2}\left\langle \left(
\frac{\partial F}{\partial\theta}-\frac{\partial E}{\partial\theta}\right)
^{2}\right\rangle \text{.} \label{pjuly12}%
\end{equation}
To further simplify Eq. (\ref{pjuly12}), let us rewrite $\partial
F/\partial\theta$ in a convenient manner. From Eqs. (\ref{july13}) and
(\ref{jul12}), we obtain%
\begin{align}
\frac{\partial F}{\partial\theta}  &  =-k_{B}T\frac{\partial}{\partial\theta
}\left\{  \log\left[  \int e^{-\beta E\left(  x,\theta\right)  }dx\right]
\right\} \nonumber\\
&  =-k_{B}T\frac{\frac{\partial}{\partial\theta}\left[  \int e^{-\beta
E\left(  x,\theta\right)  }dx\right]  }{\int e^{-\beta E\left(  x,\theta
\right)  }dx}\nonumber\\
&  =-k_{B}T\frac{\int\left(  -\beta\frac{\partial E}{\partial\theta}\right)
e^{-\beta E\left(  x,\theta\right)  }dx}{\int e^{-\beta E\left(
x,\theta\right)  }dx}\nonumber\\
&  =\int\frac{\partial E}{\partial\theta}\left(  \frac{e^{-\beta E\left(
x,\theta\right)  }}{\int e^{-\beta E\left(  x,\theta\right)  }dx}\right)
dx\nonumber\\
&  =\int\frac{\partial E}{\partial\theta}p\left(  x|\theta\right)
dx\nonumber\\
&  =\left\langle \frac{\partial E}{\partial\theta}\right\rangle \text{,}%
\end{align}
that is,%
\begin{equation}
\frac{\partial F}{\partial\theta}=\left\langle \frac{\partial E}%
{\partial\theta}\right\rangle \text{.} \label{pjuly12b}%
\end{equation}
Combining Eqs. (\ref{pjuly12}) and (\ref{pjuly12b}), the Fisher
information\textbf{ }$\mathcal{F}\left(  \theta\right)  $\textbf{ }becomes
\begin{equation}
\mathcal{F}\left(  \theta\right)  =\beta^{2}\left\langle \left(  \left\langle
\frac{\partial E}{\partial\theta}\right\rangle -\frac{\partial E}%
{\partial\theta}\right)  ^{2}\right\rangle \text{.} \label{implies}%
\end{equation}
Eq. (\ref{implies}) implies that the Fisher information of a system at thermal
equilibrium whose energy is controlled by a parameter\textbf{ }$\theta
$\textbf{ }is proportional to the variance of the infinitesimal change in
energy with respect to a change in the control parameter. In the working
hypothesis that the control parameter\textbf{ }$\theta$\textbf{ }%
equals\textbf{ }$\beta$\textbf{ }with\textbf{ }$E=E\left(  \beta\right)
$\textbf{,}%
\begin{equation}
E\left(  \beta\right)  =\left\langle E\right\rangle +\beta\left(
\frac{\partial E}{\partial\beta}-\left\langle \frac{\partial E}{\partial\beta
}\right\rangle \right)  +\mathcal{O}\left(  \beta^{2}\right)  \text{,}%
\end{equation}
it follows that the Fisher information\textbf{ }$\mathcal{F}\left(
\theta\right)  $\textbf{ }describes the size of energy fluctuations about
equilibrium,%
\begin{equation}
\mathcal{F}\left(  \theta\right)  =\sigma_{E}^{2}\overset{\text{def}}%
{=}\left\langle \left(  \left\langle E\right\rangle -E\right)  ^{2}%
\right\rangle \text{.}%
\end{equation}

\section{On the mechanical kinetic energy}

In this Appendix, we clarify the relation between the Fisher information
function and the mechanical kinetic energy.

We recall that the mechanical kinetic energy with respect to the parameter
$\theta$ is defined as,%
\begin{equation}
\mathcal{K}\left(  \theta\right)  \overset{\text{def}}{=}\sum_{m=0}%
^{N-1}\left\vert \frac{\partial\psi_{\theta}\left(  m\right)  }{\partial
\theta}\right\vert ^{2}=\sum_{m=0}^{N-1}\left\vert \dot{\psi}_{\theta}\left(
m\right)  \right\vert ^{2}\text{,} \label{kjuly20}%
\end{equation}
with $\dot{\psi}_{\theta}\left(  m\right)  \overset{\text{def}}{=}%
\partial_{\theta}\left[  \psi_{\theta}\left(  m\right)  \right]  $. The
wavefunction $\psi_{\theta}\left(  m\right)  $ is a probability amplitude
defined as,%
\begin{equation}
\psi_{\theta}\left(  m\right)  \overset{\text{def}}{=}\left\langle
m|\psi\left(  \theta\right)  \right\rangle \text{,}%
\end{equation}
with $\left\langle m|l\right\rangle =\delta_{ml}$ and the state $\left\vert
\psi\left(  \theta\right)  \right\rangle $ is given by,%
\begin{equation}
\left\vert \psi\left(  \theta\right)  \right\rangle =%
{\displaystyle\sum\limits_{m=0}^{N-1}}
\sqrt{p_{m}\left(  \theta\right)  }e^{i_{%
\mathbb{C}
}\phi_{m}\left(  \theta\right)  }\left\vert m\right\rangle \text{,}%
\end{equation}
where $p_{m}\left(  \theta\right)  \overset{\text{def}}{=}\left\vert
\psi_{\theta}\left(  m\right)  \right\vert ^{2}$. Using the polar
decomposition of the wavefunction\textbf{ }$\psi_{\theta}\left(  m\right)
$\textbf{,}%
\begin{equation}
\psi_{\theta}\left(  m\right)  =r_{\theta}\left(  m\right)  e^{i_{%
\mathbb{C}
}\phi_{\theta}\left(  m\right)  }\text{,} \label{fijuly20}%
\end{equation}
after some algebra, it follows from Eqs. (\ref{kjuly20}) and (\ref{fijuly20})
that%
\begin{equation}
\mathcal{K}\left(  \theta\right)  =\sum_{m=0}^{N-1}\left\{  \left(
\frac{\partial r_{\theta}\left(  m\right)  }{\partial\theta}\right)
^{2}+\left[  r_{\theta}\left(  m\right)  \frac{\partial\phi_{\theta}\left(
m\right)  }{\partial\theta}\right]  ^{2}\right\}  \text{.} \label{cacca3}%
\end{equation}
At this point we make two observations. First, since $p_{m}\left(
\theta\right)  =\left\vert \psi_{\theta}\left(  m\right)  \right\vert ^{2}$,
the Fisher information%
\begin{equation}
\mathcal{F}\left(  \theta\right)  \overset{\text{def}}{=}%
{\displaystyle\sum\limits_{m=0}^{N-1}}
p_{m}\left(  \theta\right)  \left(  \frac{\partial\log p_{m}\left(
\theta\right)  }{\partial\theta}\right)  ^{2}\text{,}%
\end{equation}
becomes%
\begin{equation}
\mathcal{F}\left(  \theta\right)  =4\sum_{m=0}^{N-1}\left(  \frac{\partial
r_{\theta}\left(  m\right)  }{\partial\theta}\right)  ^{2}\text{.}
\label{cacca}%
\end{equation}
Second, recalling that the quantum mechanical current density with respect to
the parameter $\theta$ is defined as,%
\begin{equation}
J_{\theta}\left(  m\right)  \overset{\text{def}}{=}\frac{1}{2i_{%
\mathbb{C}
}\left\vert \psi_{\theta}\left(  m\right)  \right\vert ^{2}}\left(
\frac{\partial\psi_{\theta}\left(  m\right)  }{\partial\theta}\psi_{\theta
}^{\ast}\left(  m\right)  -\psi_{\theta}\left(  m\right)  \frac{\partial
\psi_{\theta}^{\ast}\left(  m\right)  }{\partial\theta}\right)  \text{,}
\label{currentjuly}%
\end{equation}
inserting Eq. (\ref{fijuly20}) into Eq. (\ref{currentjuly}), we obtain
$J_{\theta}\left(  m\right)  =\partial_{\theta}\left[  \phi_{\theta}\left(
m\right)  \right]  $ and%
\begin{equation}%
{\displaystyle\sum\limits_{m=0}^{N-1}}
J_{\theta}^{2}\left(  m\right)  \left\vert \psi_{\theta}\left(  m\right)
\right\vert ^{2}=\sum_{m=0}^{N-1}\left[  r_{\theta}\left(  m\right)
\frac{\partial\phi_{\theta}\left(  m\right)  }{\partial\theta}\right]
^{2}\text{.} \label{cacca2}%
\end{equation}
Finally, from Eqs. (\ref{cacca}) and (\ref{cacca2}), we conclude that the
quantum mechanical kinetic energy $\mathcal{K}\left(  \theta\right)  $ in Eq.
(\ref{cacca3}) can be written as%
\begin{equation}
\mathcal{K}\left(  \theta\right)  =\frac{1}{4}\mathcal{F}\left(
\theta\right)  +%
{\displaystyle\sum\limits_{m=0}^{N-1}}
J_{\theta}^{2}\left(  m\right)  \left\vert \psi_{\theta}\left(  m\right)
\right\vert ^{2}\text{.}%
\end{equation}
Observe that in the case of the IG formulation of the quantum search
problem\textbf{, }$\mathcal{F}\left(  \theta\right)  =4$\textbf{, }$J_{\theta
}\left(  m\right)  =0$\textbf{ }for any\textbf{ }$0\leq m\leq N-1$\textbf{,
}and\textbf{ }$\mathcal{K}\left(  \theta\right)  $ is constantly equal to one.

\section{Optimal paths}

In this Appendix, we consider optimal paths as trajectories that minimize the
length, that is, the geodesic paths satisfying Eq. (\ref{GE}). The following
computation follows standard methods of Einstein's general relativity
\cite{weinberg}\textbf{. }

The action functional to minimize is given by,%
\begin{equation}
S\overset{\text{def}}{=}\int dS=\int\sqrt{dS^{2}}=\int\sqrt{g_{\mu\nu}\left(
\theta\right)  d\theta^{\mu}d\theta^{\nu}}\text{,}%
\end{equation}
where the classical Fisher-Rao information metric $g_{\mu\nu}\left(
\theta\right)  $ is defined as,%
\begin{equation}
g_{\mu\nu}\left(  \theta\right)  \overset{\text{def}}{=}\int dxp\left(
x|\theta\right)  \frac{\partial\log p\left(  x|\theta\right)  }{\partial
\theta^{\mu}}\frac{\partial\log p\left(  x|\theta\right)  }{\partial
\theta^{\nu}}=\int dx\frac{1}{p\left(  x|\theta\right)  }\frac{\partial
p\left(  x|\theta\right)  }{\partial\theta^{\mu}}\frac{\partial p\left(
x|\theta\right)  }{\partial\theta^{\nu}}=4\int dx\frac{\partial\sqrt{p\left(
x|\theta\right)  }}{\partial\theta^{\mu}}\frac{\partial\sqrt{p\left(
x|\theta\right)  }}{\partial\theta^{\nu}}\text{.}%
\end{equation}
We want to compute $\delta S$, that is%
\begin{equation}
\delta S=\delta\left(  \int dS\right)  =\int\delta\left(  dS\right)  \text{.}
\label{due}%
\end{equation}
Let us notice that,%
\begin{equation}
\delta\left(  dS\right)  =\frac{\delta\left(  dS^{2}\right)  }{2dS}\text{.}
\label{uno}%
\end{equation}
Thus, substituting Eq. (\ref{uno}) into Eq. (\ref{due}), we obtain%
\begin{equation}
\delta S=\int\frac{\delta\left(  dS^{2}\right)  }{2dS}\text{.} \label{tre}%
\end{equation}
Let us notice that $\delta\left(  dS^{2}\right)  $ in Eq. (\ref{tre}) may be
rewritten as,%
\begin{equation}
\delta\left(  dS^{2}\right)  =d\theta^{\mu}d\theta^{\nu}\frac{\partial
g_{\mu\nu}}{\partial\theta^{\rho}}\delta\theta^{\rho}+2g_{\mu\nu}d\theta^{\mu
}d\delta\theta^{\nu}\text{.} \label{quattro}%
\end{equation}
Substituting Eq. (\ref{quattro})\ into Eq. (\ref{tre}), the variation $\delta
S$ becomes%
\begin{equation}
\delta S=\int\left(  \frac{1}{2}\frac{d\theta^{\mu}}{dS}\frac{d\theta^{\nu}%
}{dS}\frac{\partial g_{\mu\nu}}{\partial\theta^{\rho}}\delta\theta^{\rho
}+g_{\mu\nu}\frac{d\theta^{\mu}}{dS}\frac{d\delta\theta^{\nu}}{dS}\right)
dS\text{.} \label{five}%
\end{equation}
Note that integrating by parts and imposing that $\delta\theta^{\nu}=0$ at the
extremum,\textbf{ }it happens that the second term in the integrand of Eq.
(\ref{five}) may be rewritten as,
\begin{equation}
\int g_{\mu\nu}\frac{d\theta^{\mu}}{dS}\frac{d\delta\theta^{\nu}}{dS}%
dS=g_{\mu\nu}\frac{d\theta^{\mu}}{dS}\delta\theta^{\nu}-\int\frac{d}%
{dS}\left(  g_{\mu\nu}\frac{d\theta^{\mu}}{dS}\right)  \delta\theta^{\nu
}dS=-\int\frac{d}{dS}\left(  g_{\mu\nu}\frac{d\theta^{\mu}}{dS}\right)
\delta\theta^{\nu}dS\text{.} \label{six}%
\end{equation}
Thus, substituting Eq. (\ref{six}) into Eq. (\ref{five}), the variation
$\delta S$ becomes%
\begin{equation}
\delta S=\int\left[  \frac{1}{2}\frac{d\theta^{\mu}}{dS}\frac{d\theta^{\nu}%
}{dS}\frac{\partial g_{\mu\nu}}{\partial\theta^{\rho}}\delta\theta^{\rho
}-\frac{d}{dS}\left(  g_{\mu\nu}\frac{d\theta^{\mu}}{dS}\right)  \delta
\theta^{\nu}\right]  dS\text{,}%
\end{equation}
that is,%
\begin{equation}
\delta S=\int\left[  \frac{1}{2}\frac{d\theta^{\mu}}{dS}\frac{d\theta^{\nu}%
}{dS}\frac{\partial g_{\mu\nu}}{\partial\theta^{\rho}}-\frac{d}{dS}\left(
g_{\mu\rho}\frac{d\theta^{\mu}}{dS}\right)  \right]  \delta\theta^{\rho
}dS\text{.}%
\end{equation}
Imposing $\delta S=0$ for any $\delta\theta^{\rho}$ yields,%
\begin{equation}
\frac{1}{2}\frac{d\theta^{\mu}}{dS}\frac{d\theta^{\nu}}{dS}\frac{\partial
g_{\mu\nu}}{\partial\theta^{\rho}}-\frac{d}{dS}\left(  g_{\mu\rho}%
\frac{d\theta^{\mu}}{dS}\right)  =0\text{.} \label{eq}%
\end{equation}
Let us\textbf{ }observe that,%
\begin{equation}
\frac{1}{2}\frac{d\theta^{\mu}}{dS}\frac{d\theta^{\nu}}{dS}\frac{\partial
g_{\mu\nu}}{\partial\theta^{\rho}}-\frac{d}{dS}\left(  g_{\mu\rho}%
\frac{d\theta^{\mu}}{dS}\right)  =\frac{1}{2}\frac{d\theta^{\mu}}{dS}%
\frac{d\theta^{\nu}}{dS}\frac{\partial g_{\mu\nu}}{\partial\theta^{\rho}%
}-\frac{\partial g_{\mu\rho}}{\partial\theta^{\nu}}\frac{d\theta^{\nu}}%
{dS}\frac{d\theta^{\mu}}{dS}-g_{\mu\rho}\frac{d^{2}\theta^{\mu}}{dS^{2}%
}\text{.} \label{seven}%
\end{equation}
Furthermore, note that%
\begin{equation}
\frac{\partial g_{\mu\rho}}{\partial\theta^{\nu}}\frac{d\theta^{\nu}}{dS}%
\frac{d\theta^{\mu}}{dS}=\frac{1}{2}\left(  \frac{\partial g_{\mu\rho}%
}{\partial\theta^{\nu}}+\frac{\partial g_{\nu\rho}}{\partial\theta^{\mu}%
}\right)  \frac{d\theta^{\nu}}{dS}\frac{d\theta^{\mu}}{dS}\text{.}
\label{otto}%
\end{equation}
Thus, substituting Eqs.(\ref{otto}) and (\ref{seven}) into Eq. (\ref{eq}), we
obtain%
\begin{equation}
g_{\mu\rho}\frac{d^{2}\theta^{\mu}}{dS^{2}}+\frac{1}{2}\left(  -\frac{\partial
g_{\mu\nu}}{\partial\theta^{\rho}}+\frac{\partial g_{\mu\rho}}{\partial
\theta^{\nu}}+\frac{\partial g_{\nu\rho}}{\partial\theta^{\mu}}\right)
\frac{d\theta^{\mu}}{dS}\frac{d\theta^{\nu}}{dS}=0\text{.} \label{ten}%
\end{equation}
However, recall that \cite{weinberg},%
\begin{equation}
\Gamma_{\rho\text{, }\mu\nu}\overset{\text{def}}{=}\frac{1}{2}\left(
\frac{\partial g_{\rho\mu}}{\partial\theta^{\nu}}+\frac{\partial g_{\rho\nu}%
}{\partial\theta^{\mu}}-\frac{\partial g_{\mu\nu}}{\partial\theta^{\rho}%
}\right)  \text{.} \label{nine}%
\end{equation}
Therefore substituting Eq. (\ref{nine}) into Eq. (\ref{ten}), we find%
\begin{equation}
g_{\mu\rho}\frac{d^{2}\theta^{\mu}}{dS^{2}}+\Gamma_{\rho\text{, }\mu\nu}%
\frac{d\theta^{\mu}}{dS}\frac{d\theta^{\nu}}{dS}=0\text{,}%
\end{equation}
that is,%
\begin{equation}
g^{\rho\rho}g_{\mu\rho}\frac{d^{2}\theta^{\mu}}{dS^{2}}+g^{\rho\rho}%
\Gamma_{\rho\text{, }\mu\nu}\frac{d\theta^{\mu}}{dS}\frac{d\theta^{\nu}}%
{dS}=0\text{.}%
\end{equation}
In conclusion, we obtain that the minimization of the action functional $S$
leads to the following geodesic equation,%
\begin{equation}
\frac{d^{2}\theta^{\rho}}{dS^{2}}+\Gamma_{\mu\nu}^{\rho}\frac{d\theta^{\mu}%
}{dS}\frac{d\theta^{\nu}}{dS}=0\text{,}%
\end{equation}
where $\Gamma_{\mu\nu}^{\rho}$ are the usual Christoffel connection
coefficients of the second type \cite{weinberg}.

\section{Walking on information geometric paths}

In this Appendix, we briefly describe the steps on information geometric paths
together with their size and normalization. The approximate unitarity of the
Grover operator restricted to the two-dimensional space spanned by $\left\vert
\psi_{\theta_{i}}\right\rangle $ and $U^{-1}\left\vert \psi_{\theta_{f}%
}\right\rangle $ is addressed.

\subsection{The steps}

Observe that,%
\begin{align}
\left\vert \psi_{\theta_{i+1}}\right\rangle  &  =G\left\vert \psi_{\theta_{i}%
}\right\rangle \nonumber\\
&  =\left(  1-4\left\vert U_{fi}\right\vert ^{2}\right)  \left\vert
\psi_{\theta_{i}}\right\rangle +2U_{fi}U^{-1}\left\vert \psi_{\theta_{f}%
}\right\rangle \text{.}%
\end{align}
To be explicit, we have%
\begin{align}
\left\vert \psi_{\theta_{i+1}}\right\rangle  &  =G\left\vert \psi_{\theta_{i}%
}\right\rangle \nonumber\\
&  =-\left(  1-2\left\vert \psi_{\theta_{i}}\right\rangle \left\langle
\psi_{\theta_{i}}\right\vert \right)  U^{-1}\left(  1-2\left\vert \psi
_{\theta_{f}}\right\rangle \left\langle \psi_{\theta_{f}}\right\vert \right)
U\left\vert \psi_{\theta_{i}}\right\rangle \nonumber\\
&  =-\left(  1-2\left\vert \psi_{\theta_{i}}\right\rangle \left\langle
\psi_{\theta_{i}}\right\vert \right)  U^{-1}\left[  U\left\vert \psi
_{\theta_{i}}\right\rangle -2\left\langle \psi_{\theta_{f}}|U|\psi_{\theta
_{i}}\right\rangle \left\vert \psi_{\theta_{f}}\right\rangle \right]
\nonumber\\
&  =-\left(  1-2\left\vert \psi_{\theta_{i}}\right\rangle \left\langle
\psi_{\theta_{i}}\right\vert \right)  \left[  \left\vert \psi_{\theta_{i}%
}\right\rangle -2\left\langle \psi_{\theta_{f}}|U|\psi_{\theta_{i}%
}\right\rangle \right]  U^{-1}\left\vert \psi_{\theta_{f}}\right\rangle
\nonumber\\
&  =-\left\vert \psi_{\theta_{i}}\right\rangle +2\left\vert \psi_{\theta_{i}%
}\right\rangle +2\left\langle \psi_{\theta_{f}}|U|\psi_{\theta_{i}%
}\right\rangle \left(  1-2\left\vert \psi_{\theta_{i}}\right\rangle
\left\langle \psi_{\theta_{i}}\right\vert \right)  U^{-1}\left\vert
\psi_{\theta_{f}}\right\rangle \nonumber\\
&  =\left\vert \psi_{\theta_{i}}\right\rangle +2U_{fi}U^{-1}\left\vert
\psi_{\theta_{f}}\right\rangle -4U_{fi}\left\langle \psi_{\theta_{i}}%
|U^{-1}|\psi_{\theta_{f}}\right\rangle \left\vert \psi_{\theta_{i}%
}\right\rangle \nonumber\\
&  =\left\vert \psi_{\theta_{i}}\right\rangle +2U_{fi}U^{-1}\left\vert
\psi_{\theta_{f}}\right\rangle -4U_{fi}U_{fi}^{\ast}\left\vert \psi
_{\theta_{i}}\right\rangle \nonumber\\
&  =\left(  1-4\left\vert U_{fi}\right\vert ^{2}\right)  \left\vert
\psi_{\theta_{i}}\right\rangle +2U_{fi}U^{-1}\left\vert \psi_{\theta_{f}%
}\right\rangle \text{.}%
\end{align}
Note also that,%
\begin{align}
\left\vert \psi_{\theta_{i+2}}\right\rangle  &  =G^{2}\left\vert \psi
_{\theta_{i}}\right\rangle =G\left\vert \psi_{\theta_{i+1}}\right\rangle
\nonumber\\
&  =\left(  1-4\left\vert U_{fi}\right\vert ^{2}\right)  \left\vert
\psi_{\theta_{i+1}}\right\rangle +2U_{fi}GU^{-1}\left\vert \psi_{\theta_{f}%
}\right\rangle \text{.}%
\end{align}
Furthermore, we find%
\begin{align}
\left\vert \psi_{\theta_{i+3}}\right\rangle  &  =G^{3}\left\vert \psi
_{\theta_{i}}\right\rangle =G^{2}\left\vert \psi_{\theta_{i+1}}\right\rangle
=G\left\vert \psi_{\theta_{i+2}}\right\rangle \nonumber\\
&  =\left(  1-4\left\vert U_{fi}\right\vert ^{2}\right)  \left\vert
\psi_{\theta_{i+2}}\right\rangle +2U_{fi}G^{2}U^{-1}\left\vert \psi
_{\theta_{f}}\right\rangle \text{.}%
\end{align}
Following this line of computation, we obtain%
\begin{align}
\left\vert \psi_{\theta_{i+l}}\right\rangle  &  =G^{l}\left\vert \psi
_{\theta_{i}}\right\rangle =G^{l-1}\left\vert \psi_{\theta_{i+1}}\right\rangle
=...=G\left\vert \psi_{\theta_{i+l-1}}\right\rangle \nonumber\\
&  =\left(  1-4\left\vert U_{fi}\right\vert ^{2}\right)  \left\vert
\psi_{\theta_{i+l-1}}\right\rangle +2U_{fi}G^{l-1}U^{-1}\left\vert
\psi_{\theta_{f}}\right\rangle \text{,}%
\end{align}
for any $1\leq l\leq\mathcal{N}_{s}$ with $i+\mathcal{N}_{s}\equiv f$.

\subsection{The size of the steps}

Observe that\textbf{,}%
\begin{equation}
\left\langle \psi_{\theta_{i}}|G|\psi_{\theta_{i}}\right\rangle =\left\langle
\psi_{\theta_{i+l}}|G|\psi_{\theta_{i+l}}\right\rangle \Leftrightarrow
\left\langle \psi_{\theta_{i}}|\psi_{\theta_{i+1}}\right\rangle =\left\langle
\psi_{\theta_{i+l}}|\psi_{\theta_{i+l+1}}\right\rangle \text{,}
\label{implies22}%
\end{equation}
for any $1\leq l\leq\mathcal{N}_{s}$ with $i+\mathcal{N}_{s}\equiv f$. If
$\left\langle \psi_{\theta_{i}}|G|\psi_{\theta_{i}}\right\rangle =\left\langle
\psi_{\theta_{i+l}}|G|\psi_{\theta_{i+l}}\right\rangle $, then%
\begin{equation}
\left[  ds_{\text{WY}}^{2}\right]  _{i\rightarrow i+1}=\left[  ds_{\text{WY}%
}^{2}\right]  _{i+l\rightarrow i+l+1}\overset{\text{def}}{=}4\left[
1-\left\vert \left\langle \psi_{\theta_{i+l}}|\psi_{\theta_{i+l+1}%
}\right\rangle \right\vert ^{2}\right]  \text{,} \label{dxx1}%
\end{equation}
for any $1\leq l\leq\mathcal{N}_{s}$. Let us verify that $\left\langle
\psi_{\theta_{i}}|G|\psi_{\theta_{i}}\right\rangle =\left\langle \psi
_{\theta_{i+l}}|G|\psi_{\theta_{i+l}}\right\rangle $. Note that%
\begin{equation}
G^{l}\left\vert \psi_{\theta_{i}}\right\rangle =\left\vert \psi_{\theta_{i+l}%
}\right\rangle \text{,}%
\end{equation}
implies%
\begin{equation}
\left\langle \psi_{\theta_{i+l}}\right\vert =\left\langle \psi_{\theta_{i}%
}\right\vert \left(  G^{l}\right)  ^{\dagger}=\left\langle \psi_{\theta_{i}%
}\right\vert \left(  G^{l}\right)  ^{-1}=\left\langle \psi_{\theta_{i}%
}\right\vert \left(  G^{-1}\right)  ^{l}=\left\langle \psi_{\theta_{i}%
}\right\vert G^{-l}\text{.}%
\end{equation}
Therefore,%
\begin{equation}
\left\langle \psi_{\theta_{i+l}}|G|\psi_{\theta_{i+l}}\right\rangle
=\left\langle \psi_{\theta_{i}}|G^{-l}G|\psi_{\theta_{i+l}}\right\rangle
=\left\langle \psi_{\theta_{i}}|GG^{-l}|\psi_{\theta_{i+l}}\right\rangle
=\left\langle \psi_{\theta_{i}}|G|\psi_{\theta_{i}}\right\rangle \text{.}
\label{equally}%
\end{equation}
From Eq. (\ref{implies22}) and the definition of the infinitesimal line
element in Eq. (\ref{dxx1}), we conclude that Eq. (\ref{equally}) implies that
the steps are equally spaced. We remark that while $G$ is a unitary operator
when acting on $\mathcal{H}_{2}^{n}$, it is only approximately unitary when
restricted to the two-dimensional space spanned by $\left\vert \psi
_{\theta_{i}}\right\rangle $ and\textbf{ }$U^{-1}\left\vert \psi_{\theta_{f}%
}\right\rangle $. Indeed, recalling that%
\begin{equation}
G\left\vert \psi_{\theta_{i}}\right\rangle =\left[  1-4\left\vert
U_{fi}\right\vert ^{2}\right]  \left\vert \psi_{\theta_{i}}\right\rangle
+2U_{fi}U^{-1}\left\vert \psi_{\theta_{f}}\right\rangle \text{,}%
\end{equation}
and,%
\begin{equation}
GU^{-1}\left\vert \psi_{\theta_{f}}\right\rangle =-2U_{fi}^{\ast}\left\vert
\psi_{\theta_{i}}\right\rangle +U^{-1}\left\vert \psi_{\theta_{f}%
}\right\rangle \text{,}%
\end{equation}
after some algebra, we find%
\begin{equation}
\left[  G\right]  _{\text{Span}\left\{  \left\vert \psi_{\theta_{i}%
}\right\rangle \text{, }U^{-1}\left\vert \psi_{\theta_{f}}\right\rangle
\right\}  }\overset{\text{def}}{=}\left(
\begin{array}
[c]{cc}%
\left\langle \psi_{\theta_{i}}|G|\psi_{\theta_{i}}\right\rangle  &
\left\langle \psi_{\theta_{i}}|GU^{-1}|\psi_{\theta_{f}}\right\rangle \\
\left\langle \psi_{\theta_{f}}|UG|\psi_{\theta_{i}}\right\rangle  &
\left\langle \psi_{\theta_{f}}|UGU^{-1}|\psi_{\theta_{f}}\right\rangle
\end{array}
\right)  \text{,}%
\end{equation}
that is,%
\begin{equation}
\left[  G\right]  _{\text{Span}\left\{  \left\vert \psi_{\theta_{i}%
}\right\rangle \text{, }U^{-1}\left\vert \psi_{\theta_{f}}\right\rangle
\right\}  }=\left(
\begin{array}
[c]{cc}%
1-2\left\vert U_{fi}\right\vert ^{2} & -U_{fi}^{\ast}\\
U_{fi}\left[  1-4\left\vert U_{fi}\right\vert ^{2}\right]  +2U_{fi} &
1-2\left\vert U_{fi}\right\vert ^{2}%
\end{array}
\right)  \text{.}%
\end{equation}
Observe that it is necessary to impose\textbf{ }$\left\vert U_{fi}\right\vert
\ll1$\textbf{ }in order to preserve the nature of the rotation operator
of\textbf{ }$G$,%
\begin{equation}
\det\left\{  \left[  G\right]  _{\text{Span}\left\{  \left\vert \psi
_{\theta_{i}}\right\rangle \text{, }U^{-1}\left\vert \psi_{\theta_{f}%
}\right\rangle \right\}  }\right\}  =1-\left\vert U_{fi}\right\vert
^{2}\overset{\left\vert U_{fi}\right\vert \ll1}{\approx}1\text{.}%
\end{equation}

\subsection{The normalization of the steps}

Recall that,%
\begin{equation}
\left\vert \psi_{\theta_{i+l}}\right\rangle =\left(  1-4\left\vert
U_{fi}\right\vert ^{2}\right)  \left\vert \psi_{\theta_{i+l-1}}\right\rangle
+2U_{fi}G^{l-1}U^{-1}\left\vert \psi_{\theta_{f}}\right\rangle \text{.}
\label{fiil}%
\end{equation}
Let us verify that $\left\langle \psi_{\theta_{i+l}}|\psi_{\theta_{i+l}%
}\right\rangle =1$, for any $1\leq l\leq\mathcal{N}_{s}$ with $i+\mathcal{N}%
_{s}\equiv f$. From Eq. (\ref{fiil}), we find%
\begin{align}
\left\langle \psi_{\theta_{i+l}}|\psi_{\theta_{i+l}}\right\rangle  &  =\left[
1-4\left\vert U_{fi}\right\vert ^{2}\right]  ^{2}+2U_{fi}\left[  1-4\left\vert
U_{fi}\right\vert ^{2}\right]  \left\langle \psi_{\theta_{i+l-1}}%
|G^{l-1}U^{-1}|\psi_{\theta_{f}}\right\rangle \nonumber\\
&  +2U_{fi}^{\ast}\left[  1-4\left\vert U_{fi}\right\vert ^{2}\right]
\left\langle \psi_{\theta_{f}}|UG^{1-l}|\psi_{\theta_{i+l-1}}\right\rangle
+4\left\vert U_{fi}\right\vert ^{2}\left\langle \psi_{\theta_{f}}%
|UG^{1-l}G^{l-1}U^{-1}|\psi_{\theta_{f}}\right\rangle \nonumber\\
&  =\left[  1-4\left\vert U_{fi}\right\vert ^{2}\right]  ^{2}+2U_{fi}\left[
1-4\left\vert U_{fi}\right\vert ^{2}\right]  \left\langle \psi_{\theta_{i}%
}|U^{-1}|\psi_{\theta_{f}}\right\rangle \nonumber\\
&  +2U_{fi}^{\ast}\left[  1-4\left\vert U_{fi}\right\vert ^{2}\right]
\left\langle \psi_{\theta_{f}}|U|\psi_{\theta_{i}}\right\rangle +4\left\vert
U_{fi}\right\vert ^{2}\nonumber\\
&  =\left[  1-4\left\vert U_{fi}\right\vert ^{2}\right]  ^{2}+2U_{fi}%
U_{fi}^{\ast}\left[  1-4\left\vert U_{fi}\right\vert ^{2}\right]
+2U_{fi}^{\ast}U_{fi}\left[  1-4\left\vert U_{fi}\right\vert ^{2}\right]
+4\left\vert U_{fi}\right\vert ^{2}\nonumber\\
&  =1+16\left\vert U_{fi}\right\vert ^{4}-8\left\vert U_{fi}\right\vert
^{2}+2\left\vert U_{fi}\right\vert ^{2}-8\left\vert U_{fi}\right\vert
^{4}+2\left\vert U_{fi}\right\vert ^{2}\left[  1-4\left\vert U_{fi}\right\vert
^{2}\right]  +4\left\vert U_{fi}\right\vert ^{2}\nonumber\\
&  =1+16\left\vert U_{fi}\right\vert ^{4}-8\left\vert U_{fi}\right\vert
^{2}+2\left\vert U_{fi}\right\vert ^{2}-8\left\vert U_{fi}\right\vert
^{4}+2\left\vert U_{fi}\right\vert ^{2}-8\left\vert U_{fi}\right\vert
^{4}+4\left\vert U_{fi}\right\vert ^{2}\nonumber\\
&  =1-8\left\vert U_{fi}\right\vert ^{2}+4\left\vert U_{fi}\right\vert
^{2}+4\left\vert U_{fi}\right\vert ^{2}\nonumber\\
&  =1\text{,}%
\end{align}
that is,%
\begin{equation}
\left\langle \psi_{\theta_{i+l}}|\psi_{\theta_{i+l}}\right\rangle =1\text{.}
\label{normal}%
\end{equation}
Eq. (\ref{normal}) implies that states $\left\vert \psi_{\theta_{i+l}%
}\right\rangle $ with $1\leq l\leq\mathcal{N}_{s}$ are properly normalized.
\end{document}